\documentstyle{report}

\newcommand{\beq}{\begin{equation}}
\newcommand{\eeq}{\end{equation}} 
\newcommand{\pket}{\mbox{$| \phi \rangle$}}

\newcommand{\sket}{\mbox{$| \psi \rangle$}}
\newcommand{\sbra}{\mbox{$\langle \psi | $}}
\newcommand{\lda}{\mbox{$\lambda$}}
\newcommand{\R}{\mbox{$\rm I\!R$}}

\newcommand{\V}{\mbox{$V^{\psi}_{\lda}(O)$}}
\newcommand{\Erwin}{Schr{\"o}dinger }
\newcommand{\Erwins}{Schr{\"o}dinger's }
\newcommand{\upa}{\uparrow}
\newcommand{\dwna}{\downarrow}
\newcommand{\uupa}{\!\uparrow}
\newcommand{\ddwna}{\!\downarrow}
\newcommand{\ra}{\rangle}
\newcommand{\la}{\langle}

\newcommand{\sgox}{\mbox{$\sigma^{(1)}_{x}$}}
\newcommand{\sgoy}{\mbox{$\sigma^{(1)}_{y}$}}
\newcommand{\sgtx}{\mbox{$\sigma^{(2)}_{x}$}}
\newcommand{\sgty}{\mbox{$\sigma^{(2)}_{y}$}}

\newcommand{\ks}{Kochen and Specker }
\newcommand{\kss}{Kochen and Specker's }
\newcommand{\gl}{Gleason }
\newcommand{\gls}{Gleason's }

\newcommand{\bm}{\mbox{\boldmath $\mu$}}
\newcommand{\hil}{\mbox{${\cal H}$}}

\begin{document}

\title{\textbf{Hidden Variables and Nonlocality
in Quantum Mechanics}}
\author{Douglas L. Hemmick}

\date{October 1996}

\maketitle

\begin{abstract}
\renewcommand{\thepage}{\roman{page}}
\setcounter{page}{2}
At the present time, most physicists continue to hold a 
skeptical attitude toward the proposition of a `hidden variables' 
interpretation of quantum theory, in spite of David Bohm's successful 
construction of such a 
theory and John S. Bell's  strong arguments in favor of the idea. Many are 
convinced either that it is impossible to interpret 
quantum theory in this way, or that such an interpretation would actually 
be irrelevant. There are essentially two reasons behind such 
doubts. The first concerns certain mathematical theorems (von Neumann's, 
Gleason's, Kochen and Specker's, and Bell's) which can be applied to the 
hidden variables issue. These theorems are often credited with proving 
that hidden variables are indeed `impossible', in the sense that they cannot 
replicate the predictions of quantum mechanics. Many who do not draw 
such a strong conclusion nevertheless accept that hidden variables have been 
shown to exhibit prohibitively complicated features. The second reason
hidden variables are disregarded is that the most 
sophisticated example of a hidden 
variables theory---that of David Bohm---exhibits {\em 
nonlocality}, i.e., it can happen in this theory that the consequences of 
events at one place propagate to other places instantaneously. 
However, as we shall show in the present work, neither the mathematical 
theorems in question nor the attribute of nonlocality serve to detract from 
the importance of a hidden variables interpretation of quantum theory.
The theorems imply neither that hidden variables are impossible
nor that they must be overly complex. As regards nonlocality, 
this feature is present in quantum mechanics itself, and is a
required characteristic of {\em any} theory that agrees with the
quantum mechanical predictions. 

In the present work, the hidden variables issue is addressed in the 
following ways. We first discuss the earliest analysis
of hidden variables---that of von Neumann's theorem---and review
John S. Bell's refutation of von Neumann's `impossibility proof'. 
We recall and elaborate on Bell's arguments 
regarding the theorems of Gleason, and Kochen and Specker.
According to Bell, these latter theorems do not imply that hidden variables 
interpretations are untenable, but instead that such theories must
exhibit {\em contextuality}, i.e., they must allow for the dependence of 
measurement results on
the characteristics of both measured system and measuring apparatus.
We demonstrate a new way to understand the implications of both Gleason's 
theorem and Kochen and Specker's theorem 
by noting that they prove a result we call ``spectral 
incompatibility''. We develop further insight into the concepts
involved in these two theorems by investigating 
a special quantum mechanical experiment which was 
first described by David Albert. We review the 
Einstein--Podolsky--Rosen paradox, Bell's theorem, and Bell's
later argument 
that these imply that quantum mechanics is irreducibly
nonlocal. 
We present this discussion in a somewhat more 
gradual fashion than does Bell so that the logic of the argument 
may be more transparent.

The paradox of Einstein, Podolsky, and Rosen was generalized by Erwin
\Erwin in the same paper where his famous `cat paradox' appeared. 
We develop several new results regarding this generalization. 
We show that \Erwins conclusions can be derived using a simpler 
argument---one which
makes clear the relationship between the quantum state and the
`perfect correlations' exhibited by the system. We use 
\Erwins EPR analysis 
to derive a wide variety of new quantum nonlocality proofs. These
proofs share two important features with that of Greenberger, Horne, and 
Zeilinger. First, they are of a deterministic character, i.e., 
they are `nonlocality without inequalities' proofs. Second, as we 
shall show, the quantum
nonlocality results we develop may be experimentally verified 
in such a way that one need only observe 
the `perfect correlations' between the appropriate observables; no 
further tests are required. This latter feature serves to contrast 
these proofs with EPR/Bell nonlocality, the
laboratory confirmation of which demands not only the observation of perfect 
correlations, but also an additional set of observations, namely those
required to
test whether `Bell's inequality' is violated. 
The `Schr{\"o}dinger
 nonlocality' proofs we give differ from the GHZ proof in that they 
apply to two-component composite 
systems, while the latter involves a composite system of
at least three-components. In addition, some of the
\Erwin proofs involve 
classes of observables larger than that addressed in the GHZ proof.

\end{abstract}

\renewcommand{\thepage}{\roman{page}} 

 \vspace*{0.5 in} 

\begin{center}
{\bf Acknowledgments }
\end{center}

 \vspace*{0.2 in}
 
 I would like to express my profound gratitude to my advisor, Sheldon Goldstein, for his very kind and generous guidance. It is impossible to imagine a research advisor who is more inspiring to work with than Shelly. I would also like to express thanks to Joel Lebowitz for his support and encouragement. It has been a most rewarding experience to work within the friendly and stimulating environment of the Rutgers mathematical physics group. For many helpful discussions, I thank some of my peers with whom I worked at Rutgers: Karin Berndl, Martin Daumer, Mahesh Yadav, Nadia Topor, and Subir Ghoshal. My thanks are extended also to Richard McKenzie, Andrew Pica, Asif Shakur, David Kanarr, and Gail Welsh of Salisbury State University for their interest in this work, and for providing an opportunity to present talks on quantum mechanics to physics student audiences. I would like to thank especially Asif Shakur, who has helped proofread the manuscript and has given valuable advice toward the presentation of the material. For their continuous support and zealous help with the proofreading, I thank my parents. Finally, I would like to thank Melissa Thomas, Tom Hemmick, and Raju Farouk for aiding me with the computer communications techniques which made it possible to complete this work while living away from campus. This work was supported by a Rutgers University Excellence Fellowship.
\renewcommand{\thepage}{\roman{page}} 

 \vspace*{0.5 in} 

\begin{center}
{\bf Dedication }
\end{center}

 \vspace*{0.2 in} 
 
 This dissertation is dedicated to the memory of Geordie Williams, whose energy and enthusiasm made him an inspiration to all of his friends from the old Joppatowne group.
\vspace*{0.5 in}
\begin{quotation}
``Attention has recently been called to the obvious but very disconcerting
fact that even though we restrict the disentangling measurements to {\em
one} system, the representative obtained for the {\em other} system is by
no means independent of the particular choice of observations which
we select for that purpose and which by the way are {\em entirely}
arbitrary.''
\end{quotation}
 Erwin Schr{\"o}dinger \cite{Camb1}

\renewcommand{\thepage}{\roman{page}}

\renewcommand{\thepage}{\roman{page}}
\tableofcontents
\chapter{Introduction} 
\section{The issue of hidden variables}
\subsection{Contextuality, nonlocality, and hidden variables}
\renewcommand{\thepage}{\arabic{page}}
\setcounter{page}{1}
The aim of this thesis is to contribute to the issue of hidden variables as a
viable interpretation of quantum mechanics.  Our efforts will be directed 
toward the examination of certain mathematical theorems that are relevant to 
this issue. We  shall examine the relationship these theorems bear toward 
hidden variables and the lessons they provide regarding quantum mechanics 
itself. The theorems in question include those of John von Neumann
\cite{von Neumann}, A. M. Gleason \cite{Gleason}, J.S. Bell \cite{Bell's
theorem}, and S. Kochen and E. P. Specker \cite{Kochen Specker}.  Arguments 
given by John Stewart 
Bell\footnote{See references by Bell: 
\cite{Bell Eclipse, Bell Imposs Pilot} on the theorems of von Neumann, Gleason,
and Kochen and Specker. Bell's works 
emphasize the conclusion that these theorems would place no serious 
limitations on a hidden variables theory, and he declares that
\cite{Bell Imposs Pilot} ``What is proved by impossibility proofs [these 
theorems] \ldots \ldots is lack of imagination.'' A recent work by
Mermin 
\cite{Mermin Review} 
also addresses the impact of these theorems on hidden variables. Mermin 
does not accept Bell's conclusions, however, and states ``Bell \ldots is 
unreasonably dismissive of the importance of \ldots impossibility 
proofs [theorems]. \ldots [Bell's] criticism [of these theorems] 
undervalues the importance of defining limits to what speculative 
theories can or cannot be expected to accomplish.'' 
In the present work, we argue for Bell's position on these 
matters.}
demonstrate that the prevailing 
view that these results disprove hidden 
variables\footnote{The following references found in
the bibliography present discussions of all four theorems along with their
relationship to hidden variables: \cite {1974 Belinfante Book, Hughes Simple
Math, Jammer Philosophy, Mermin Review}. Von Neumann's theorem is claimed to
disprove hidden variables in \cite{Albertson, Born, Jauch-Piron, von Neumann}.
This conclusion is reached for both Gleason's theorem and the 
theorem of Kochen and Specker in \cite{Kochen Specker, Reality Marketplace}. 
The following references state that Bell's theorem may be regarded in this way:
\cite{Bethe} ,\cite[p. 172]{Gell-Mann}, \cite{Wigner Big Red}.} 
is actually a false one. According
to Bell, what is 
shown\footnote{The theorem due to von Neumann is not as 
closely
connected with these concepts as the others. We include it for the sake of
completeness and as introduction to the discussion of the other theorems. In
addition, we will point out the presence in a work by Schr{\"o}dinger
\cite{Present Sit} of an analysis leading to essentially the same conclusion as
von Neumann's.} 
by these theorems is that hidden variables must allow for two
important and fundamental quantum features: {\em contextuality} and {\em
nonlocality}.

To allow for contextuality, one must consider the results of a
measurement to depend on the 
attributes of both the system, and the measuring apparatus. The concept 
of contextuality is evidently in contrast to the tendency to consider 
the quantum system in isolation, without considering the configuration 
of the experimental apparatus\footnote{See
papers in the recent book by Bell \cite{Bell Eclipse, Bell Imposs Pilot} for a
good discussion of contextuality.}. 
Such a tendency conflicts with the lesson of
Niels Bohr concerning \cite[page 210]{Einstein impeachment} ``the 
impossibility
of any sharp separation between the behavior of atomic objects and the
interaction with the measuring instruments which serve to define the conditions
under which the phenomena appear.''  In any attempt to
develop a theory of hidden variables, one must take contextuality into
account.

The concept of contextuality lies at the heart of both Gleason's 
theorem and the theorem of Kochen and Specker. However, typical 
expositions\footnote{The exception to this is two
works by Bell: bibliography references \cite{Bell Eclipse, Bell Imposs Pilot}. 
In these works, contextuality is made quite clear, and it is concluded
that neither theorem proves the impossibility of hidden variables. 
References that
discuss the Kochen and Specker theorem {\em without} treating contextuality are
\cite{Kochen Specker, Mermin's Gleason, Reality Marketplace}. The following
references 
briefly discuss the relationship of Gleason's and Kochen and Specker's
theorems to contextuality, but fail to convey its 
importance:
\cite{1974 Belinfante Book, Hughes Simple Math, Jammer Philosophy}.} 
of these 
arguments give either no discussion of the concept, or else a very
brief treatment
that fails to convey its full meaning and importance. The conclusion drawn by
authors giving no discussion of contextuality is that the mathematical 
impossibility of hidden variables has been proved.
 References which give only a brief mention of
the concept tend to leave their readers with the impression that it is not of
much significance and is somewhat artificial. Readers of these works would be
led to the conclusion that Gleason's and Kochen and Specker's theorems 
demonstrate that the prospect of hidden variables is a dubious one, at best.

The concept of nonlocality is well known through Bell's famous mathematical
theorem \cite{Bell's theorem} in which he addresses the problem of the
Einstein--Podolsky--Rosen paradox\footnote{The EPR paper appears in \cite{EPR},
and it is reprinted in \cite{Big Red}.}. In a system exhibiting nonlocality, 
the consequences of events at one place propagate to other 
places instantaneously. Although Einstein, Podolsky and Rosen were
attempting to demonstrate a different conclusion (the incompleteness of the
quantum theory), their analysis served to point out the conditions under which
(as would become evident after Bell's work) nonlocality arises.  The
existence of
such conditions in quantum mechanics was regarded by Erwin Schr{\"o}dinger as
being of great significance, and in a work \cite{Camb1} in which he generalized
the EPR argument, he called these conditions ``{\em the} 
characteristic trait of
quantum mechanics, the one that forces its entire departure from classical
lines.'' Bell's work essentially completed the proof that the
quantum phenomenon discovered by EPR and elaborated by Schr{\"o}dinger truly
does entail nonlocality.

Although the EPR paradox and Bell's theorem are quite well known,
there exists a misperception regarding the relationship 
these arguments bear to nonlocality. Some authors 
(mistakenly) conclude\footnote{See for example,
\cite[p 48]{Ghost Interviews}, \cite{PhysWorld, Mermin Review, Reality 
Marketplace}. Clauser and Shimony \cite{Clauser-Shimony}
regard the EPR paradox and Bell's theorem as proof that quantum mechanics 
must conflict with any `realistic' local theory. Their view seems
to fall short of Bell's (see Bell \cite{Bell cascade photons,
Bertlemann's Socks}) contention that it is 
{\em locality itself} which leads to this clash with quantum theory.} 
that the EPR paradox and Bell's theorem imply that 
a conflict exists {\em only} between local theories of hidden 
variables and quantum mechanics, when a more general 
conclusion than this follows. When taken together, the EPR 
paradox\footnote{Bohm's spin singlet version \cite{Bohm's Famous Text}
of EPR.} and Bell's theorem imply that 
{\em any local theoretical explanation whatsoever must disagree with 
quantum mechanics}. In Bell's words: \cite{Cosmologists} 
``It now seems that the non-locality is deeply rooted in
quantum mechanics itself and will persist in any completion.''
We can conclude from the EPR paradox and Bell's theorem that the
quantum theory is irreducibly nonlocal\footnote{See the following:
\cite{Albert, Bell cascade photons, Bertlemann's Socks, Intuitive Nonlocality, 
Baby Snakes, Tim, Shadows}.}.

As we have mentioned, Erwin Schr{\"o}dinger regarded the situation brought out 
in the Einstein, Podolsky, Rosen paper as a very important problem. His own 
work published just after the EPR paper \cite{Present Sit} provided a
generalization
of this analysis. This work contains Schr{\"o}dinger's presentation of the 
well known
paradox of `Schr{\"o}dinger's cat'. However, a close look at the paper 
reveals much more. Schr{\"o}dinger's work contains a very extensive and 
thought provoking analysis of
quantum theory. He begins with a statement of the nature of
theoretical modeling and a comparison of this to the framework of
quantum theory. He continues with the cat paradox, the measurement
problem, and
finally his generalization of the EPR paradox and its implications. In
addition,
he derives a result quite similar to that of von Neumann's
theorem\footnote{Schr{\"o}dinger does not explicitly mention von Neumann's 
work, however.}.

Subsequent to our discussion of the issues of the EPR paradox,
Bell's theorem, and nonlocality, we will discuss a new
form of quantum nonlocality proof\footnote{A nonlocality proof similar to 
that presented in this thesis was developed by Heywood and Redhead
\cite{Lugubrious Fellow} and by Brown and Svetlichny \cite{Tom's Paper}.
These authors present arguments addressed to one particular example of 
the general class of maximally entangled states we consider. Their 
argumentation does not bring out all the conclusions we do in
the present work, as they do not draw the {\em general} conclusion 
of nonlocality, but instead only the nonlocality of the particular
class of hidden variables they address.} based on Schr{\"o}dinger's 
generalization 
of the EPR paradox. The type of argument we shall present falls into the 
category of a ``nonlocality without
inequalities'' proof\footnote{See Greenberger, Horne, Shimony and Zeilinger in
\cite{GHZ AJP}}. Such analyses---of which the 
first\footnote{See Mermin in \cite{Mermin GHZ}} 
to be developed was that of
Greenberger, Horne, and Zeilinger \cite{GHZ}---differ from
the nonlocality proof derived from the EPR paradox and Bell's
theorem in that they do not possess the statistical 
character of the latter.

\subsection{David Bohm's theory of hidden variables \label{Bohm}} It must be
mentioned that the problem of whether hidden variables are possible is
more than just conjectural. Louis de Broglie's 1926 `Pilot Wave'
theory\footnote{See \cite{de Broglie 1926, de Broglie 1927, de Broglie 1930}.
A summary of the early development of the theory is
given by de Broglie in \cite{de Broglie 1953}.}
 is, in fact, a viable interpretation of
quantum phenomena. The theory was more systematically presented 
by David Bohm in 1952\footnote{See \cite{Bohmian Mechanics}. See also
Bell in \cite{Bell Imposs Pilot, Bell Six Worlds} A recent book
on Bohmian mechanics is \cite{Holland}.}. This
encouraged de Broglie 
himself to take up the idea again. We shall refer to the theory as ``Bohmian 
mechanics'', after David Bohm.

The validity of Bohmian mechanics is a result of the first importance,  since 
it restores {\em objectivity} to the description of quantum
phenomena\footnote{It is for this reason that Bohm refers to his theory as an
`Ontological Interpretation' of quantum mechanics: 
\cite[page 2]{Bohm 1993 Book}  
``Ontology is concerned
primarily with that which {\em is} and only secondarily with how we obtain our
knowledge about this. We have chosen as the subtitle of our book 
``An Ontological
Interpretation of Quantum Theory'' because it gives the clearest and most
accurate description of what the book is about....the primary question is 
whether
we can have an adequate conception of the reality of a quantum system, be this
causal or be it stochastic or be it of any other nature.''}. 
The quantum 
formalism addresses only the properties of quantum
systems as they appear when the system is {\em measured}. An objective
description, by contrast, must give a measurement-independent
description of the
system's properties. The lack of objectivity in quantum theory is something
which Albert Einstein disliked, as is clear from his writings: 
\cite[page 667]{Einstein impeachment}

\begin{quotation}
What does not
satisfy me, from the standpoint of principle, is (quantum theory's)
attitude toward that which appears to be the programmatic aim of all
physics: the
complete description of any (individual) real situation (as it
supposedly exists
irrespective of any act of observation or substantiation).
\end{quotation}
 Moreover, Einstein saw quantum theory as
{\em incomplete}, in the sense that it omits essential elements from its 
description of state: \cite[page 666]{Einstein impeachment} 
``I am, in fact, firmly convinced that the essentially statistical
character of contemporary quantum theory is solely to be ascribed to the fact
that this (theory) operates with an incomplete description of physical 
systems.`' Bohmian mechanics essentially fulfills Einstein's 
goals for quantum physics: to complete the quantum description 
and to restore objectivity to the theoretical picture. 

Besides the feature of objectivity, another reason the success of Bohmian
mechanics is important is that it restores {\em determinism}. Within this
theory one may predict, from the present state of the system, its form at any
subsequent time. In particular, the results of quantum measurements are so
determined from the present state. Determinism is the feature with which the
hidden variables program has been traditionally 
associated\footnote{This is the
motivation mentioned by von Neumann in his no-hidden variables proof. See
reference \cite{von Neumann}}.

The importance of the existence of a successful theory of hidden variables in 
the form of Bohmian mechanics is perhaps expressed most succinctly by J.S. 
Bell:  \cite{Bell Imposs Pilot}

\begin{quotation} 
Why is the pilot wave picture [Bohmian mechanics] ignored in text 
books? Should it not be taught, not as the only way, but as an 
antidote to the prevailing complacency?  To show that
vagueness, subjectivity, and indeterminism are not forced upon us by 
experimental facts, but by deliberate theoretical choice? 
\end{quotation}  

\section{Original results to be presented} 
The analysis we present will consist of
two main parts: one devoted to contextuality and the theorems of 
Gleason, and Kochen and Specker, and the other devoted to nonlocality, Bell's 
theorem, and
Schr{\"o}dinger's generalization of the Einstein--Podolsky--Rosen 
paradox. We will review
and elaborate on the reasons why these theorems do not disprove hidden
variables, but should instead be regarded as illustrations of
contextuality and nonlocality. Our presentation will link topics that have most
often been addressed in isolation, as we relate the various mathematical 
theorems to one another. Our analysis of contextuality and the theorems of 
\gl and \ks goes
beyond that of other authors\footnote{Bell \cite{Bell Eclipse, Bell Imposs
Pilot} and Mermin \cite{Mermin Review}.} in the following ways. First, we will
show that the theorems of Gleason, and Kochen and Specker imply a result we
call ``spectral-incompatibility''. Regarded as proofs of this result, the
implications of these theorems toward hidden variables become much more
transparent. Second, through discussion of an experimental procedure first
given by David Albert \cite{Albert}, we show that contextuality provides
an indication
that the role traditionally ascribed to the Hermitian operators in the
description of a quantum system is not tenable\footnote{A work by
Daumer, D{\"u}rr, Goldstein, and Zangh{\'i} (\cite{Martin})
argues for the same conclusion, but does not do so by focussing on
the Albert experiment.}. 

After reviewing the EPR paradox, Bell's theorem, and
the argument for quantum nonlocality that follows, we then present what
is perhaps the most distinctive feature of the thesis, as we focus  
on Erwin Schr{\"o}dinger's generalization of the EPR paradox. First, we show 
that the conclusions of `\Erwins paradox' may be reached using a more 
transparent method than \Erwins which allows one to relate the form of 
the quantum state to the perfect correlations it exhibits. Second, we
use Schr{\"o}dinger's paradox to 
derive a broad spectrum of new instances of
quantum nonlocality. Like the proof given by Greenberger, Horne, and
Zeilinger, these `Schr{\"o}dinger nonlocality'
proofs are of a deterministic character, i.e., they are `nonlocality
without inequalities' proofs.
The quantum nonlocality result based on the EPR paradox and Bell's
theorem is, on
the other hand, essentially statistical, since it concerns the average of the
results from a series of quantum measurements. 
We demonstrate that just as is true for the GHZ nonlocality, Schr{\"o}dinger 
nonlocality differs from the EPR/Bell demonstration in that it
may be experimentally confirmed simply by
observing the `perfect correlations' between the relevant 
observables. To verify the EPR/Bell nonlocality result, on the other
hand, one must
observe not only that the perfect correlations exist, but also that 
Bell's inequality is violated\footnote{See \cite{Aspect 1, Aspect 2,
Aspect 3}.}.
The test of Schr{\"o}dinger nonlocality is simpler than that of GHZ in
that the perfect correlations in question are between observables of two
subsystems, rather than between those of three or more.
Moreover, some of the Schr{\"o}dinger
nonlocality proofs are stronger than the GHZ proof in that they
involve larger classes of
observables than that addressed by the latter.

\section{Review of the formalism of quantum mechanics \label{formalism}} 
\subsection{The state and
its evolution} The formalism of a physical theory typically consists of two
parts. The first describes the representation of the state of the system; the
second, the time evolution of the state. The quantum formalism contains besides these a prescription\footnote{It is the existence of such rules in
the quantum mechanical formalism that marks its departure from an objective
theory, as we discussed above.} for the results of measurement.

We first discuss the quantum formalism's representation of state.  The quantum
formalism associates with every system a Hilbert space ${\cal H}$ and
represents
the state of the system by a vector $\psi$ in that Hilbert space. The vector
$\psi$ is referred to as the {\em wave function}. This is in contrast to the
classical description of state which for a system of particles is
represented by
the coordinates $\{q_{1},q_{2},q_{3},...\}$ and momenta
$\{p_{1},p_{2},p_{3},...\}$. We shall often use the Dirac notation: a Hilbert
space vector may be written as $\sket$, and the inner product of two vectors 
$|\psi _{1} \rangle, \mid \psi_{2} \rangle$ as $\langle \psi_{1} \mid \psi_{2}
\rangle$. The quantum mechanical state $\psi$ in the case of a non-relativistic
$N$-particle system is given\footnote{In the absence of spin.} by a
complex-valued function on configuration space, $\psi({\bf q})$, which is an
element of $L_{2}$---the Hilbert space of square-integrable functions. Here 
${\bf q}$ represents a point $\{q_{1},q_{2},...\}$ in the configuration space
\R$^{3N}$. The fact that $\psi$ is an element of $L_{2}$ implies  
\beq
\sbra \psi \rangle = \mid \mid \psi \mid \mid^{2}=\int \mbox{ {\bf dq} } \psi
^{*} ({\bf q}) \psi({\bf q}) < \infty   \label{eq:L2} 
\eeq 
where
$\mbox{{\bf dq}}=(\mbox{dq}_{1} \mbox{dq}_{2}...)$. The physical state
is defined
only up to a multiplicative constant, i.e., $c \sket$ and $\sket$
(where $c \neq
0$) represent the same state.  We may therefore choose to normalize $\psi$ so
that \ref{eq:L2} becomes 
 \beq 
\langle \psi \mid \psi \rangle=1
\label{eq:psi-norm} \mbox{.} 
\eeq

The time evolution of the quantum mechanical wave function is governed by
\linebreak 
Schr{\"o}dinger's equation: 
 \beq 
i\hbar\frac{\partial \psi }{
\partial t} = H \psi 
\mbox{,}
\eeq 
where $H$ is an operator whose form depends
on the nature of the system and in particular on whether or not it is
relativistic. To indicate its dependence on time, we write the wave function as
$\psi_{t}$. For the case of a non-relativistic system, and in the absence of
spin, the Hamiltonian takes the form  
\beq
 H=-\frac{1}{2} \hbar ^{2} \sum_{j}
\frac{\nabla_{j}^{2}}{m_{j}} + V({\bf q}) 
\eeq 
where $V({\bf q})$ is the
potential energy\footnote{We ignore here the possibility of an
external magnetic
field.} of the system.  With the time evolution so specified, $\psi_{t}$ may be
determined from its form $\psi_{t_{0}}$ at some previous time.

\subsection{Rules of measurement} As stated above, the description of the state
and its evolution does not constitute the entire quantum formalism. The wave
function provides only a formal description and does not by itself make contact
with the properties of the system. Using {\em only} the wave function and its
evolution, we cannot make predictions about the typical systems in which we are
interested, such as the electrons in an atom, the conduction electrons of a
metal, and photons of light. The connection of the wave function to
any physical
properties is made through the rules of measurement. Because the physical
properties in quantum theory are defined through measurement, or observation,
they are referred to as `observables'. The quantum formalism represents the
observables by Hermitian operators\footnote{In the following
description, we use
the terms observable and operator interchangeably.} on the
system Hilbert space.

Associated with any observable $O$ is a set of {\em eigenvectors} and {\em
eigenvalues}. These quantities are defined by the relationship  \beq O
\mid \phi \rangle = \mu \mid \phi \rangle \label{eq:Eigenvalue} \mbox{,}
\eeq where $\mu$ is a real constant. We refer to $| \phi \rangle$ as
the eigenvector corresponding to the eigenvalue $\mu$. We label the set of
eigenvalues so defined as $\{\mu_{a}\}$. To each member of this set, there
corresponds a set of eigenvectors, all of which are elements of a {\em
subspace} of ${\cal H}$, sometimes referred to as the {\em eigenspace}
belonging
to $\mu_{a}$. We label this subspace as ${\cal H}_{a}$. Any two such distinct
subspaces are orthogonal, i.e., every vector of ${\cal H}_{a}$ is orthogonal to
every vector of ${\cal H}_{b}$ if $\mu_{a} \neq \mu_{b}$.

It is important to develop explicitly the example of a {\em commuting set} of
observables $(O^{1}, O^{2},...)$ i.e., where each pair $( O^{i}, O^{j})$ of 
observables has commutator zero: 
 \beq [O^{i}, O^{j}]=
O^{i} O^{j}- O^{j} O^{i}=0 \eeq  For this case, we have a series of
relationships of the form \ref{eq:Eigenvalue} \begin{eqnarray} O^{i} \pket & = &
\mu^{i} \pket \label{eq:Joint-Eig} \\ i & = & 1,2,...  \nonumber \end{eqnarray}
defining the set of {\em simultaneous} eigenvalues and eigenvectors, or {\em
joint-eigenvalues} and {\em joint-eigenvectors}. Note that \ref{eq:Eigenvalue}
may be regarded as a vector representation of the set of equations
\ref{eq:Joint-Eig} where $O=(O^{1},O^{2},\ldots)$ and $\bm
=(\mu^{1},\mu^{2},\ldots)$ are seen respectively as ordered sets of operators 
and numbers.  We use the same symbol, $\mu$, to denote a 
joint-eigenvalue, as was used to designate an eigenvalue.
 We emphasize that for a 
commuting set $\bm$ refers to an {\em ordered
set} of numbers. The correspondence of the joint-eigenvalues to the
joint-eigenvectors is similar to that between the eigenvalues and eigenvectors
discussed above; to each $\bm_{a}$ there corresponds a set of
joint-eigenvectors
forming a subspace of ${\cal H}$ which we refer to as the {\em
joint-eigenspace}
belonging to $\bm_{a}$.

We denote by $P_{a}$ the projection operator onto the eigenspace belonging to
$\mu_{a}$. It will be useful for our later discussion to express
$P_{a}$ in terms
of an orthonormal basis of ${\cal H}_{a}$. If ${{ \phi_{k} }}$ is such a basis,
we have  \beq P_{a}= \sum_{k} \mid \phi_{k} \rangle \langle \phi_{k}
\mid \mbox{,} \label{eq:Psum} \eeq which means that the operator which
projects onto a subspace is equal to the sum of the projections onto the
one-dimensional spaces defined by any basis of the subspace. If we label the
eigenvalues of the observable $O$ as $\mu_{a}$ then $O$ is represented in terms
of the $P_{a}$ by  \beq O=\sum_{a}\mu_{a} P_{a} \label{eq:Dirac}
\mbox{.} \eeq

Having developed these quantities, we now discuss the rules of measurement. The
first rule concerns the possible outcomes of a measurement and it states that
they are restricted to the eigenvalues $\{\mu_{a}\}$ for the measurement of a
single observable\footnote{A measurement may be classified either as
{\em ideal}
or {\em non-ideal}. Unless specifically stated, it will be assumed whenever we
refer to a measurement process, what is said will apply just as well to {\em
either} of these situations.} $O$ or {\em joint}-eigenvalues $\{\bm_{a}\}$ for
measurement a commuting set of observables $(O^{1},O^{2},...)$. 

The second rule
provides the probability of the measurement result equaling one particular
eigenvalue or joint-eigenvalue: \begin{eqnarray} P(O=\mu_{a}) & = & \sbra P_{a}
\sket \label{eq:Prob} \\ P((O^{1},O^{2},...) = \bm_{a}) & = & \sbra P_{a} \sket
\nonumber \end{eqnarray} where the former refers to the measurement of a single
observable and the latter to a commuting set. As a consequence of the
former, the
expectation value for the result of a measurement of $O$ is given by \beq
E(O)=\sum_{a}  \mu_{a}  \sbra P_{a} \sket = \sbra O \sket \label{eq: Exp} \eeq
where the last equality follows from \ref{eq:Dirac}.

The third rule governs the effect of measurement on the system's wave function.
It is here that the measurement is governed by a different rule depending on
whether one performs an ideal or non-ideal measurement. An ideal measurement is
defined as one for which the wave function's form after measurement is given by
the (normalized) projection of $\psi$ onto the eigenspace  ${\cal H}_{a}$ of the
measurement result ${\bf \mu}_{a}$ \beq P_{a} \psi / \mid \mid P_{a} \psi \mid
\mid \label{eq:NIC} \eeq The case of a non-ideal measurement arises when an ideal
measurement of a commuting set of observables $(O^{1}, O^{2},...)$ is regarded 
as
a measurement of an {\em individual} member of the set. An ideal measurement of
the {\em set} of observables leaves the wave function as the projection of 
$\psi$ onto their {\em joint}-eigenspace. For an
arbitrary vector $\psi$, the 
projection onto the eigenspace of an individual
member of the set does not generally equal the projection onto the set's
joint-eigenspace. Therefore an ideal measurement of a commuting set cannot be 
an ideal measurement of an individual member of the set. We refer to the 
procedure as a
{\em non-ideal} measurement of the observable. This completes the presentation 
of
the general rules of measurement. We discuss below two special cases for which
these rules reduce to a somewhat simpler form.

The first special case is that of a {\em non-degenerate} observable. The 
eigenspaces of such an observable are one-dimensional and are 
therefore spanned by a single normalized vector called an eigenvector. We label
 the eigenvector corresponding to eigenvalue $\mu_{a}$ as $|a\ra$. The operator
  which projects onto such a one-dimensional space
is written as:  \beq P_{a} = \mid
a \rangle \langle a \mid \mbox{.} \eeq  It is easy to see that the projection 
given in \ref{eq:Psum} reduces to this form.
 The form of the observable given in \ref{eq:Dirac} then reduces to:
 \beq O=\sum_{a} \mu_{a} \mid a \rangle \langle a \mid \mbox{.}
\eeq The probability of a measurement result being equal to $\mu_{a}$
is then  \beq P(O=\mu_{a})= | \langle a \mid \psi \rangle |^{2}
\mbox{,} \label{eq:ProbNon} \eeq in which case the wave function
subsequent to the measurement is given by $| a \rangle$. These rules
are perhaps the most familiar ones since the non-degenerate case is usually
introduced first in presentations of quantum theory. 

Finally, there is a case 
for which the measurement rules are essentially the same as these, and this is 
that of a set of commuting observables which form a {\em complete set}. The
joint-eigenspaces of a complete set
are one-dimensional. The rules governing  the probability of the measurement
result $\bm_{i}$ and the effect of  measurement on the wave function are
perfectly analogous to those of the non-degenerate observable. This completes 
our discussion of the rules of measurement.

\section{Von Neumann's theorem and hidden variables \label{Von}}
\subsection{Introduction \label{Vint}} The quantum formalism contains features
which may be considered objectionable\footnote{J.S. Bell goes further
than this:
he refers to quantum mechanics as  {\em unprofessional} (\cite{Bell 
In Bohm} and \cite[p. 45]{Ghost
Interviews}) in its lack of clarity.} by some. These are its {\em subjectivity}
and {\em indeterminism}. The aim of the development of a hidden variables
theory\footnote{See the discussion of Bohmian mechanics above. The great
success of this theory is that it explains quantum phenomena {\em without} such
features. The general issue of hidden variables is, of course, discussed in
several references. Bell's work is the most definitive: see \cite{Bell Eclipse,
Determine Bell, Bell Imposs Pilot} in  \cite{Speakable}. A recent review was 
published by N.D.  Mermin, who has done much to
popularize Bell's theorem through articles in {\em Physics Today} and through
popular lectures. Discussions may be also be found in Bohm 
\cite{Bohmian Mechanics, Bohm 1993 Book}, Belinfante \cite{1974
Belinfante Book},  Hughes \cite{Hughes Simple Math} and Jammer \cite{Jammer
Philosophy}.} is to give a formalism which, while being empirically equivalent 
to
the quantum formalism, does not possess these features. In the present section 
we
shall present and discuss one of the earliest works to address the hidden
variables question, which is the 1932 analysis of John von Neumann\footnote{The
original work is \cite{von Neumann}. Discussions of von Neumann's hidden
variables analysis may be found within \cite{Albertson, Bell Eclipse, Bell 
Imposs Pilot,  Jauch-Piron}}. We shall also review and elaborate on J.S. Bell's
analysis \cite{Bell Eclipse} of this work, in which he made clear its
limitations.

Von Neumann's hidden variables analysis appeared in his now classic book {\bf
Mathematical Foundations of Quantum Mechanics}. This book is notable 
both for its exposition of the mathematical structure of quantum theory, 
and as one
of the earliest works\footnote{Within chapter $4$ of the present
work, we shall discuss a 1935 analysis by Erwin Schr{\"o}dinger \cite{Present
Sit}. This is the paper in which the `Schr{\"o}dinger's cat' paradox first
appeared, but it is not generally appreciated that it contains other results of
equal or perhaps greater significance, such as Schr{\"o}dinger's generalization
of the Einstein--Podolsky--Rosen paradox. We believe that this remarkable paper
could have done much to advance the study of the foundations of quantum
mechanics, had these latter features been more widely appreciated.} 
to
systematically address both the hidden variables issue and the measurement
problem. The quantum formalism presents us with two different types of state
function evolution: that given by the Schr{\"o}dinger equation and that which
occurs during a measurement. The latter evolution appears in the formal rule
given above in equation \ref{eq:NIC}. The measurement problem is the
problem of the
reconciliation of these two types of evolution.

In his analysis of the hidden variables problem, von Neumann proved a
mathematical result now known as von Neumann's theorem and then argued
that this theorem implied the very strong conclusion that no hidden
variables theory can provide empirical agreement with quantum
mechanics: (preface p. ix,x) `` \ldots such an explanation (by `hidden
parameters') is incompatible with certain qualitative fundamental
postulates of quantum mechanics.''  The author further states:
\cite[p.  324]{von Neumann} ``It should be noted that we need not go
further into the mechanism of the `hidden parameters,' since we now
know that the established results of quantum mechanics can never be
re-derived with their help.''  The first concrete
demonstration\footnote{See Jammer's book
\cite[page 272]{Jammer Philosophy} for a discussion of an early work by Grete
Hermann that addresses the impact of von Neumann's theorem.} that this
claim did
not follow was given in 1952 when David Bohm constructed a viable theory of
hidden variables \cite{Bohmian Mechanics}. Then in 1966, J.S. Bell \cite{Bell
Eclipse} analyzed von Neumann's argument against hidden variables and showed it
to be in error. In this section, we begin by discussing an essential
concept of von Neumann's analysis: the state representation of
a hidden variables theory. We then present von Neumann's theorem and
no-hidden-variables argument. Finally, we show where the flaw in this argument
lies.

The analysis of von Neumann is concerned with the description of the
{\em state}
of a system, and the question of the {\em incompleteness} of the quantum
formalism's description. The notion of the incompleteness of the quantum
mechanical description was particularly emphasized by Einstein, as we noted in
section \ref{Bohm}. The famous Einstein--Podolsky--Rosen paper was
designed as a
proof of such incompleteness and the authors concluded this work with the
following statement: \cite{EPR}
``While we have thus shown that the wave function does not provide a
complete description of the physical reality, we left open the question of
whether or not such a description exists. We believe, however, that such a 
theory is possible.''  
The hidden variables program, which is an endeavor to
supplement the state description, is apparently exactly the type of program
Einstein was advocating. A complete state description might be
constructed to remove some of the objectionable features of the quantum
theoretical description.

The particular issue which von Neumann's analysis addressed was the
following: is it possible to restore {\em determinism} to the
description of physical systems by the introduction of hidden
variables into the description of the state of a system. The quantum
formalism's state representation given by $\psi$ does not generally
permit deterministic predictions regarding the values of the physical
quantities, i.e. the observables. Thus results obtained from
performance of measurements on systems with {\em identical} state
representations $\psi$ may be expected to {\em vary}. (The statistical
quantity called the {\em dispersion} is used to describe this
variation quantitatively.). While it generally does not provide
predictions for each individual measurement of an observable $O$, the
quantum formalism does give a prediction for its average or {\em
expectation} value\footnote{When generalized to the case of a mixed
state, this becomes
\beq 
E(O)=\mbox{Tr}(UO) 
\label{eq:Eqfms}
\eeq
where $U$ is a positive operator with the property  $\mbox{Tr}(U)=1$.
Here $U$ is known as the ``density matrix''. See for example, 
\cite[p. 378]{Schiff}}: 
\beq 
E(O)=\langle \psi \mid O \mid \psi \rangle \label{eq:Eqf} 
\mbox{.}
\eeq 

Von Neumann's analysis addresses the question of whether the lack of
determinism in the quantum formalism may be ascribed to the fact that
the state description as given by $\psi$ is incomplete. If this were
true, then the complete description of state---consisting of both
$\psi$ and an additional parameter we call $\lda$---should allow one
to make predictions regarding individual measurement results for each
observable. Note that such predictability can be expressed
mathematically by stating that for every $\psi$ and $\lda$, there must
exist a ``value map'' function, i.e. a mathematical function assigning
to each observable its value. We represent such a value map function
by the expression $V^{\psi}_{\lda}(O)$.  Von Neumann referred to a
hypothetical state described by the parameters $\psi$ and $\lda$ as a
``dispersion-free state'', since results obtained from measurements on
systems with identical state representations in $\psi$ and $\lda$ are
expected to be identical and therefore exhibit no dispersion.

Von Neumann's theorem is concerned with the general form taken by a
function $E(O)$, which assigns to each observable its expectation
value. The function addressed by the theorem is considered as
being of sufficient generality that the expectation function of 
quantum theory, or of any empirically equivalent theory must assume the 
form von Neumann derived. In the case of quantum theory, $E(O)$ should take 
the form
of the quantum expectation formula \ref{eq:Eqfms}. In the case of a
dispersion-free state, the average over a series of $E(O)$ should
return the value of each observable. When one analyzes the form of
$E(O)$ developed in the theorem, it is easy to see that it cannot be a
function of the latter type. Von Neumann went on to conclude from this
that no theory involving dispersion-free states can agree with quantum
mechanics.  However, since the theorem places an unreasonable
restriction on the function $E(O)$, this conclusion does {\em not}
follow.

\subsection{Von Neumann's theorem} 
The assumptions regarding the function $E(O)$ are as follows. First,
the value $E$ assigns to the `identity observable' ${\bf 1}$ is equal
to unity: \beq E({\bf 1}) = 1 \label{eq:1} \eeq The identity
observable is the projection operator associated with the entire
system Hilbert space. All vectors are eigenvectors of ${\bf 1}$ of
eigenvalue $1$. The second assumption is that $E$ of any real-linear
combination of observables is the same linear combination of the
values $E$ assigns to each individual observable: \beq E(aA+bB+...) =
aE(A)+bE(B)+...  \label{eq:Silly}
\eeq
where $(a,b,...)$ are real numbers and $(A,B,...)$ are
observables. Finally, it is assumed that $E$ of any projection
operator $P$  must be non-negative:
 \beq 
E(P) \geq 0 .     \label{eq:Pos} 
\eeq
 For
example, in the case of the value map function $V^{\psi}_{\lda}$, $P$
must be assigned
either $1$ or $0$, since these are its possible values. According to
the theorem, these premises imply that $E(O)$ must be
given by the form 
\beq 
E(O)=\mbox{Tr}(UO) 
\eeq
where $U$ is a positive operator with the property  $\mbox{Tr}(U)=1$.

The demonstration\footnote{A proof of the theorem may be found in von Neumann's
original work \cite{von Neumann}. Albertson presented a simplification of this
proof in 1961 \cite{Albertson}. What we present here is a further
simplification.} of this conclusion is straightforward. We begin by noting that
any operator $O$ may be written as a sum of Hermitian operators. 
Define $A$ and $B$ by the relationships $A=\frac{1}{2}(O+O^{\dag})$ and
$B= \frac{1}{2i}(O-O^{\dag})$, where $O^{\dag}$ is the Hermitian conjugate of 
$O$. Then it is easily seen that $A$ and $B$ are Hermitian and that
 \beq
O=A+iB 
\mbox{.}
\eeq 
We define the function $E^{*}(O)$
by 
 \beq 
E^{*}(O)=E(A)+iE(B) \label{eq:E*Def} 
\mbox{,} 
\eeq
where $E(O)$ is von Neumann's $E(O)$, and $A$ and $B$ are defined as
above. From the equations \ref{eq:E*Def} and \ref{eq:Silly} we have
that $E^{*}(O)$ has the property of complex linearity. Note that
$E^{*}(O)$ is a generalization of von Neumann's $E(O)$: the latter is
a {\em real}-linear function on the Hermitian operators, while the
former is a {\em complex}-linear function on {\em all} operators. The
general form of $E^{*}$ will now be investigated for the case of a
finite-dimensional operator expressed as a matrix in terms of some
orthonormal basis. We write the operator $O$ in the form 
\beq
O=\sum_{m,n} |m\ra \la m | O | n \ra \la n |
\mbox{,}
\eeq 
where the sums over $m,n$ are finite sums. This form of $O$ is a linear
combination of the operators $|m\ra \la n|$, so that the complex linearity 
of $E^{*}$ implies that 
 \beq 
 E^{*}(O)=\sum_{m,n} \la m | O | n \ra  E^{*}(|m\ra \la n|) 
 \label{eq:E}
\mbox{.}
\eeq 
Now we define the operator $U$ by the relationship 
$U_{nm}=E^{*}(|m\ra \la n|)$ and the \ref{eq:E} becomes 
 \beq 
E^{*}(O)=\sum_{m,n} O_{mn} U_{nm}
=\sum_{m}(UO)_{mm}=\mbox{Tr}(UO) 
\label{eq:Trace*} 
\mbox{.}
\eeq 
Since von
Neumann's $E(O)$ is a special case of $E^{*}(O)$, \ref{eq:Trace*} implies that
 \beq 
E(O)=\mbox{Tr}(UO) \label{eq:Trace} \mbox{.} 
\eeq 
We now show that $U$ is a positive operator. It is a premise of the
theorem that $E(P) \geq 0$ for any projection operator $P$. Thus we
write $E(P_{\chi}) \geq 0$ where $P_{\chi}$ is a one-dimensional
projection operator onto the vector $\chi$.  Using the form of $E$
found in \ref{eq:Trace}, we have\footnote{The equality
$\mbox{Tr}(UP_{\chi})= \langle \chi \mid U \mid \chi \rangle$ in
\ref{eq:easy} is seen as follows. The expression
$\mbox{Tr}(UP_{\chi})$ is independent of the orthonormal basis
${\phi_{n}}$ in terms of which the matrix representations of $U$ and
$P_{\chi}$ are expressed, so that one may choose an orthonormal basis
of which $| \chi\rangle$ itself is a member. Since $P_{\chi} = \mid
\chi \rangle
\langle \chi |$, and $P_{\chi} \mid \phi_{n} \rangle = 0$  for all $|
\phi_{n}
\rangle$ except $| \chi \rangle$, we have
$\mbox{Tr}(UP_{\chi})=\langle \chi | U
| \chi \rangle$.} \beq \mbox{Tr}(UP_{\chi})=\langle \chi \mid U \mid
\chi \rangle
\geq 0 \mbox{.} \label{eq:easy} \eeq Since $\chi$ is an arbitrary vector, it
follows that $U$ is a positive operator. The relation $\mbox{Tr}(U)=1$
is shown as follows: from the first assumption of the theorem
\ref{eq:1} together with the form of $E$ given by \ref{eq:Trace} we
have $\mbox{Tr}(U)=\mbox{Tr}(U{\bf 1})=1$.  This completes the
demonstration of von Neumann's theorem.

\subsection{Von Neumann's impossibility proof \label{Vonimp}} We now
present von
Neumann's argument against the possibility of hidden
variables. Consider the function $E(O)$ evaluated on the
one-dimensional projection operators $P_{\phi}$.  For such
projections, we have the relationship \beq P_{\phi}=P^{2}_{\phi}
\label{eq:Pr} \mbox{.} \eeq As mentioned above, in the case when $E(O)$ is to
correspond to a dispersion free state represented by some $\psi$ and
$\lda$, it must map the observables to their values. We write
$V^{\psi}_{\lda}(O)$ for the value map function corresponding to the
state specified by $\psi$ and $\lda$. Von Neumann noted
$V^{\psi}_{\lda}(O)$ must obey the relation: \beq
f(V^{\psi}_{\lda}(O))=V^{\psi}_{\lda}(f(O)) \label{eq:11} \mbox{,}
\eeq where $f$ is any mathematical function. This is easily seen by
noting that the quantity $f(O)$ can be measured by measuring $O$ and
evaluating $f$ of the result. This means that the value of the
observable $f(O)$ will be $f$ of the value of $O$.  Thus, if
$V^{\psi}_{\lda}(O)$ maps each observable to a value then we must have
\ref{eq:11}. Hence
$V^{\psi}_{\lda}(P^{2}_{\phi})=(V^{\psi}_{\lda}(P_{\phi}))^{2}$ which together
with \ref{eq:Pr} implies \beq
V^{\psi}_{\lda}(P_{\phi})=(V^{\psi}_{\lda}(P_{\phi}))^{2} \eeq This last
relationship implies that $V^{\psi}_{\lda}(P_{\phi})$ must be equal to
either $0$
or $1$.

Recall from the previous section the relation $E(P_{\phi})=\langle
\phi | U |
\phi \rangle$. If $E(O)$ takes the form of a value map function such as
$V^{\psi}_{\lda}(P_{\phi})$, then it follows that the quantity
$\langle \phi | U | \phi \rangle$ is equal to either $0$ or
$1$. Consider the way this quantity depends on vector $|\phi\ra$. If
we vary $|\phi\ra$ continuously then $\langle \phi | U | \phi \rangle$
will also vary continuously. If the only possible values of $\langle
\phi | U | \phi \rangle$ are $0$ and $1$, it follows that this
quantity must be a {\em constant}, i.e. we must have either $\langle
\phi | U | \phi \rangle=0 \mbox{ } \forall \phi \in \hil$, or $\langle
\phi | U |
\phi \rangle=1 \mbox{ } \forall \phi \in \hil$. If the former holds true, then
it must be that $U$ itself is zero. However, with this and the form
\ref{eq:Trace}, we find that $E({\bf 1})=0$; a result that conflicts
with the theorem
assumption that $E({\bf 1})=1$ (\ref{eq:1}). Similarly, if $\langle
\phi | U |
\phi \rangle=1$ for all $\phi \in {\cal H}$ then it follows that $U=1$. This
result also conflicts with the requirement \ref{eq:1}, since it leads
to $E({\bf 1})=\mbox{Tr}(1)=n$ where $n$ is the dimensionality of
${\cal H}$.

From the result just obtained, one can conclude that any function
$E(O)$ which satisfies the constraints of von Neumann's theorem (see
\ref{eq:1},
\ref{eq:Silly}, and \ref{eq:Pos}) must fail to satisfy the relationship
\ref{eq:11}, and so cannot be a value map function on the
observables\footnote{It
should be noted that the same result may be proven without use of
\ref{eq:11} since the fact that $V^{\psi}_{\lda}(P_{\phi})$ must be
either $0$ or $1$ follows simply from the observation that these are
the eigenvalues of $P_{\phi}$.}. From this result, von Neumann
concluded that it is impossible for a deterministic hidden variables
theory to provide empirical agreement with quantum theory:
\cite[p. 325]{von Neumann} 

\begin{quotation} It is therefore not, as is often assumed, a question
of a re-interpretation of quantum mechanics --- the present system of quantum
mechanics would have to be objectively false in order that another description
of the elementary processes than the statistical one be possible. 
\end{quotation}   

\subsection{Refutation of von Neumann's impossibility proof
\label{RefVon}} While
it is true that the mathematical theorem of von Neumann is a valid one, it is 
not
the case that the impossibility of hidden variables follows. The
invalidity of von Neumann's argument against hidden variables was
shown by Bohm's
development \cite{Bohmian Mechanics} of a successful hidden variables
theory (a
{\em counter-example} to von Neumann's proof) and by J.S. Bell's explicit
analysis \cite{Bell Eclipse} of von Neumann's proof. We will now present the
latter.

The no hidden variables demonstration of von Neumann may be regarded as
consisting of two components: a mathematical theorem and an analysis of its
implications toward hidden variables. As we have said, the theorem itself is
correct when regarded as pure mathematics. The flaw lies in the
analysis connecting this
theorem to hidden variables. The conditions prescribed for the
function $E$ are found in equations \ref{eq:1}, \ref{eq:Silly}, and
\ref{eq:Pos}.
The theorem of von Neumann states that from these assumptions follows the
conclusion that the form of $E(O)$ must be given by \ref{eq:Trace}. When one
considers an actual physical situation, it becomes apparent that the second of
the theorem's conditions is not at all a reasonable one. As we shall see, the
departure of this condition from being a reasonable constraint on $E(O)$ is
marked by the case of its application to non-commuting observables.

We wish to demonstrate why the assumption \ref{eq:Silly} is an
unjustified constraint on $E$. To do so, we first examine a particular
case in which such a relationship is reasonable, and then contrast
this with the case for which it is not. The assumption itself calls
for the real-linearity of $E(O)$, i.e. that $E$ must satisfy $E(aA+bB+
\ldots)=aE(A)+bE(B)+\ldots$ for any observables $\{A,B,\ldots\}$ and
real numbers $\{a,b,\ldots\}$. This is in fact, a sensible requirement
for the case where $\{A,B,\ldots\}$ are {\em commuting}
observables. Suppose for example, the observables $O_{1}$, $O_{2}$,
$O_{3}$ form a commuting set and that they obey the relationship
$O_{1}=O_{2}+O_{3}$. We know from the quantum formalism that one may
measure these observables simultaneously and that the the the
measurement result $\{o_{1},o_{2},o_{3}\}$ must be a member of the
joint-eigenspectrum of the set. By examining the relation
\ref{eq:Joint-Eig} which defines the joint-eigenspectrum, it is easily
seen that any member of the joint eigenspectrum of $O_{1},O_{2},O_{3}$
must satisfy $o_{1}=o_{2}+o_{3}$. This being the case, one might well
expect that the function $E(O)$---which in the case of a dispersion
free state must be a value map $V^{\psi}_{\lda}(O)$ on the
observables---should be required to satisfy
$E(O_{1})=E(O_{2})+E(O_{3})$.  On the other hand, suppose we consider
a set $\{O,P,Q\}$ satisfying $O=P+Q$, where the observables $P$ and
$Q$ {\em fail to commute}, i.e. $[P,Q] \neq 0$. It is easy to see that
$O$ commutes with neither $P$ nor $Q$. It is therefore impossible to
perform a simultaneous measurement of any two of these
observables. Hence, measurements of these observables require three
{\em distinct} experimental procedures. This being so, there is {\em
no justification} for the requirement that $E(O)=E(P)+E(Q)$ for such
cases.

As an example one may consider the case of a
spin $\frac{1}{2}$ particle. Suppose that the components of the spin given by
$\sigma_{x}$, $\sigma_{y}$ and $\sigma^{'}$ where 
\beq
\sigma^{'}=\frac{1}{\sqrt{2}}(\sigma_{x}+\sigma_{y}) 
\label{eq:spins} 
\mbox{,}
\eeq 
are to be examined. The
measurement procedure for any given component of the spin of a particle is
performed by a suitably oriented Stern-Gerlach magnet. For example,
to measure the $x$-component, the magnet must be oriented along the $x$-axis; 
for the $y$-component it must be oriented along the $y$-axis. A measurement of
$\sigma^{'}$ is done using a Stern-Gerlach magnet oriented along an axis 
in yet another direction\footnote{It is not difficult to show that $\sigma^{'}$
 defined in this way
is the spin component along an axis which is in the $x,y$ plane and lies at
$45^{\circ}$ from both the $x$ and $y$ axis.}. The relationship
\ref{eq:spins} cannot be a reasonable demand to place on the expectation
function $E(O)$ of the observables $\sigma_{x}, \sigma_{y}, \sigma^{'}$, since
these quantities are measured using completely distinct procedures.

Thus von Neumann's hidden variables argument is seen to be an unsound one. That
it is based on an unjustified assumption is sufficient to show this. It should
also be noted that the presence of the real-linearity postulate discussed above
makes von Neumann's entire case against hidden variables into an argument of a
rather trivial character. Examining the above example involving the three spin
components of a spin $\frac{1}{2}$ particle, we find that the eigenvalues of
these observables $\pm \frac{1}{2}$ do not obey \ref{eq:spins}, i.e. \beq \pm
\frac{1}{2} \neq \frac{1}{\sqrt{2}}(\pm \frac{1}{2} \pm \frac{1}{2}) \eeq Since
$E(O)$ by hypothesis must satisfy \ref{eq:spins}, it cannot map the observables
to their eigenvalues. Hence, with the real-linearity assumption one can almost
immediately `disprove' hidden variables. It is therefore apparent that Von
Neumann's case against hidden variables rests essentially upon the arbitrary
requirement that $E(O)$ obey real linearity---an assumption that gives immediate
disagreement with the simple and natural demand that $E$ agree with quantum
mechanics in giving the eigenvalues as the results of measurement.

\subsection{Summary and further remarks \label{Von Summ}} In our discussion of
von Neumann's no hidden variables argument, we found that the argument may be
regarded as consisting of two components: a theorem which concerns the general
form for an expectation function $E(O)$ on the observables, and a
proof that the
function $E(O)$ so developed cannot be a value map function. Because the assumption
of the real-linearity of $E(O)$ is an unjustified one, the work of von Neumann
does {\em not} imply the general failure of hidden variables. Finally, we noted
that assuming {\em only} the real-linearity of $E$, one may easily arrive at the
conclusion that such a function cannot be a map to the observables'
eigenvalues. Ultimately, the lesson to be learned from von Neumann's
theorem is simply that
there exists no mathematical function from observables to their values 
obeying the requirement of real-linearity.

Abner Shimony has reported \cite{Shimony} that Albert Einstein was aware of
both the von Neumann analysis itself and the reason it fails as a hidden
variables impossibility proof. The source of Shimony's report was a personal
communication with Peter G. Bergmann. Bergmann reported that during a 
conversation with Einstein
regarding the von Neumann proof, Einstein opened von Neumann's
book to the page where the proof is given, and pointed to the linearity
assumption. He then said that there is no reason why this premise should 
hold in a state not acknowledged by quantum mechanics, if the observables are 
not simultaneously measurable. Here the ``state not acknowledged by quantum
mechanics''  seems to refer to von Neumann's dispersion-free state, i.e. the
state specified by $\psi$,and $\lda$. It is almost certain that Erwin 
Schr{\"o}dinger
would also have realized the error in von Neumann's impossibility proof, since 
in
his 1935 paper \cite{Present Sit} he gives a derivation which is equivalent to
von Neumann's theorem in so far as hidden variables are concerned, yet he does
not\footnote{In fact, if \Erwin {\em had} interpreted his result this way,
this---in light of his own generalization of the EPR paradox presented in the
same paper--would have allowed him to reach a further and very striking
conclusion. We shall discuss this in chapter 4.} 
arrive at von Neumann's conclusion of the impossibility of hidden variables. 
We shall discuss \Erwins
derivation in the next section. In view of the scarcity\footnote{See
Max Jammer's
book \cite[p. 265]{Jammer Philosophy}. Jammer mentions that not only
was there
very little response to von Neumann's impossibility proof, but the book itself
was never given a review before 1957, with the exception of two brief works by
Bloch and Margenau (see the Jammer book for these references).} of early
responses to von Neumann's proof, it is valuable to have such evidence of
Einstein's and \Erwins awareness of the argument and its shortcomings. In
addition, this affirms the notion that Einstein regarded the problem
of finding a
complete description of quantum phenomena to be of central importance (see
again the Einstein quotes given in sections \ref{Bohm} and \ref{Vint}).

In our introduction to von Neumann's theorem, we stated that the existence of a
deterministic hidden variables theory led to the result that for each $\psi$ and
$\lda$, there exists a value map on the observables. We represented such value
maps by the expression $V^{\psi}_{\lda}(O)$. If one considers the question of
hidden variables more deeply, it is clear that the agreement of their
predictions with those of quantum mechanics requires an additional criterion 
that
beyond the existence of a value map for each $\psi$ and $\lda$: it requires
agreement with the {\em statistical} predictions of the quantum formalism.
To make
possible the empirical agreement of quantum theory, in which only statistical
predictions are generally possible with the deterministic description of a hidden
variables theory, we regard their descriptions of a quantum system in the
following way. The quantum mechanical state given by $\psi$ corresponds to a {\em
statistical ensemble} of the states given by $\psi$ and $\lda$; the members of
the ensemble being described by the same $\psi$, but differing in 
$\lda$. The variation in measurement results found for a series of quantum
mechanical systems with identical $\psi$ is to be accounted for by the  
variation in parameter $\lda$ among the ensemble of $\psi,\lda$ states.
For precise agreement in this regard, we require that for all $\psi$ and $O$, 
the
following relationship must hold: 
\beq
 \int^{\infty}_{-\infty} d \lda \rho (\lda)
\V  = \sbra O \sket \mbox{,} \label{eq:Expl} 
\eeq 
where $\rho(\lda)$ is the
probability distribution over $\lda$.

We have seen from von Neumann's result and from our simple examination of the
spin $\frac{1}{2}$ observables $\sigma_{x},\sigma_{y},
\frac{1}{\sqrt{2}}(\sigma_{x}+\sigma_{y})$, 
that it is impossible to develop a linear function mapping the observables to 
their 
eigenvalues. We shall find also that an impossibility proof 
may be developed showing that the criterion of agreement with the quantum 
statistics, i.e., the agreement with \ref{eq:Expl}, cannot be met by  
functions of the form $\V$. Bell's theorem is, in fact, such a proof. We will 
present the proof of
Bell's result along with some further discussion in chapter $3$.

\subsection{\Erwins derivation of von Neumann's `impossibility proof' 
\label{Schrvon}} 
As mentioned above, in
his famous ``cat paradox'' paper \cite{Present Sit}, \Erwin presented an
analysis which, as far as hidden variables are concerned, was essentially 
equivalent to the von Neumann proof. \Erwins study of the problem was
motivated by the results of his generalization of the 
Einstein--Podolsky--Rosen paradox. While EPR had concluded definite values on 
the position and momentum observables only, \Erwin was able to show that such 
values must exist for {\em all} observables of the state considered by EPR. 
We will discuss both the original
Einstein--Podolsky--Rosen analysis, and \Erwins generalization thereof in much
more detail in chapter 4. To probe the possible relationships which might 
govern the values assigned to the various observables, \Erwin then gave a brief 
analysis
of a system whose Hamiltonian takes the form 
\beq H=p^{2}+a^{2}q^{2} 
\mbox{.}
\label{eq:HrO} 
\eeq 
We are aware from the well-known solution of the harmonic
oscillator problem, that this Hamiltonian's eigenvalues are given by the set 
$\{a
\hbar, 3a \hbar,5a \hbar, 7a \hbar, \ldots \}$. Consider a mapping $V(O)$ from
observables to values. If we require that the assignments $V$ makes to the
observables $H,p,q$ satisfy  \ref{eq:HrO} then we must have  \beq
V(H)=(V(p))^{2}+a^{2}(V(q))^{2} \mbox{,} \label{eq:HrOV} \eeq which implies \beq
((V(p))^{2}+ a^{2}(V(q))^{2})/a\hbar= \mbox{an odd integer} \mbox{.}
\eeq This latter relationship cannot generally be satisfied by the
eigenvalues of $q$, and $p$---each of which may be any real number---and an
arbitrary positive number $a$.

The connection of this result to the von Neumann argument is immediate. In the
discussion of section \ref{Vonimp}, we noted that the value of the observable
$f(O)$ will be $f$ of the value of $O$, so that any value map must satisfy
$f(V(O))=V(f(O))$, as given in equation \ref{eq:11}. Here $f$ may be any
mathematical function. It follows from this that \ref{eq:HrOV} is equivalent to a
relation between the observables $H$, $p^{2}$, and $q^{2}$ given by: \beq
V(H)=V(p^{2})+a^{2}V(q^{2}) \mbox{.} \label{eq:HrOVE} \eeq With the known
eigenvalues of $H$, this leads to 
\beq (V(p^{2})+a^{2}V(q^{2}))/a\hbar= \mbox{an
odd integer} \mbox{,} \eeq which cannot generally be satisfied by the 
eigenvalues
of $q^{2}$, and $p^{2}$---each of which may be any positive real number---and 
an arbitrary positive number $a$. We have here another example leading to a 
demonstration of von Neumann's result that there is no linear value map on
the observables (Recall the example of the spin component
observables---$\sigma_{x}$,$\sigma_{y}$ and
$\sigma^{'}=\frac{1}{\sqrt{2}}(\sigma_{x}+\sigma_{y})$ given above.). If we
consider the von Neumann function $E(O)$, the real-linearity
assumption requires
it to satisfy \ref{eq:HrOVE}. Therefore, $E(O)$ cannot map the observables to
their eigenvalues. \Erwin did not regard this as proof of the impossibility of
hidden variables, as von Neumann did, but concluded only that the relationships
such as \ref{eq:HrO} will not
necessarily be satisfied by the value assignments made to the 
observables constrained by such a relation.  Indeed, if \Erwin
{\em had} made von Neumann's error of interpretation, this would contradict
results he had developed previously according to which such hidden variables 
must
exist. Such a contradiction would have allowed \Erwin to reach a further
conclusion which is quite striking, and which we shall discuss in chapter 4.

\setcounter{chapter}{1} 
\chapter{Contextuality} We have seen in the
previous section that the analysis of von Neumann has little impact
on the question of whether a viable hidden variables theory may be
constructed. However, further mathematical results were
developed  by Andrew M. Gleason \cite{Gleason} in 1957 and by Simon
Kochen and E.P. Specker in 1967 \cite{Kochen Specker}, which were
claimed by some\footnote{See \cite{Kochen Specker, Mermin's  Gleason, Reality
Marketplace}.} to imply the impossibility of hidden
variables. In the words of Kochen and Specker \cite [p. 73]{Kochen 
Specker}: ``If a physicist X believes in hidden variables... ...the
prediction of X contradicts the  prediction of
quantum mechanics''. The Gleason, and Kochen and Specker arguments are in 
fact, {\em stronger} than von Neumann's in that
they assume linearity only for {\em commuting} observables. Despite 
this, a close analysis reveals that the impossibility proofs
of Gleason and of Kochen and Specker share with
von Neumann's proof the neglect of the possibility of a hidden variables 
feature called {\em contextuality} \cite{Bell Eclipse, Bell Imposs Pilot}.
We will find that this shortcoming
makes these theorems inadequate as proofs of the impossibility of 
hidden variables.

We begin this chapter
with the presentation of Gleason's theorem and the theorem of Kochen and 
Specker. We then discuss a more recent theorem discovered by 
Mermin\footnote{See Mermin in 
\cite{Mermin's Gleason, Mermin Review}.},
which is similar to these, and yet admits a much simpler proof. 
This will be followed by a discussion of 
contextuality and its relevance to these analyses. We will make clear
in this discussion why the theorems in question fail as impossibility
proofs. As far as the question of what conclusions {\em do} follow, 
we show that these theorems' implications can be expressed in a simple and 
concise fashion, which we refer to as ``spectral-incompatibility''.
We conclude the chapter with the discussion of an experimental 
procedure first discussed by Albert\footnote{See D. Albert \cite{Albert}.} 
which provides further insight into contextuality.
 
\section{Gleason's theorem} Von Neumann's theorem
addressed the question of the form taken by a function $E(O)$ of the
observables. Gleason's theorem essentially addresses the same 
question\footnote{The original form presented by A.M. Gleason referred
to  a probability measure on the subspaces of a Hilbert space, but the
equivalence of such a construction with a value map on the projection
operators is simple and immediate.  This may be seen by considering
that there is a one-to-one correspondence of the subspaces and
projections of a Hilbert space and that the values taken by the
projections are $1$ and $0$, so that a function mapping projections to
their eigenvalues  is a special case of a probability measure on these
operators.}, the most significant difference being that the linearity
assumption is relaxed to the  extent that it is demanded that $E$ be
linear on only {\em commuting} sets of observables. Besides
this, Gleason's theorem involves a function $E$ on only the projection
operators of the system, rather than on all observables.
Finally, Gleason's  theorem contains the assumption that the system's
Hilbert space is at least  three dimensional. As for the conclusion of
the theorem, this is identical to von Neumann's: $E(P)$ takes the form
$E(P)=\mbox{Tr}(UP)$ where $U$ is a positive operator and
$\mbox{Tr}(U)=1$.

Let us make the requirement of linearity
on the commuting observables somewhat more explicit. First we note
that any
set of projection $\{P_{1},P_{2}, \ldots\}$ onto mutually orthogonal
subspaces  $\{{\cal H}_{1},{\cal H}_{2},\ldots\}$ will form a commuting
set. Furthermore, if $P$ projects onto the direct sum ${\cal H}_{1}
\oplus  {\cal H}_{2} \oplus \ldots$ of these subspaces, then
$\{P,P_{1},P_{2},\ldots\}$  will also form a commuting set. It is in the
case of this latter type of set that the linearity requirement comes
into play, since these observables obey the relationship 
\beq
P=P_{1}+P_{2}+\ldots 
\label{eq:GlPsum} \mbox{.} 
\eeq 
The condition on
the function $E$ is then 
\beq E(P)=E(P_{1})+E(P_{2})+\ldots \mbox{.}
\label{eq:comm lin} 
\eeq  
The formal statement of Gleason's theorem is
expressed as follows. For any quantum system whose Hilbert space is at
least three dimensional, any expectation function  $E(P)$ obeying
the conditions \ref{eq:comm lin}, $0 \leq E(P) \leq 1$, and $E({\bf
1})=1$ must take the form 
\beq
E(P)=\mbox{Tr}(UP)
\label{eq:Gltracedude}
\mbox{,}
\eeq
 where $\mbox{Tr}(U)=1$
and $U$ is a positive operator. We do not present the proof of this 
result\footnote{See Bell \cite{Bell Eclipse}. Bell proves that any
function
 $E(P)$ satisfying the conditions of Gleason's theorem cannot map 
the projection operators to their eigenvalues.} here. In the next section,
we give the outline the proof of Kochen and Specker's theorem. The same 
impossibility result derived from \gls theorem
follows from this theorem.
.

It is straightforward to demonstrate that
the function $E(P)$ considered within Gleason's theorem cannot be a value map 
function on these observables.
To demonstrate this, one may argue in the same fashion as was done by von 
Neumann, since the form developed here for $E(P)$ is the same as that 
concluded 
by the latter. (See section \ref{Vonimp}). We recall that if $E(O)$ is
to represent a dispersion free
state specified by $\psi$ and $\lda$, it {\em must} take the form
of such a value map,
and $E(O)$ evidently cannot be the expectation function
for such a state. It is on this basis that the impossibility of hidden
variables has been claimed to follow from Gleason's theorem.

\section{The theorem of Kochen and Specker}
As with Gleason's theorem, the essential assumption
of Kochen and Specker's theorem is
that the expectation function $E(O)$ must be linear on commuting sets of 
observables. It differs from the former only in the set of observables
considered. Gleason's theorem was addressed to
the projection operators on a Hilbert space of arbitrary dimension $N$.
Kochen and Specker consider the squares
$\{s^{2}_{\theta,\phi}\}$ of the spin components of a spin $1$ particle. 
One may note that these
observables are formally identical to projection
operators on a three-dimensional Hilbert space.
Thus, the \ks observables are a 
subset\footnote{The set of projections on a three-dimensional space
is actually a larger class of observables.} 
of the ``$N=3$'' case of the \gl 
observables. Among 
the observables $\{s^{2}_{\theta,\phi}\}$ any subset  
$\{s^{2}_{x},s^{2}_{y},s^{2}_{x}\}$ corresponding
to mutually orthogonal component directions $x,y,z$ will be a commuting set.
Each such set obeys the relationship
\beq
s^{2}_{x}+s^{2}_{y}+s^{2}_{z}=2
\label{eq:KS comm}
\mbox{.} 
\eeq
For every such subset, Kochen and Specker 
require
that $E(s^{2}_{\theta,\phi})$ must obey 
\beq
E(s^{2}_{x})+E(s^{2}_{y})+E(s^{2}_{z})=2
\mbox{.}
\label{eq:KS proj lin}
\eeq
Kochen and Specker theorem states that there exists no
 function $E(s^{2}_{\theta,\phi})$ on the squares of the spin components  
of a spin $1$ particle which maps each observable to either $0$ or $1$ and
which satisfies \ref{eq:KS proj lin}.

We now make some comments regarding the nature of this theorem's proof. 
The problem becomes somewhat simpler to discuss when formulated in 
terms of a geometric model. Imagine a sphere of unit radius surrounding 
the origin in $\R^{3}$. It is easy to see that
each point on this sphere's surface corresponds to a direction in space,
which implies that each point is associated with one observable of the set
$\{s^{2}_{\theta,\phi}\}$. 
With this, $E$ may be regarded as a function on
the surface of the unit sphere. Since the eigenvalues
of each of these spin observables are $0$ and $1$, it
must be that $E(O)$ must take on these values, if it
is to assume the form of a value map function. Satisfaction
of \ref{eq:KS proj lin} requires that for each set of mutually orthogonal 
directions, $E$ must 
assign to one of them $0$ and to the other directions $1$. To gain some 
understanding\footnote{We follow here the argument given in Belinfante
\cite[p. 38]{1974 Belinfante Book}}
 of why such an assignment of values must fail, we proceed as follows.
To label the points on the sphere, we imagine
that each point on the sphere to which the number $0$ is assigned is
painted red, and each point assigned $1$ is painted blue. We label each
direction as by the unit vector $\hat{n}$. Since one
direction of every mutually perpendicular set is assigned red, then in
total we require that one-third of the sphere is painted red.
If we consider the components of the spin in {\em opposite} directions
$\theta,\phi$, and $180^{\circ}-\theta,180+\phi$, these values are
always opposite, i.e., if $s_{x}$ takes the value $+1$, then $s_{-x}$
takes the value $-1$. This implies that the values of $s^{2}_{\theta,\phi}$
and $s^{2}_{180^{\circ}-\theta,180+\phi}$ will be {\em equal}.
There, we must have that points on the sphere lying
directly opposite each other, i.e. the ``antipodes'', must receive
the same assignment from $E$. Suppose that one direction and its antipode 
are painted red. These points form the two poles of a great circle, and all points along this circle must then be painted blue,
since all such points represent directions orthogonal to the directions
of our two `red' points. For every such pair of red points on the sphere,
there must be many more blue points introduced, and we will find this makes
it impossible to make one-third of the sphere red, as would be necessary
to satisfy \ref{eq:KS proj lin}.

Suppose we paint the entire first octant of the sphere red. In terms of
the coordinates used by geographers, this is similar to
the region in the Northern hemisphere between $0$ degrees and $90$
degrees longitude. If the point at the north pole is painted red then
the great circle at the equator must be blue. Suppose that
the $0$ degree and $90$
degree meridians are also painted red. Then the octant which is the antipode
of the first octant
must also be painted red. This octant would be within the `southern
hemisphere' between $180$ degrees and $270$ degrees longitude. If we now
apply the condition that for every point painted red, all points lying
on the great circle defined with the point at its pole, we find that all
remaining points of the sphere must then be colored blue, thereby
preventing the addition of more red points. This assignment fails to
meet the criterion that {\em less} than one-third of the sphere is
blue, so there are some sets of mutually
perpendicular directions which are all colored blue. An example of such a
set of directions is provided by the points on the sphere's surface
lying in the mutually orthogonal directions represented by 
$(.5,.5,-.7071)$, $(-.1464,.8535,.5)$,
$(.8535,-.1464,.5)$. Each of these points lies in a quadrant to which we
have assigned the color blue by the above scheme. This assignment of
values to the spatial directions must therefore fail to meet the criteria
demanded of the function  $E(s^{2}_{\theta,\phi})$ in Kochen and Specker's premises: that $E(s^{2}_{\theta,\phi})$ satisfies 
 \ref{eq:KS proj lin} and maps the observables to their
values. 

In their proof, Kochen and Specker show that for a discrete
set of $117$ different directions in space,
it is impossible to give appropriate value assignments to the
corresponding spin observables\footnote{Since
their presentation, the proof has been simplified by Peres in 1991
\cite{Peres' Gleason} whose proof is based on examination of 33 such
$\hat{n}$ vectors. We also note that a proof presented by Bell
\cite{Bell Eclipse} may be shown to lead easily to a proof of Kochen and
Specker. See Mermin in this connection \cite[p. 806]{Mermin Review}.}.
 Kochen and Specker then assert that hidden 
variables cannot agree
 with the predictions of quantum mechanics. Their conclusion 
is that if some physicist `X', mistakenly
decides to accept the validity of hidden variables then
``the prediction of X (for some measurements) contradicts the 
prediction of quantum mechanics.'' \cite[p 73]{Kochen Specker}
The authors cite a particular system on which one
can perform an experiment
they claim reveals the failure of the hidden variables prediction.
We will demonstrate in the next section of this work that what
follows from Kochen and Specker's
theorem is only that a {\em non-contextual} hidden
variables theory will conflict with quantum mechanics, so that
the general possibility of hidden variables has not been disproved.
Furthermore, we show that if the type of experiment envisioned 
by these authors is considered in more detail, it does not indicate
where hidden variables must fail, but instead serves as an illustration
indicating that the requirement of contextuality is a quite natural one.

\subsection{Mermin's theorem \label{Mermin}}
We would like to discuss here a much more recent theorem discovered by
N. David 
Mermin\footnote{See \cite{Mermin's Gleason} and \cite{Mermin Review}}
 in 1990, which is of the same character as those considered
above. Mermin makes essentially the same assumption regarding the 
function $E(O)$
as did Gleason, and Kochen and Specker: that
$E(O)$ must obey all 
relationships among the commuting sets of observables.
This theorem is more straightforward in its
proof and simpler in form than those of \gl and of Kochen and Specker.

The system addressed by Mermin's theorem is that of a pair of spin
$\frac{1}{2}$
particles. The observables involved are the $x$ and $y$ components of these
spins, and six other observables which are defined in terms of these four.
We begin with the derivation of the expression
\beq
\sgox\sgty\sgoy\sgtx\sgox\sgtx\sgoy\sgty=-1
\label{eq:PM}
\mbox{,}
\eeq
since this is actually quite crucial to the theorem.
For simplicity, we normalize the spin eigenvalues from $\pm \frac{1}{2}$
to $\pm 1$. To demonstrate \ref{eq:PM}, we make use of the \linebreak[0] 
commutation \linebreak[0] rules \linebreak[0] for \linebreak[0] 
the \linebreak[0] $x$ \linebreak[0] and \linebreak[0] $y$ \linebreak[0] 
 \linebreak[0] components  \linebreak[0] of \linebreak[0] two \linebreak[0]
 spin $\frac{1}{2}$ \linebreak[0] particles. \linebreak[0] Any \linebreak[0] 
 pair \linebreak[0] of \linebreak[0] such \linebreak[0]
observables \linebreak[0] associated \linebreak[0] with
 \linebreak[0] different \linebreak[0] particles \linebreak[0] will
 \linebreak[0] commute, \linebreak[0] so \linebreak[0] that
 \linebreak[0] we \linebreak[0] have \linebreak[0] $[\sgox,\sgtx]=0$
 \linebreak[0] and \linebreak[0] $[\sgox,\sgty]=0$, \linebreak[0] for 
 \linebreak[0] example. \linebreak[0] Any \linebreak[0] pair
 \linebreak[0] which \linebreak[0] involves \linebreak[0] the \linebreak[0]
 same \linebreak[0] component \linebreak[0] will \linebreak[0] also  
\linebreak[0] 
commute, \linebreak[0] thus \linebreak[0] we \linebreak[0] have \linebreak[0]
$[\sgox,\sgox]=0$ \linebreak[0] and \linebreak[0]
$[\sgtx,\sgtx]=0$. \linebreak[0]
 Note \linebreak[0] that \linebreak[0] commutation \linebreak[0] of 
 \linebreak[0] two  \linebreak[0]
observables \linebreak[0] $O_{1},O_{2}$ \linebreak[0] implies \linebreak[0]
 that \linebreak[0] $O_{1}O_{2}=O_{2}O_{1}$.
 \linebreak[0] Those \linebreak[0]
 \linebreak[0] pairs \linebreak[0] associated \linebreak[0] with  \linebreak[0]
the  \linebreak[0] same   \linebreak[0] particle   \linebreak[0] but   
\linebreak[0] {\em different}   \linebreak[0] components   \linebreak[0] do
  \linebreak[0] not   \linebreak[0] commute,   \linebreak[0] but
\linebreak[0] 
{\em anticommute}.   \linebreak[0] For   \linebreak[0] two   \linebreak[0] 
anticommuting   \linebreak[0] observables   \linebreak[0] $O_{1},
O_{2}$,   \linebreak[0] it   \linebreak[0] follows   \linebreak[0] that   
\linebreak[0]   \linebreak[0] their   \linebreak[0] anticommutator 
  \linebreak[0] $[O_{1},O_{2}]^{+}=
O_{1}O_{2}+O_{2}O_{1}$ is equal to $0$. This implies that
$O_{1}O_{2}=-O_{2}O_{1}$.
Using these rules, we may manipulate the expression 
on the left hand side 
of \ref{eq:PM} by sequentially interchanging the first $\sgox$ with the 
operator appearing to its right. If we exchange $\sgox$ with $\sgty$,
$\sgoy$ and $\sgtx$ the expression becomes $-\sgty\sgoy\sgtx\sgox\sgox
\sgtx\sgoy\sgty$.
 The overall minus sign results from its interchange with $\sgoy$. At this
point, it is straightforward to simplify the expression using that
the square of any component of the spin has the value $1$, i.e. 
$(\sigma^{(i)}_{j})^{2}=1$. If we apply this to the expression in question, we
can easily see that the entire expression reduces to $-1$, thereby 
verifying \ref{eq:PM}.

Motivated by the expression \ref{eq:PM}, we introduce six additional 
observables 
\beq
\{A,B,C,X,Y,Z\}
\mbox{.}
\eeq
 If we group the 
observables in the left hand side of \ref{eq:PM} by pairs,
we can rewrite this relationship as $ABXY=-1$ where $A,B,X,Y$ are defined as
\begin{eqnarray}
A & = & \sgox\sgty \label{eq:Pduct} \\
B & = & \sgoy\sgtx \nonumber \\ 
X & = & \sgox\sgtx \nonumber \\
Y & = & \sgoy\sgty \nonumber 
\mbox{.}
\end{eqnarray}
Defining observables $C$ and $Z$ as
\begin{eqnarray}
C & = & AB \label{eq:SPduct}  \\
Z & = & XY \nonumber
\end{eqnarray}
then allows one to write \ref{eq:PM} as 
\beq
CZ=-1
\mbox{.}
\label{eq:IPduct}
\eeq
It is important to note that the equation \ref{eq:IPduct} 
is equivalent to \ref{eq:PM}, given that $A,B,C,X,Y,Z$ are defined as given
above in \ref{eq:SPduct} and \ref{eq:Pduct}.

If we examine the first equation in \ref{eq:Pduct}, we see that
the three observables
involved are a commuting set: $[\sgox,\sgty]=0$, $[\sgox,
A]=[\sgox,\sgox\sgty]=0$ and 
$[\sgty,A]=[\sigma^{(2)}_{y},\sigma^{(1)}_{x}\sgty]=0$. Examination of
the other equations in \ref{eq:Pduct} reveals that 
the same holds true for 
these, i.e. the observables in each form a commuting set. Repeated
application of the commutation rules reveals that the sets $\{C,A,B\}$, 
$\{Z,X,Y\}$, and $\{C,Z\}$ are also commuting sets. As was done in \gls
theorem and \kss theorem, we 
consider the question of a function $E(O)$ on the observables
$\sgox,\sgoy,\sgtx,\sgty,A,B,C,X,Y,Z$ which returns for each observable
its value. We require that $E(O)$ satisfy all constraining relationships
on each commuting set of observables.
The relationships in question differ from those of Gleason 
(\ref{eq:comm lin}), and of Kochen and Specker (\ref{eq:KS proj lin}) 
only in that they involve {\em products} of observables, rather than 
just linear combinations. However, this distinction is not a 
significant one---the essential
feature of all of these theorems is simply that they require $E(O)$ to 
satisfy the relationships constraining each {\em commuting} set; no 
relations on non-commuting observables are regarded as necessary 
constraints. The relationships in question in the present analysis 
are the defining equations in \ref{eq:Pduct}, \ref{eq:SPduct} and
\ref{eq:IPduct}. Given that $E(O)$ satisfies these, the Mermin
theorem implies that this function cannot map the observables
to their eigenvalues. 

We will now present the proof of the theorem. Since $\sigma^{(i)}_{j} = \pm 
1 \mbox{ } \forall i,j$, the eigenvalues
of each of the ten observables will be $\pm 1$. This is readily
seen for the observables $A,B,X,Y$ defined by \ref{eq:Pduct}: each of 
these is the product of two commuting observables whose eigenvalues 
are $1$ and $-1$. With this, it follows similarly that
$C$ and $Z$ each have eigenvalues of $\pm 1$. We therefore
require that $E(O)$ must assign either $-1$ or $1$ to each observable. 
Now let us recall an important result which we saw above: the relationship
$CZ=-1$ is {\em equivalent} to the relationship 
\ref{eq:PM}, provided that the observables $A,B,C,X,Y,Z$ are defined
by \ref{eq:SPduct}, and 
\ref{eq:Pduct}. 

Consider the function $E(O)$. Since
the assignments made by this function have been required to satisfy all the 
commuting
relationships \ref{eq:IPduct}, \ref{eq:SPduct}, and \ref{eq:Pduct},
it follows that they must also satisfy
\ref{eq:PM}. However, if we examine \ref{eq:PM}, it follows that no
assignment of the values $-1$ and $+1$ can possibly be made which will
satisfy this equation. This follows
since each of the spin observables appears {\em twice} in the left hand 
side of
\ref{eq:PM}, so that any such assignment must give a value of $1$ to the
entire expression, but the right hand side of \ref{eq:PM} is $-1$. 
Therefore, there is no 
function $E(O)$ which maps each observable
of the set $\sgox,\sgoy,\sgtx,\sgty,A,B,C,X,Y,Z$ to an eigenvalue, if we insist
that $E(O)$ on every commuting set must obey all the constraint
equations. This 
completes the proof of Mermin's theorem.

\section{Contextuality and the refutation of the impossibility proofs
of Gleason, Kochen and Specker, and Mermin}
We have seen that the theorems of Gleason, Kochen and Specker, and 
Mermin all demonstrate the impossibility of value maps on
some sets of observables such that the constraining relationships
on each commuting set of observables are obeyed.
One might be at first inclined to conclude with Kochen and 
Specker that such results imply the impossibility of a hidden 
variables theory. However, if we consider that there exists
a successful theory of hidden variables, namely Bohmian mechanics
\cite{Bohmian Mechanics} (see section \ref{Bohm}), we see that
such a conclusion is in error. Moreover, an explicit analysis
of Gleason's theorem has been carried out by Bell \cite{Bell Eclipse, Bell 
Imposs Pilot} and its inadequacy as an impossibility proof was shown. 
Bell's argument may easily be 
adapted\footnote{As we have mentioned, since the \ks observables are formally 
equivalent to projections on a three-dimensional Hilbert space, this 
theorem is actually a special case of Gleason's. Therefore, Bell's 
argument essentially addresses the Kochen and Specker theorem as well as 
Gleason's.} 
to provide a similar demonstration regarding Kochen and Specker's theorem.
The key concept underlying Bell's argument is that of {\em contextuality}, and 
we now present a discussion of this notion.

 Essentially, contextuality refers 
to the dependence of measurement results on the detailed experimental 
arrangement being employed. In discussing this notion,
we will find that an inspection of the quantum formalism suggests 
that contextuality is a natural feature to expect in a theory explaining the 
quantum phenomena. Furthermore we shall find that the concept
is in accord with Niels Bohr's remarks regarding the fundamental
principles of quantum mechanics. According to Bohr,
\cite{Bohr Nature} ``a closer examination reveals that the procedure of 
measurement has an 
essential influence on the conditions on which the very definition of the 
physical quantities in question rests.'' In addition, he 
stresses 
\cite[p. 210]{Einstein impeachment} ``the impossibility of any sharp
distinction 
between the
behavior of atomic objects and the interaction with the measuring
instruments which serve to define the conditions under which the phenomena
appear.'' 
The concept of contextuality represents a concrete manifestation of
the quantum theoretical aspect to which Bohr refers. We will first
explain the concept itself in detail,
and then focus on its relevance to the theorems 
of Gleason, Kochen and Specker, and Mermin.

We begin by recalling a particular feature of the quantum formalism.
 In the presentation of this
formalism given in chapter one, we discussed the 
representation of the 
system's state, the rules for the state's time evolution, and
the rules governing the measurement of an observable. The measurement
rules are quite crucial, since it is only through 
measurement 
that the physical significance of the abstract quantum state (given 
by $\psi$) is made manifest. Among these rules,  
one finds that any commuting set
of observables may be measured simultaneously. With a little 
consideration, one is led to observe 
that the possibility exists for {\em a variety of different  
experimental procedures} to measure a single observable. 
Consider for example, an observable 
$O$ which is a member of the commuting set 
$\{O,A_{1},A_{2},\ldots\}$. We label this set as ${\cal C}$. A 
simultaneous measurement of the set ${\cal C}$ certainly
gives among its results a value for $O$ and thus may be regarded as 
providing a measurement of $O$. It is possible that $O$ may
be a member of another commuting set of
observables ${\cal C}^{'}=\{O,B_{1},B_{2}, \ldots\}$, so that a 
simultaneous measurement of ${\cal C}^{'}$
\linebreak[0] also provides a measurement of $O$. 
Let us suppose further that the members of set $\{A_{i}\}$ fail 
to commute with those of $\{B_{i}\}$. It is then clear that experiments
measuring ${\cal C}$ and ${\cal C}^{'}$ are quite different, and hence 
must be distinct. A concrete difference appears for example,
in the effects of such experiments on the system wave function. The
measurement rules tell us that the wave function subsequent to an
ideal measurement of a commuting set 
is prescribed by the equation \ref{eq:NIC}, according to which the
post-measurement wave function is calculated from the 
pre-measurement wave function by taking the projection of the latter
into the joint-eigenspace of that set.
Since the members of ${\cal C}$ and ${\cal C}^{'}$ fail to
commute, the joint-eigenspaces of the two are necessarily different, and 
the system wave function will not
generally be affected in the same way by the two experimental 
procedures. Apparently 
\linebreak[0] the concept of `the measurement of an
\linebreak[0] observable' is 
\linebreak[0] {\em ambiguous}, since there can be distinct experimental
 procedures for the measurement of a single observable.

There are, in fact, more subtle distinctions between different procedures 
for measuring the same observable, and these also may be important.
We will see such an example in studying Albert's experiment at the conclusion
of this chapter. To introduce the experimental procedure of
measurement into
our formal notation, we shall write ${\cal E}(O)$, ${\cal E}^{'}(O)$, 
etc., to represent experimental 
procedures used to measure the observable $O$.
From what we have seen here, it is quite natural to expect
that a hidden variables theory should allow for the possibility 
that {\em different
experimental procedures}, e.g., ${\cal E}(O)$ and ${\cal E}^{'}(O)$, for the 
measurement of an observable
might yield {\em different results} on an individual system. This is 
contextuality.

Examples of observables for which there exist incompatible measurement 
procedures are found among the observables addressed in each of the
theorems of Gleason, Kochen and Specker, and Mermin.
Among observables addressed by Gleason are the 
one-dimensional projection operators $\{P_{\phi}\}$
on an $N$-dimensional Hilbert space ${\cal H}_{N}$. 
Consider a one-dimensional projection $P_{\phi}$ where $\phi$ belongs 
to two sets of orthonormal vectors given by 
$\{\phi,\psi_{1},\psi_{2},\ldots\}$ and 
$\{\phi,\chi_{1},\chi_{2},\ldots\}$. Note that the sets $\{\psi_{1},\psi_{2},
\ldots\}$ and $\{\chi_{1},\chi_{2},\ldots\}$ are constrained only in that 
they span ${\cal H}_{\phi}^{\perp}$ (the
orthogonal complement of the one-dimensional space spanned by $\psi$). 
Given this, there exist examples of such sets for which some 
members of  $\{\psi_{1},\psi_{2},\ldots\}$
are distinct from and not orthogonal to the vectors in $\{\chi_{1},\chi_{2},
\ldots\}$.
Since any distinct vectors that are not orthogonal correspond to projections 
which fail to commute,
 the experimental 
procedures ${\cal E}(P_{\phi})$ measuring $\{P_{\phi},P_{\psi_{1}},
P_{\psi_{2}},\ldots\}$ and ${\cal E}^{'}(P_{\phi})$
measuring $\{P_{\phi},P_{\chi_{1}},P_{\chi_{2}},\ldots\}$ are 
incompatible. The argument just given applies also to the
Kochen and Specker observables (the squares of the 
spin components of a spin $1$ particle), since these  
are formally identical to projections on a three-dimensional Hilbert 
space. To be explicit, 
the observable $s^{2}_{x}$ is a member
of the commuting sets $\{s^{2}_{x},s^{2}_{y},s^{2}_{z}\}$ and 
$\{s^{2}_{x},s^{2}_{y^{'}},
s^{2}_{z^{'}}\}$ where the $y^{'}$ and $z^{'}$ are oblique relative to
the $y$ and $z$ axes. In this case, $s^{2}_{y^{'}},s^{2}_{z^{'}}$ do not 
commute with $s^{2}_{y},s^{2}_{z}$. Thus, the experimental procedures to 
measure these sets are incompatible.
If we examine the Mermin observables, we find that here also,
each is a member of two incompatible commuting sets.
For example $\sgox $ belongs to $\{\sgox ,\sgty ,A\}$,
and to $\{\sgox ,\sgtx ,X\}$. Here, the observables $\sgty ,A$ do not 
commute with $\sgtx ,X$, so that the 
experimental procedures ${\cal E}(\sgox)$ and ${\cal E}^{'}(\sgox)$
that entail respectively the measurement of $\{\sgox ,\sgty ,A\}$ and 
$\{\sgox ,\sgtx ,X\}$ are incompatible.

While it is true that the arguments against hidden
variables derived from these theorems are superior to von Neumann's, since
they require agreement only with operator relationships among 
commuting sets, these arguments nevertheless possess
the following shortcoming. Clearly, the mathematical functions 
considered in each
case, $E(P)$,$E(s^{2}_{\theta,\phi})$ and $E(\sigma^{(1)}_{x},
\sigma^{(1)}_{y},\sigma^{(2)}_{x},\sigma^{(2)}_{y},
A,B,C,X,Y,Z)$ do {\em not} allow for the
possibility that the results of measuring each observable using different
and possibly {\em incompatible} procedures may lead to different results. 
 What the theorems demonstrate is that 
no hidden variables formulation {\em based on assignment of a unique
value to each observable} can possibly agree with quantum
mechanics. But this is a result we might well have expected from the
fact that the quantum formalism allows the possibility of incompatible 
experimental procedures for the measurement of an 
observable. For this reason, none of the theorems
here considered---Gleason's theorem, 
Kochen and Specker's theorem and Mermin's theorem---imply the
impossibility of hidden variables, since they fail to account for
such a fundamental feature of the quantum formalism's rules of measurement.

\subsubsection{Discussion of a procedure to measure the Kochen and Specker
observables}
In a discussion of the implications of their theorem, Kochen and Specker
mention a system for which 
well-known techniques of atomic spectroscopy
may be used to measure the relevant spin observables. 
Although these authors mention this 
experiment to support their case against hidden variables, the
examination of such an experiment actually reinforces the 
assertion that one should allow for contextuality---the
very concept that {\em refutes} their argument against hidden variables.

Kochen and Specker note\footnote{One can derive the analogous first-order 
perturbation term 
arising for a charged particle of orbital angular momentum $L=1$ in
such an electric field  using the fact that the joint-eigenstates of 
$L^{2}_{x},L^{2}_{y},L^{2}_{y}$ are the eigenstates of the potential energy 
due to the field. This latter result is shown in Kittel \cite[p. 
427]{Kittel}.} that for an atom of 
{\em orthohelium}\footnote{Orthohelium and parahelium are two species
of helium which are distinguished by 
the total spin $S$ of the two electrons: for the former we have $S=1$,
and for the latter $S=0$. There is a rule of atomic spectroscopy 
which prohibits atomic transitions for which $\Delta S=1$, so that no 
transitions from one form to the other can occur spontaneously.}
which is subjected to an
electric field of a certain configuration, the first-order
effects of this field on the electrons may be accounted for by 
adding a term of the form 
$aS^{2}_{x}+bS^{2}_{y}+cS^{2}_{z}$ to the electronic Hamiltonian. 
Here $a,b,c$ are distinct constants, and $S^{2}_{x},S^{2}_{y},S^{2}_{z}$
are the squares of the components of the total spin of the two electrons 
with respect to
the Cartesian axes $x,y,z$. The Cartesian axes are defined
by the orientation of the applied external field. 
For such a system, an
experiment measuring the energy of the electrons also measures the squares
of the three spin components. To see this, note that the value of the 
perturbation energy will be $(a+b)$, $(a+c)$, or $(b+c)$ if the 
joint values of the set $S^{2}_{x},S^{2}_{y},S^{2}_{z}$ equal 
respectively $\{1,1,0\}$, $\{1,0,1\}$, or $\{0,1,1\}$. 

To understand why the external electric field affects the 
orthohelium electrons in this way, consider the ground state of 
orthohelium\footnote{Using spectroscopic notation, this state
would be written as the `$2^{3}$S' state of orthohelium. The `$2$'
refers to the fact that the principle quantum number $n$ of the state 
equals $2$, `S' denotes that the total orbital angular momentum is zero, 
and the `3' superscript means that it is a spin triplet state. 
Orthohelium has no state of principle quantum number $n=1$, since
the Pauli exclusion principle forbids the `$1^{3}$S' state.}.
The wave function of this state is given by a spatial part
$\phi(r_{1},r_{2})$, depending only 
on $r_{1},r_{2}$ (the radial coordinates of the electrons), multiplied by the 
spin part, which is a linear combination of
the eigenvectors $\psi_{+1},\psi_{0},\psi_{-1}$ of $S_{z}$ , corresponding 
to $S_{z}=+1,0,-1$, respectively. Thus, the ground state may be represented by
any vector in the three-dimensional Hilbert space spanned by the vectors
$\phi(r_{1},r_{2})\psi_{+1},\phi(r_{1},r_{2})\psi_{0},
\phi(r_{1},r_{2})\psi_{-1}$. The external electric field will have the 
effect of ``lifting the degeneracy'' of the state, i.e. the new 
Hamiltonian will not be degenerate in this space, but its eigenvalues
will correspond to three unique orthogonal vectors.
Suppose that we consider a particular set of Cartesian axes $x,y,z$.
We apply an electric field which is of orthorhombic 
symmetry\footnote{Orthorhombic symmetry is defined by the criterion that
rotation about either the
$x$ or $y$ axis by $180^{\circ}$ would bring such a field back to itself.}
with respect to these axes. It can be
shown\footnote{A straightforward way to see this is by analogy with
a charged particle of orbital angular momentum $L=1$. The effects
of an electric or magnetic field on a charged particle of spin $1$
are analogous to the effects of the same field on a charged particle
of orbital angular momentum $1$. To calculate the first-order effects of an 
electric field of orthorhombic symmetry for such a particle, one can 
examine the spatial dependence of the $L_{z}=1,0,-1$ states 
$\psi_{-1},\psi_{0},\psi_{+1}$, together with
the spatial dependence of the perturbation potential $V({\bf r})$,
to show that the states $1/\sqrt(\psi_{1}-\psi_{-1})$, $1/\sqrt{2}(\psi_{1}+
\psi_{-1})$, and $\psi_{0}$ are the eigenstates of such a perturbation.
A convenient choice of $V$ for this purpose is $V=Ax^{2}+By^{2}+Cz^{2}$. 
See Kittel in \cite[p. 427]{Kittel}.} 
that the eigenvectors of the Hamiltonian due to this
field are $v_{1}=1/\sqrt{2}((\psi_{1}-\psi_{-1})$, $v_{2}=1/\sqrt{2}(\psi_{1}+
\psi_{-1})$ and $v_{3}=\psi_{0}$. 
We drop the factor $\phi(r_{1},r_{2})$ for convenience of expression.
These vectors are {\em also} the joint-eigenvectors
of the observables $S^{2}_{x},S^{2}_{y},S^{2}_{z}$, as we can easily show.
When expressed as a matrix in
the $\{\psi_{+1},\psi_{0},\psi_{-1}\}$ basis, 
the vectors $v_{1},v_{2},v_{3}$ take the form 
\begin{eqnarray}
v_{1}  & = &
\left(
\begin{array}{c}
\frac{1}{\sqrt{2}} \\
0  \\
-\frac{1}{\sqrt{2}}
\end{array}
\right) \\
v_{2}  & = &
\left(
\begin{array}{c}
\frac{1}{\sqrt{2}} \\
0  \\
\frac{1}{\sqrt{2}}
\end{array}
\right) \\
v_{3}  & = &
\left(
\begin{array}{c}
0 \\
1  \\
0
\end{array}
\right) 
\mbox{.}
\end{eqnarray}
If we then express $S^{2}_{x},S^{2}_{y},S^{2}_{z}$ as matrices in 
terms of the same basis, then 
by elementary matrix multiplication, one can show that
$v_{1}$ corresponds to the joint-eigenvalue $\bm=\{0,1,1\}$,
$v_{2}$ corresponds to $\bm=\{1,0,1\}$,
and $v_{3}$ corresponds to $\bm=\{1,1,0\}$.
Thus, the eigenvectors $v_{1},v_{2},v_{3}$ of 
the Hamiltonian term $H^{'}$ which arises from a perturbing 
electric field (defined with respect to $x,y,z$) are 
also the joint-eigenvectors of the set $\{S^{2}_{x},S^{2}_{y},
S^{2}_{z}\}$. This implies that we can represent $H^{'}$ by the 
expression $aS^{2}_{x}+bS^{2}_{y}+cS^{2}_{z}$, where $H^{'}$'s eigenvalues are
$\{(b+c), (a+c),(a+b)\}$.

All of this leads to the following conclusion regarding the
measurement of the spin of the orthohelium ground state electrons.
Let the system be subjected to an electric field with orthorhombic
symmetry with respect to a given set of Cartesian
axes $x,y,z$. Under these circumstances, the measurement of the
total Hamiltonian will yield a result equal
(to first-order approximation) to the (unperturbed) ground state 
energy plus one of the perturbation corrections $\{(b+c), (a+c),(a+b)\}$.
If the measured value of the 
perturbation energy is $(a+b)$, $(a+c)$, or $(b+c)$ then the 
joint values of the set $\{S^{2}_{x},S^{2}_{y},S^{2}_{z}\}$ are given  
respectively  by $\{1,1,0\}$, $\{1,0,1\}$, or $\{0,1,1\}$.  

Suppose we consider two such 
experiments\footnote{The type of experiments mentioned by Kochen and 
Specker are actually {\em non-ideal} measurements of sets such
as $\{S^{2}_{x},S^{2}_{y},S^{2}_{z}\}$, since the post-measurement
wave function of the electrons is not equal to the projection of 
the pre-measurement wave function into an eigenspace of the set 
$\{S^{2}_{x},S^{2}_{y},S^{2}_{z}\}$ (see section \ref{formalism}).
In particular, they envision a spectroscopic analysis of the atom, i.e.,
the observation of photons emitted during transitions between the
stationary states of the electrons. The fact that they consider 
non-ideal measurements of each set $\{S^{2}_{x},S^{2}_{y},S^{2}_{z}\}$ 
rather than ideal only serves to widen the range of possibilities for different 
experimental procedures, and so strengthens our point that 
distinct procedures exist for what Kochen and Specker consider
to be a `measurement of $S^{2}_{x}$'.}, 
one which measures
the set $\{S^{2}_{x},S^{2}_{y},S^{2}_{z}\}$ and the other of which measures
$S^{2}_{x},S^{2}_{y^{'}},S^{2}_{z^{'}}$, i.e. the squares of the components
in the $x,y^{'},z^{'}$ system. The former experiment involves an
electric field with orthorhombic symmetry with respect to $x,y,z$,
while the latter involves an electric field {\em with a different
orientation in space}, since it must have symmetry with respect
to the axes $x,y^{'},z^{'}$. Although both procedures can be regarded
as measurements of $S^{2}_{x}$, they involve quite different experiments. 

It is quite apparent from this example that it would be 
unreasonable to require that a hidden variables theory must assign a 
single value to $S^{2}_{x}$, independent of the experimental procedure.

\section{Contextuality theorems and spectral-incompatibility 
\label{Spec}}
We saw in our discussion of von Neumann's theorem that
its implications toward hidden variables amounted to the assertion that 
there can be no
mathematical function $E(O)$ that is linear on the observables and 
which maps them to their eigenvalues. 
This was neither a surprising,
nor particularly enlightening result, since it follows also from
a casual observation of some example of linearly related non-commuting 
observables, as we saw in examining the observables 
$\frac{1}{\sqrt{2}}(\sigma_{x}+\sigma_{y}),\sigma_{x},\sigma_{y}$ of a 
spin $\frac{1}{2}$ particle. The theorems of Gleason, Kochen and 
Specker, and Mermin imply a somewhat less obvious type of 
impossibility: there exists no function $E(O)$ mapping the observables to their 
eigenvalues, which obeys all relationships constraining {\em commuting} 
observables. What we develop here is a somewhat simpler expression
of the implication of these theorems. We will 
find that they imply
the {\em spectral-incompatibility} of the value map function:
there exists no mathematical function that assigns to each commuting
set of observables a joint-eigenvalue of that set.

We begin by recalling the notions of joint-eigenvectors and 
joint-eigenvalues of a commuting set of observables $(O^{1},O^{2},\ldots)$.
For a commuting set, the eigenvalue equation \ref{eq:Eigenvalue}
$O \sket = \mu \sket $ is replaced by a
set of relationships \ref{eq:Joint-Eig}: 
$O^{i}\sket=\mu^{i}\sket \mbox{ } i=1,2,\ldots$, one for each 
member of the commuting set. If a given $\sket$ satisfies this relationship
for {\em all} members of the set, it is referred to as a 
{\em joint-eigenvector}. The set of numbers $(\mu^{1},\mu^{2},\ldots)$
that allow the equations to be satisfied for this vector are 
collectively referred to as the {\em joint-eigenvalue} corresponding
to this eigenvector, and the symbol ${\bm}=(\mu_{1},\mu_{2},\ldots)$
is used to refer to this set. The set of all joint-eigenvalues 
$\{\bm_{a}\}$
is given the name `joint-eigenspectrum'.

In general, the members of any given commuting set of observables 
might not be independent, i.e., they may be constrained by mathematical 
relationships. We label the relationships for any given 
commuting set $\{O^{1},O^{2},\ldots\}$ as 
\begin{eqnarray}
f_{1}(O^{1},O^{2},\ldots) & = & 0 \label{eq:comm relat} \\
f_{2}(O^{1},O^{2},\ldots) & = & 0 \nonumber \\
 \vdots & & \nonumber
\mbox{.}
\end{eqnarray}
The equations \ref{eq:GlPsum} in Gleason's theorem, \ref{eq:KS comm} 
in Kochen and Specker's theorem, and \ref{eq:Pduct}, \ref{eq:SPduct}
and \ref{eq:IPduct} in Mermin's theorem, are just such relations. We 
now demonstrate the following two results. First, that every member of 
the joint-eigenspectrum must satisfy
all relationships \ref{eq:comm relat}. Second, that any set of numbers 
$\xi_{1},\xi_{2},\ldots$ satisfying all of these relationships
is a joint-eigenvalue. 

To demonstrate the first of these, we suppose that ${\bm}=
(\mu^{1},\mu^{2},\ldots)$ is a
joint-eigenvalue of the commuting set $\{O^{1},O^{2},\ldots\}$, with 
joint-eigenspace ${\cal H}$. We then consider the operation of
$f_{i}(O^{1},O^{2},\ldots)$ on a vector
$\psi \in {\cal H}$ where $f_{i}(O^{1},O^{2},\ldots)=0$ is one 
of the relationships constraining the commuting set. We find 
\beq
f_{i}(O^{1},O^{2},\ldots) \psi=
f_{i}(\mu_{1},\mu_{2},\ldots) \psi=0
\mbox{.}
\eeq
The second equality implies that $f_{i}(\mu_{1},\mu_{2},
\ldots)=0$. Since $f_{i}(O^{1},O^{2},\ldots)=0$ is an 
arbitrary member of the relationships \ref{eq:comm 
relat} for the commuting set $\{O^{1},O^{2},\ldots\}$, it 
follows that every joint-eigenvalue $\bm$ of the set must satisfy
 {\em all} such relationships. 

We now discuss the demonstration of the second point. Suppose that
the numbers $\{\xi_{1},\xi_{2},\ldots\}$ satisfy 
all of \ref{eq:comm relat} for some commuting set $\{O^{1},O^{2},\ldots\}$.
We consider the following relation: 
\beq
\left([(O^{1}-\mu^{1}_{1})^{2}+
(O^{2}-\mu^{2}_{1})^{2}+\ldots][(O^{1}-\mu^{1}_{2})^{2}+
(O^{2}-\mu^{2}_{2})^{2} + \ldots] \ldots\right) \psi=0
\mbox{.}
\label{eq:jointprod}
\eeq
Here, we operate on the vector $\psi$ with a product whose factors each
consist of a sum of various operators.
The product is taken over all joint-eigenvalues $\{\bm_{i}\}$. We represent
each joint-eigenvalue ${\bm}_{i}$ by a set $(\mu^{1}_{i},\mu^{2}_{i},
\ldots)$. The validity of \ref{eq:jointprod}
is easily seen. Since the joint-eigenspaces of any commuting set are 
complete, the vector $\psi$ must lie within such a space. Suppose
 that $\psi \in {\cal H}_{i}$, where ${\cal H}_{i}$ is the joint-eigenspace 
 corresponding to $\bm_{i}$. Then the $i$th factor in the product of operators 
 in \ref{eq:jointprod} must give zero when operating on $\psi$. Therefore, the 
 entire product operating
on $\psi$ must also give zero. Since $\psi$ is an arbitrary vector, it follows
that 
\beq
[(O^{1}-\mu^{1}_{1})^{2}+
(O^{2}-\mu^{2}_{1})^{2}+\ldots][(O^{1}-\mu^{1}_{2})^{2}+
(O^{2}-\mu^{2}_{2})^{2} + \ldots] \ldots=0
\label{eq:Communi}
\mbox{.}
\eeq 
Note that \ref{eq:Communi} is itself
a constraining relationship on the commuting set, so that
 it must be satisfied by the numbers $(\xi_{1},\xi_{2},\ldots)$. This can only
  be true if 
these numbers form a joint-eigenvalue of 
$\{O^{1},O^{2},\ldots\}$, and this is the result we were to prove.

From this, we can discern a simple way to regard the implications
of the theorems of Gleason,
Kochen and Specker, and Mermin toward the question of a value map.
The requirement \ref{eq:comm lin} Gleason's theorem 
places on the function $E(P)$ can be re-stated as the requirement 
that for each commuting set, $E(P)$ must satisfy all the relationships 
constraining its members. From the above argument, it follows that this 
assumption is equivalent to the constraint that $E(P)$ must assign to
each commuting set a joint-eigenvalue. 
The same is also true of \kss assumption that $E(s^{2}_{\theta,\phi})$ satisfy 
equation \ref{eq:KS proj lin}, and the Mermin requirement that $E$ satisfy
\ref{eq:Pduct}, \ref{eq:SPduct} and \ref{eq:IPduct}.

Thus, all three of these theorems can be regarded as proofs of the 
impossibility of a function mapping the observables to their values
such that each commuting set is assigned a joint-eigenvalue. An 
appropriate name for such a proof would seem to be 
`spectral-incompatibility theorem.'
 
\section{Albert's example and contextuality \label{UncleAlb}} 
In thinking about any given physical phenomenon, it is natural to 
try to picture to oneself the properties of the system
being studied. In using the quantum formalism to develop
such a picture, one may tend to regard the `observables' of this formalism, 
i.e., the Hermitian operators (see section \ref{formalism}), as representative 
of these properties.
However, the central role played by the {\em experimental procedure} 
${\cal E}(O)$ 
in the measurement of any given observable $O$ seems to
suggest that such a view of the operators 
may be untenable. We describe an experiment originally discussed by David 
Albert \cite{Albert} 
that indicates that this is indeed the case:
the Hermitian operators cannot be regarded as 
representative of the properties of the system\footnote{This 
idea has been propounded by Daumer, D{\"u}rr, Goldstein, 
and Zangh{\'i} in \cite{Martin}. See also Bell in 
\cite{Last of Bell}}. 
Albert considers two laboratory procedures that may be used
to measure the $z-$component of the spin of a spin $\frac{1}{2}$
particle. Although the two procedures are quite similar to one another, they 
cannot not be regarded as identical when considered in light of
the hidden variables theory known as Bohmian mechanics. This is a particularly 
striking instance of contextuality, and it indicates the inadequacy of 
the conception that the spin operator $\sigma_{z}$  
represents an intrinsic property of the
particle. From Albert's example, one can clearly see that the outcome 
of the $\sigma_{z}$ measurement depends not only on the parameters 
of the particle itself, but also on the complete experimental setup.

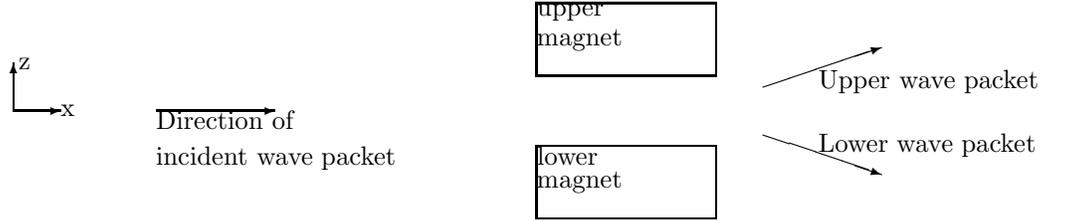
\begin{figure}
\begin{picture}(360,118)
\put(247.5,17){\framebox(67.5,27)[tl]{lower}}
\put(247.5,30.5){\makebox(0,0)[l]{magnet}}
\put(247.5,71){\framebox(67.5,27)[tl]{upper}}
\put(247.5,84.5){\makebox(0,0)[l]{magnet}}
\put(49.5,57.5){\vector(0,1){18}}
\put(51.5,75.5){\makebox(0,0)[l]{z}}
\put(49.5,57.5){\vector(1,0){18}}
\put(67.5,57.5){\makebox(0,0)[l]{x}}
\put(103.5,57.5){\vector(1,0){45}}
\put(103.5,57.5){\makebox(0,0)[tl]{Direction of}}
\put(103.5,44){\makebox(0,0)[tl]{incident wave packet}}
\put(333,66.5){\vector(3,1){45}}
\put(354,73){\makebox(0,0)[tl]{Upper wave packet}} 
\put(333,48.5){\vector(3,-1){45}}
\put(354,42){\makebox(0,0)[bl]{Lower wave packet}}
\end{picture}
\caption{Geometry of the Stern-Gerlach Experiment \label{Stern}}
\end{figure}

The Albert example is concerned with the 
measurement of 
spin\footnote{As is usual in discussions of Stern-Gerlach 
experiments, we consider only those effects relating to the
interaction of the magnetic field with the magnetic moment of the particle. 
We consider the electric charge of the particle to be zero.}  
as performed using a Stern-Gerlach magnet. 
The schematic diagram given in figure \ref{Stern} exhibits the 
configuration used in both of the measurement procedures to be 
described here. Note that we use a Cartesian coordinate system for which the 
$x$-axis lies along the horizontal 
direction with positive $x$ directed toward the right, and the 
$z$-axis lies along the vertical direction with
positive $z$ directed upward. The $y$-axis (not shown) is perpendicular to the 
plane of the figure, and---since we use a 
right-handed coordinate system---positive $y$ points into this plane. 
The long axis of the Stern-Gerlach magnet system 
is oriented along the $x$-axis, as shown. The upper and lower magnets 
of the apparatus are located in the directions of positive $z$ and 
negative $z$. 
We define the Cartesian system further by requiring that  
$x$-axis (the line defined by $y=0,z=0$) passes through the center of the
Stern-Gerlach magnet system. 
In each experiment, the spin $\frac{1}{2}$ particle to be measured is 
incident on the apparatus along the positive $x-$axis.
In the region of space the particle occupies before
entering the Stern-Gerlach apparatus, its wave function is of the form
\beq
\psi_{t}({\bf r}) = \varphi_{t}({\bf r}) (|\uparrow \rangle + 
| \downarrow \rangle)
\label{eq:Bef}
\mbox{,}
\eeq
where the vectors $|\uparrow \rangle$ and $|\downarrow\rangle$ are the 
eigenvectors of $\sigma_{z}$
corresponding to eigenvalues $+\frac{1}{2}$ and $-\frac{1}{2}$, respectively.
Here $\varphi_{t}({\bf r})$ is a localized wave packet moving 
in the positive $x$ direction toward the magnet. 

We wish to consider two experiments that differ only
in the orientation of the magnetic field inside the Stern-Gerlach 
apparatus.
In the experiment $1$, the upper magnet has a strong magnetic north pole
toward the region of particle passage, while the lower has a somewhat 
weaker magnetic south pole toward this region. In the experiment $2$, the
magnets are such that the gradient of the field points in the {\em opposite} 
direction, i.e., the
upper magnet has a strong magnetic {\em south} pole toward the region 
of passage, while 
the lower has a weak magnetic north pole towards it.
After passing the Stern-Gerlach apparatus, 
the particle will be described by a wave function of one of the 
following forms:
\begin{eqnarray}
\psi^{1}_{t}({\bf r}) & = &  
\frac{1}{\sqrt{2}}(\phi_{t}^{+}({\bf r}) | \uparrow \rangle
+ \phi_{t}^{-}({\bf r}) | \downarrow \rangle)
\label{eq:Aft}  \\
\psi_{t}^{2} ({\bf r}) & = & \frac{1}{\sqrt{2}}(\phi_{t} ^{-}({\bf r})| 
\uparrow \rangle + 
\phi_{t} ^{+}({\bf r})| \downarrow \rangle) \nonumber
\mbox{.}
\end{eqnarray}
Here $\psi^{1}_{t}({\bf r})$ corresponds to experiment $1$,
and $\psi_{t} ^{2} ({\bf r})$ corresponds to experiment $2$. In both 
cases, the function $\phi_{t} ^{+}({\bf r})$ represents a 
localized wave packet moving obliquely upward and
$\phi_{t} ^{-}({\bf r})$ represents a localized wave packet moving obliquely 
downward. To measure $\sigma_{z}$, one places detectors in the
paths of these wave packets. Examination of the first equation of 
\ref{eq:Aft} shows that for experiment $1$, if the particle is 
detected in the upper path, the result of 
our $\sigma_{z}$ measurement is $+\frac{1}{2}$. If the particle is detected in 
the 
lower path, the result is $-\frac{1}{2}$.
For experiment $2$, the second equation of \ref{eq:Aft} 
leads to the conclusion that similar detections are associated with results
opposite in sign to those of experiment $1$. Thus, for experiment $2$, detection
in the upper path implies $\sigma_{z}=-\frac{1}{2}$, while detection in the 
lower 
implies $\sigma_{z}=+\frac{1}{2}$.

We make here a few remarks regarding the symmetry of the system. 
We constrain the form of the wave packet $\varphi_{t}({\bf r})$
of \ref{eq:Bef}, by demanding that it has no dependence on $y$, and 
that it exhibits reflection symmetry through the plane defined by $z=0$, 
i.e., $\varphi_{t} ({x,z})=
\varphi_{t} ({x,-z})$. Moreover, the vertical extent of this wave packet 
is to be the same size as the vertical spacing between the 
upper and lower magnets of the apparatus. As regards the wave packets
$\phi_{t} ^{+}({\bf r})$ and $\phi_{t} ^{-}({\bf r})$ of \ref{eq:Aft}, 
if the magnetic field within the apparatus is such that 
$\partial B_{z}/\partial z$ is 
constant\footnote{The term added to the 
particle's Hamiltonian to account for a magnetic field is 
$g{\bf s} \cdot {\bf B}$, where ${\bf s}$ is the spin, ${\bf B}$
is the magnetic field and $g$ is the gyromagnetic ratio. 
To determine the form of this term in the case of a Stern-Gerlach
apparatus, we require the configuration of the magnetic field.
A Stern-Gerlach magnet apparatus has a ``long axis'' which for the example
of figure \ref{Stern} lies along the $x$-axis. Since the component of the 
magnetic field along this axis will
{\em vanish} except within a small region before and after the apparatus, 
the effects of $B_{x}$ may be neglected. Furthermore, $B_{y}$ and $B_{z}$ 
within the apparatus may be regarded as being independent of $x$. 
The magnetic field in the $x,z$ plane between the magnets lies in the 
$z$-direction, i.e., ${\bf B}(x,0,z)=B_{z}(z)\hat{k}$. Over the region of 
incidence of the particle, the field is such that $\frac{\partial B_{z}}
{\partial z}$ is constant. See for example, Weidner and Sells \cite{Weidner} 
for a more detailed discussion of the Stern-Gerlach apparatus. The motion of 
the particle in the $y$-direction is of no importance to us, and so we do not
consider the effects of any Hamiltonian terms involving 
only $y$ dependence. The results we discuss in the present 
section are those which arise from taking account of the magnetic field
by adding to the Hamiltonian term of the form 
$g\sigma_{z}B_{z}(z=0)+g\sigma_{z}(\frac{\partial B_{z}}{\partial z}(z=0))z$.}, 
then for both experiments, these packets move at
equal angles above and below the horizontal (See figure \ref{Stern}). 
Thus, the particle is described both before and after 
it passes the Stern-Gerlach magnet, by a wave function
which has reflection symmetry through the plane defined by $z=0$.

\subsubsection{Bohmian Mechanics and Albert's example}
We have mentioned that the hidden variables theory developed by David Bohm
\cite{Bohmian Mechanics} gives an explanation of quantum phenomena which is 
empirically equivalent
to that given by the quantum formalism. Bohmian mechanics allows us to regard 
any given system as a set of
particles having well-defined (but distinctly non-Newtonian 
\cite{Quantum Equilibrium}) trajectories. Within Bohmian mechanics, it
is the the {\em configuration} of the system ${\bf q}=(q_{1},q_{2},q_{3},...)$
which plays the role of the hidden variables parameter $\lda$.
Thus, the state description in this theory consists of both $\psi$ and 
${\bf q}$. Bohmian mechanics does not involve a change in the mathematical 
form of $\psi$: just as in the quantum formalism, $\psi$ is a 
vector in the Hilbert space associated with the system,
and it evolves with time according to the Schr{\"o}dinger equation:
\begin{equation} 
i\hbar \frac{\partial \psi}{\partial t} = H \psi 
\label{eq:erwin}
\mbox{.}
\end{equation} 
The system configuration ${\bf q}$ is governed by the equation:
\beq 
\frac{d{\bf q}}{dt}= (\hbar/m) \mbox{Im}(\frac{\psi^{*}{\bf \nabla}
\psi}{\psi^{*} \psi})
\label{eq:Bohm}
\mbox{.}
\eeq
In the case of a particle with spin, we make use of the spinor inner-product
in this equation. For example, in the case of a spin $\frac{1}{2}$ particle 
whose wave function is $\psi=\chi_{+}({\bf r})|\! \uparrow \rangle + 
\chi_{-}({\bf r})|\! \downarrow \rangle$, the equation \ref{eq:Bohm} 
assumes the form 
\beq
\frac{d{\bf q}}{dt}=(\hbar/m)\mbox{Im}(\frac{\chi_{+}^{*}({\bf r})
{\bf \nabla}\chi_{+}({\bf r})+\chi_{-}^{*}({\bf r})
{\bf \nabla}\chi_{-}({\bf r})}
{\chi_{+}^{*}({\bf r})\chi_{+}({\bf r}) + \chi_{-}^{*}({\bf r})
\chi_{-}({\bf r})})
\mbox{.}
\label{eq:Bohmsp}
\eeq
As we expect from the fact that this theory is in empirical agreement with 
quantum theory, Bohmian mechanics does {\em not} generally provide, 
given $\psi$ and ${\bf q}$, 
a mapping from the observables to their values. In other words, it does
not provide a non-contextual value map for each state.
As we shall see, the choice of {\em experimental procedure} plays such a  
pronounced role in the Bohmian mechanics description of Albert's spin 
measurements, that one cannot possibly regard the spin operator as 
representative of an objective property of the particle.

We first discuss the Bohmian mechanics description of the
Albert experiments. Since the wave function and its evolution
are the same as in quantum mechanics, the 
particle's wave function $\psi$
is taken to be exactly as described above. As far as the 
configuration ${\bf q}$ is concerned, there
are two important features of the Bohmian evolution equations to be considered:
the uniqueness of the trajectories and the {\em equivariance}
of the time evolution. The first feature refers to the fact that each initial 
$\psi$ and ${\bf q}$ leads to a {\em unique} 
trajectory. Since the particle being measured has a 
fixed initial wave function, its
initial conditions are defined solely by its initial position. The 
equivariance of the system's time evolution is a more complex property. 
Suppose that at some time $t$, the probability that the
system's configuration is within the region $d{\bf q}$ about ${\bf q}$
obeys the relationship:
\beq
P({\bf q}^{'}\in d{\bf q})=|\psi({\bf q})|^{2}d{\bf q}
\mbox{.}
\label{eq:equ}
\eeq
According to equivariance, this relationship will continue to hold 
for all later times $t^{'}:\mbox{ }t^{'} >t$. In considering the 
Bohmian mechanics description of a system, we assume that
the particle initially obeys \ref{eq:equ}. 
By equivariance, we then have that for all later times the particle
will be guided to ``follow'' the motion 
of the wave function. Thus, after it
passes through the Stern-Gerlach apparatus, the particle will
enter either the upward or downward moving packet. From consideration of the 
uniqueness of the trajectory and the equivariance of the time evolution,
it follows that the question of {\em which branch} of the wave function the 
particle enters {\em depends solely on its initial position}.
 
If we consider the situation in a little more detail, we find
a simple criterion on the initial position of the particle
that determines which branch of the wave function it will enter. 
Recall that the initial \ref{eq:Bef} and final
\ref{eq:Aft} wave function have no dependence on the $y$ coordinate,
and that they exhibit reflection symmetry through the $z=0$ plane.
From this symmetry together with the uniqueness of Bohmian trajectories, 
it follows that the particle cannot cross the $z=0$ plane. 
In conjunction with equivariance, this result implies that if the particle's 
initial $z$ coordinate is greater than zero, it must enter the upper 
branch, and if its initial $z$ is less than zero the particle must enter the 
lower branch. 

If we now consider the above described spin measurements, we find a 
somewhat curious situation. For any given initial configuration ${\bf q}$, 
the question of whether
the measurement result is $\sigma_{z} =+\frac{1}{2}$ or $\sigma_{z}=-\frac{1}{2}$ 
depends on the 
configuration of the experimental apparatus. Suppose that the particle 
has an initial $z>0$, so that according to the results just shown,
it will enter the upper branch of the wave function. If the magnetic 
field inside the Stern-Gerlach apparatus is 
such that $\partial B_{z}/\partial z < 0$, then the particle's final 
wave function is given by the first equation in \ref{eq:Aft}, and its
detection in the upper branch then implies that $\sigma_{z} =+\frac{1}{2}$. If, 
on the other 
hand, the magnetic field of the Stern-Gerlach magnet has the {\em opposite} 
orientation, 
i.e., $\partial B_{z}/\partial z > 0$, then the second equation in 
\ref{eq:Aft} obtains and the detection in the upper branch implies
that $\sigma_{z}=-\frac{1}{2}$. Thus, we arrive at the conclusion that
the ``measurement of $\sigma_{z}$'' gives a {\em different} result for 
two situations that differ only in the experimental configuration.

The quantum formalism's rules for the 
measurement of observables strongly suggest that
the Hermitian operators represent objective properties 
of the system. Moreover, such a conception is a common 
element of the expositions given in quantum mechanics 
textbooks. On the other hand, the fact that the result
of the ``measurement'' of $\sigma_{z}$ can
depend on properties of {\em both} system {\em and} apparatus 
contradicts this conception. In general, one must consider the 
results of the 
``measurement of an observable'' to be a joint-product of system and measuring 
apparatus. Recall Niels Bohr's comment that \cite{Bohr Nature} ``a 
closer examination 
reveals that the procedure of measurement has an essential influence
on the conditions on which the very definition of the physical quantities
in question rests.'' For further discussion of the role of
Hermitian operators in quantum theory, the reader is directed to 
Daumer, D{\"u}rr, Goldstein, and Zangh{\'i} in \cite{Martin}. 
According to these authors: 
``the basic problem 
with quantum theory \ldots more fundamental than the measurement problem and 
all the 
rest, is a naive realism about operators \ldots by (this) we refer
to various, not entirely sharply defined, ways of taking too seriously the 
notion of 
operator-as-observable, and in particular to the all too casual talk about 
`measuring 
operators' which tends to occur as soon as a physicist enters quantum mode.''

\chapter{The Einstein--Podolsky--Rosen paradox and nonlocality}  
In contrast to contextuality, nonlocality is quite an unexpected
and surprising feature to meet with in the quantum phenomena.
As seen in the previous chapter, contextuality is essentially
the dependence of the hidden variables predictions on
the different possible experimental procedures
for the measurements of the observable.
Although the nonlocality of a quantum system is not something one would 
regard 
as natural, it may be proved by a mathematical
analysis that this feature is inevitable. The demonstration
in question arises from the well-known Einstein--Podolsky--Rosen
paradox \cite{EPR}, in combination with Bell's theorem \cite{Bell's 
theorem}. 
There exists, however, a
misperception among some authors that what follows from these analyses
is {\em only} that local
hidden variables theories must conflict with quantum
mechanics. In fact, a more careful examination shows that
the conjunction of the EPR paradox with Bell's theorem
implies that {\em any} local theoretical explanation 
must disagree with quantum mechanics. To demonstrate this conclusion, we
now review the EPR paradox and Bell's theorem. We begin with a discussion 
of the spin singlet version of the EPR 
argument. This will lead us to the presentation of Bell's theorem and
its proof. We then discuss the conclusion that follows from
the conjunction of the EPR paradox and Bell's theorem. 

\section{Review of the Einstein-Podolsky-Rosen analysis \label{EPRB}} 
\subsection{Rotational invariance of the spin singlet state and 
perfect correlations between spins}
The well-known work of Einstein, Podolsky, and Rosen, first published in 
1935, 
was not designed to address the possibility of nonlocality, as such. 
 The title of the paper was
``Can Quantum Mechanical Description of Physical Reality be Considered 
Complete?'', and the goal of these authors was essentially the opposite 
of authors such as von Neumann: Einstein, Podolsky,
and Rosen wished to demonstrate that the addition of
hidden variables to the description of state is {\em necessary} for
a complete description of a quantum system. According to these authors, 
the quantum mechanical
state description given by $\psi$ is {\em incomplete}, i.e. it cannot
account for all the objective properties of the
system. This conclusion is stated in the paper's closing
remark: ``While we have thus shown that the wave function does
not provide a complete description of the physical reality, we left open
the  question of whether such a description exists. We believe, however,
that such a theory is possible.'' 

Einstein, Podolsky, and Rosen arrived at this conclusion
 having shown that for the system
they considered, each of the particles must have position and
momentum as simultaneous ``elements of reality''. Regarding the 
completeness of a physical theory, the authors state: \cite{EPR}
 (emphasis due to 
EPR) ``Whatever the meaning
assigned to the term {\em complete}, the following requirement for a 
complete
theory seems to be a necessary one: {\em every element of the physical
reality must have a counterpart in the physical theory}''. This 
requirement leads them to conclude that the quantum theory is
incomplete, since it does not account for the possibility of
position and momentum as being simultaneous elements of reality.
To develop this conclusion for the position and
momentum of a particle, the authors make use of the following 
{\em sufficient condition}
for a physical quantity to be considered as an element of reality:
 ``If without in any way disturbing a system, we
can predict with certainty (i.e. with probability equaling unity) the
value of a physical quantity, then there exists an element of physical
reality corresponding to this quantity''. 

What we shall present 
here\footnote{The Bohm spin singlet version and the original version 
of the EPR paradox differ essentially in the states and observables 
with which they are concerned. We shall consider the original
EPR state more explicitly in chapter 4, section \ref{EPR orig}.} 
is a form of the EPR paradox which was developed in 
1951, 
by David Bohm\footnote{\cite[p. 611-623]{Bohm's Famous Text}. A recent
reprint appears within \cite[p. 356-368]{Big Red}}. Bohm's version 
involves the properties of the {\em spin singlet
state} of a pair of spin $\frac{1}{2}$ particles. Within his argument, 
Bohm 
shows that various components of the spin of
a pair of particles must be elements of reality in the same sense as the
position and momentum were for Einstein, Podolsky, and Rosen.  We begin 
with a discussion 
of the formal properties of the spin singlet state, and then
proceed with our presentation of Bohm's version of the EPR 
argument.             

Because the spin and its components all commute with the 
observables associated with the system's spatial properties, 
one may analyze a particle with spin by separately analyzing 
the spin observables and the spatial observables. The spin observables 
may be analyzed in terms of a Hilbert space ${\cal H}_{s}$ (which is
a two-dimensional space in the case of a spin $\frac{1}{2}$ particle) 
and the spatial observables in terms of $L_{2}(\R)$. The full Hilbert 
space of the system is then given by 
{\em tensor product} ${\cal H}_{s} \otimes L(\R)_{2}$ of these spaces.
Hence we proceed to discuss the spin observables only, without explicit
reference to the spatial observables of the system. We denote each 
direction in space by its 
$\theta$ and $\phi$ coordinates in spherical polar coordinates, 
and (since we consider spin $\frac{1}{2}$ particles) the symbol 
$\sigma$ denotes the spin. To represent
the eigenvectors of $\sigma_{\theta,\phi}$ corresponding to the 
eigenvalues  
$+\frac{1}{2}$ and $-\frac{1}{2}$, we write 
$|\!\uparrow \theta,\phi \rangle$,
and $|\!\downarrow \theta,\phi \rangle $, respectively. For the 
eigenvectors
of $\sigma_{z}$, we write simply $|\!\uparrow \rangle$ and 
$|\! \downarrow \rangle$. Often vectors and observables 
are expressed in terms of the basis formed by the eigenvectors of 
$\sigma_{z}$.
The vectors $|\!\uparrow \theta,\phi \rangle$,
and $|\!\downarrow \theta,\phi \rangle$ when expressed
in terms of these are 
\begin{eqnarray}
|\!\uparrow \theta,\phi \rangle  & = & 
\cos(\theta/2) \, |\uupa\ra \,\,
+ \,\, \sin(\theta/2) e^{i\phi} \, |\ddwna\ra  \label{eq:thetap} \\
|\!\downarrow \theta,\phi \rangle  & = & 
\sin(\theta/2) e^{-i\phi} \, |\! \uparrow \rangle \,\,
- \,\, \cos(\theta/2) \, |\! \downarrow \rangle
\nonumber
\mbox{.}
\end{eqnarray}

We now discuss the possible states for a system consisting of {\em two}
spin $\frac{1}{2}$ particles\footnote{See for example, Messiah 
\cite{Messiah}, and Shankar \cite{Shankar}.}. For such a system, the
states are often classified in terms of the
total spin $S=\sigma^{(1)}+\sigma^{(2)}$ of the particles, where 
$\sigma^{(1)}$
is the spin of particle $1$ and $\sigma^{(2)}$ is the spin of particle 
$2$. The 
state
characterized by $S=1$ is known as the {\em spin triplet state}, 
$\psi_{ST}$; it
is so named because it consists of a combination of the three
eigenvectors of $S_{z}$---the z-component of the total spin---which 
correspond to the eigenvalues $-1,0,1$ . The 
{\em spin singlet state}, in which we shall be interested, is 
characterized 
by $S=0$. The name given to the state
reflects that it contains just one eigenvector of the
z-component $S_{z}$ of the total spin: that corresponding to
the eigenvalue $0$. In fact, as we shall demonstrate,
the spin singlet state is an eigenvector 
of {\em all} components of the total spin
with an eigenvalue $0$. We 
express\footnote{Note that a term such as $|a\ra^{(1)}|b\ra^{(2)}$ 
represents a {\em tensor product} 
of the vector $|a\ra^{(1)}$ of the Hilbert space associated with the first 
particle with the vector $|b\ra^{(2)}$ of the Hilbert space associated with
the second. The formal way of writing such a quantity is as:
 $|a\ra^{(1)}\otimes |b\ra^{(2)}$. For simplicity of 
expression, we shall omit the symbol `$\otimes$' here.} the state in terms 
of the eigenvectors $|\uupa\ra^{(1)},|\ddwna\ra^{(1)}$ of 
$\sigma^{(1)}_{z}$ and 
$|\uupa\ra^{(2)},|\ddwna\ra^{(2)}$ of $\sigma^{(2)}_{z}$, as 
follows: 
\beq 
\psi_{ss}=
|\! \uparrow \rangle^{(1)}\, |\!\downarrow \rangle^{(2)} \,\,
- \,\, |\! \downarrow \rangle^{(1)} \, |\! \uparrow \rangle^{(2)}
\label{eq:zss} 
\mbox{.}
\eeq    
For simplicity, we have suppressed the normalization factor
$\frac{1}{\sqrt{2}}$.
Note that each of the two terms consists of a product of
an eigenvector of $\sigma^{(1)}_{z}$ with an eigenvector of 
$\sigma^{(2)}_{z}$ 
such that the corresponding eigenvalues are the negatives of each other. 

If we invert the relationships \ref{eq:thetap} we 
may then re-write the spin singlet state \ref{eq:zss} in 
terms of the eigenvectors $|\!\upa \theta,\phi\rangle$
and $|\!\dwna \theta,\phi \rangle$
of the component of spin in the $\theta,\phi$ direction. Inverting 
\ref{eq:thetap} gives:
\begin{eqnarray}
|\!\uparrow \rangle  & = & 
\cos(\theta/2) \, |\uupa \theta ,\phi \ra \,\,
+ \,\, \sin(\theta/2) e^{i\phi} \, |\ddwna \theta ,\phi \ra  
\label{eq:zthetap} \\
|\!\downarrow  \rangle  & = & 
\sin(\theta/2) e^{-i\phi} \, |\! \uparrow \theta,\phi \rangle \,\,
- \,\, \cos(\theta/2) \, |\! \downarrow \theta,\phi \rangle
\nonumber
\mbox{.}
\end{eqnarray}
Now let us consider re-writing the spin singlet state \ref{eq:zss} 
in the following manner. We 
substitute for $|\!\upa \rangle^{(2)}$ and $|\!\dwna \rangle^{(2)}$ 
expressions
involving $|\!\uparrow \theta_{2},\phi_{2}\rangle^{(2)}$ and 
$|\!\dwna \theta_{2},\phi_{2}\rangle^{(2)}$ by making use of 
\ref{eq:zthetap}. Doing so gives us a new expression for the spin singlet
state:
\begin{eqnarray}
\psi_{ss} & = & |\!\upa \rangle^{(1)}
\left[
\sin(\theta_{2}/2)e^{-i\phi_{2}} \, |\! \upa \theta_{2},\phi_{2} 
\rangle^{(2)}
 \,\, - \,\, \cos(\theta_{2}/2) \, |\!\dwna \theta_{2},\phi_{2} 
\rangle^{(2)}
\right] \\
 & & -|\!\dwna \rangle^{(1)}
\left[
\cos(\theta_{2}/2) \, |\! \upa \theta_{2},\phi_{2}
\rangle^{(2)}
 \,\, + \,\, \sin(\theta_{2}/2)e^{i \phi_{2}} \, 
|\!\dwna \theta_{2},\phi_{2} \rangle^{(2)}
\right] 
\nonumber \\
 & = &  -\left[\cos(\theta_{2}/2) \, |\!\upa \rangle^{(1)} \,\, + \,\,
\sin(\theta_{2}/2)e^{i\phi_{2}} \, |\!\dwna \rangle^{(1)}\right] 
\, |\!\dwna\theta_{2},\phi_{2} \rangle^{(2)}
\nonumber
\\
 & & + \left[\sin(\theta_{2}/2)e^{-i\phi_{2}}\, |\!\upa \rangle^{(1)} \,\, 
-
\,\, \cos(\theta_{2}/2)\, |\!\dwna \rangle^{(1)}\right] \, 
|\!\upa\theta_{2},
\phi_{2}\rangle^{(2)} 
\nonumber
\mbox{.}
\end{eqnarray} 
Examining the second of these relationships, we see from
\ref{eq:thetap} that $\psi_{ss}$ 
reduces\footnote{If we multiply a wave function by any constant factor 
$c$, 
where $c \neq 0$ the resulting wave function represents the same physical 
state. We multiply $\psi_{ss}$ by $-1$ to facilitate 
comparison with \ref{eq:zss}}:
\beq
\psi_{ss}=
|\!\upa\theta_{2},\phi_{2} \rangle^{(1)} 
\, |\!\dwna \theta_{2},\phi_{2} \rangle^{(2)}
\,\,  - \,\, |\!\dwna\theta_{2},\phi_{2} 
\rangle^{(1)}\, |\!\upa \theta_{2},\phi_{2} \rangle^{(2)} 
\label{eq:thetassp}
\mbox{.}
\eeq
Note
that this form is similar to that given in \ref{eq:zss}. Each of 
its terms is a product of an eigenvector of 
$\sigma^{(1)}_{\theta_{2},\phi_{2}}$ 
and an 
eigenvector of $\sigma^{(2)}_{\theta_{2},\phi_{2}}$ such that the factors 
making up 
the product 
correspond to eigenvalues that are just the opposites of each other. We 
drop the `$2$' from 
$\theta$ and $\phi$ giving
\beq
\psi_{ss}=
|\!\upa \theta,\phi \rangle^{(1)} \,
|\!\dwna \theta,\phi \rangle^{(2)}
  \,\, -\,\, |\!\dwna \theta, \phi \rangle^{(1)}
\, |\!\upa \theta, \phi \rangle^{(2)} 
\label{eq:thetass}
\mbox{.}
\eeq

Suppose that we consider the implications of \ref{eq:thetass} for
individual measurements of $\sigma^{(1)}_{\theta,\phi}$ of particle $1$
and the same spin component of particle $2$. In each term of this
spin singlet form, we have
a product of eigenvectors of the two spin components 
such that the eigenvalues are simply the negatives of one another.
Therefore, it follows that
{\em measurements of} $\sigma^{(1)}_{\theta,\phi}$ 
{\em and} $\sigma^{(2)}_{\theta,\phi}$  {\em always give 
results that sum to zero}, i.e. if measurement of 
$\sigma^{(1)}_{\theta,\phi}$ gives the result $\pm \frac{1}{2}$, then
the measurement of $\sigma^{(2)}_{\theta,\phi}$ always gives 
$\mp \frac{1}{2}$. We say that these observables exhibit {\em perfect 
correlation}. Note that this holds true for pairs of spin observables
$\sigma^{(1)}_{\theta,\phi}$ and $\sigma^{(2)}_{\theta,\phi}$ where
$\theta,\phi$ refer to an {\em arbitrary} direction.
 
\subsection{Incompleteness argument}
Consider a situation in which the two particles described by
the spin singlet state are spatially separated from one another,
and spin-component measurements are to be carried out on each. Since
there exist perfect correlations, it is possible to predict with 
certainty the result of a measurement of $\sigma^{(1)}_{x}$ from
the result of a previous measurement of $\sigma^{(2)}_{x}$. Suppose 
for example, if we measure $\sigma^{(2)}_{x}$ and find the result 
$\sigma^{(2)}_{x}=\frac{1}{2}$. A subsequent measurement of 
$\sigma^{(1)}_{x}$ must give $\sigma^{(1)}_{x}=-\frac{1}{2}$. If
we assume {\em locality} then the measurement of $\sigma^{(2)}_{x}$
cannot in any way disturb particle $1$, which is spatially separated from
particle $2$. Using the Einstein-Podolsky-Rosen criterion, it follows 
that $\sigma^{(1)}_{x}$ is an element of reality. 

Similarly, one can predict with certainty the result of a measurement 
of $\sigma^{(1)}_{y}$ from a previous measurement of $\sigma^{(2)}_{y}$.
Again, locality implies that measurement of $\sigma^{(2)}_{y}$ does 
not disturb particle $1$, and by the EPR criterion, 
$\sigma^{(1)}_{y}$ must be an element of reality. 
In total we have shown that both $\sigma^{(1)}_{x}$ and 
$\sigma^{(1)}_{y}$ are elements of reality.
On the other hand, from the quantum formalism's description of state given 
by $\psi$, one can deduce, at most, one of two such non-commuting 
quantities. Therefore, 
we may conclude that this description of state is {\em incomplete}. 
Moreover, since we have perfect correlations between all components
of the spins of the two particles, a similar argument may be given for
any component of $\sigma^{(1)}$. Therefore,
particle $1$'s spin component in an {\em arbitrary} direction 
$\theta,\phi$ 
must be an element of reality. 

Note that we can 
interchange the roles of particle $1$ and $2$ in this argument to show 
the same conclusion for all components of particle $2$'s spin, as well.    

The EPR paradox thus has the following structure. The analysis 
begins with the observation that the spin singlet state exhibits perfect
correlations such that for any direction $\theta,\phi$, measurements of 
$\sigma^{(1)}_{\theta,\phi}$ 
and $\sigma^{(2)}_{\theta,\phi}$ always give results which sum to zero. 
Together with the assumption of locality, this leads to the conclusion 
that all components of the spins of both 
particles must be elements of reality. Since the quantum state description 
given by $\psi$ does not allow for the simultaneous reality of 
non-commuting quantities such as $\sigma^{(1)}_{x}$ and 
$\sigma^{(1)}_{y}$, 
it follows that this description is incomplete. 

\section{Bell's theorem \label{bellinequ}}
The incompleteness of the quantum mechanical state description 
concluded by EPR implies that
one must consider a theoretical description of state consisting of $\psi$
and some additional parameter, in order to fully account for a system's 
properties. In terms of such a state description, one ought to be
able to mathematically represent the 
definite values concluded by EPR using a 
function\footnote{Since we are considering a system of {\em fixed} $\psi$, 
namely that of the spin singlet state, no $\psi$ dependence need be
included in $V$} 
$V_{\lda}(O)$ mapping each
component of the spin of each particle to a value. 
Here we have denoted the 
supplemental state parameter as $\lda$ in 
accordance with the discussion of von Neumann's theorem in chapter $1$.  
In 1965, John S. Bell presented a famous theorem which 
addressed the possibility of just such a function on the spin observables.
Bell was able to show that this formulation must {\em conflict} with 
the statistical predictions of 
quantum mechanics for various spin measurements. 
We now present Bell's theorem.

Let us first fix our notation. To denote directions in space,
we write unit vectors such as $\hat{a},\hat{b},\hat{c}$ 
instead of $\theta$ and $\phi$. Rather than using the form 
$V_{\lda}(O)$, we shall write $A(\lda,\hat{a})$ and 
$B(\lda,\hat{b})$ to represent functions on
the spin components of particles $1$ and $2$, respectively.
Since the two particles are each of spin $\frac{1}{2}$,
we should have $A=\pm \frac{1}{2}$ and $B=\pm \frac{1}{2}$, however for 
simplicity we rescale these to $A=\pm 1 $ and $B=\pm 1$.

\subsection{Proof of Bell's theorem}
The key feature of the spin singlet version of the EPR paradox was its 
analysis of the perfect correlations arising when the two particles
of a spin singlet pair are subject to measurements of the same spin
component. Thus, it may not be surprising that Bell's theorem is
concerned with a {\em correlation function}, which is essentially a
measure of the
statistical correlation between the results of spin component measurements
of the two particles. 
The correlation function is to be determined 
as follows: we set the apparatus measuring particle $1$ to
probe the component in the $\hat{a}$ direction, and the apparatus
measuring $2$ is set for the $\hat{b}$ direction. We make a series of
measurements of spin singlet pairs using this configuration, recording the
{\em product} $\sigma^{(1)}_{\hat{a}}\sigma^{(2)}_{\hat{b}}$ of the results on 
each trial. The average of these products over the series of measurements
is the value of the correlation function. 

In general, we expect the value of the average
determined in this way to depend on the 
directions $\hat{a}, \hat{b}$ with respect to which the
spin components are measured. 
According to the quantum formalism, we may predict the average, or
{\em expectation value} of any observable using the formula 
$E(O)=\la\psi|O\psi\ra$. For the series of experiments just described,
we take the expectation value of product of the appropriate spin component 
observables, giving:   
\begin{equation}
P_{QM}({\hat a},{\hat b}) = 
\la\sigma^{(1)}_{a} \sigma^{(2)}_{b}\ra  = -{\hat a} \cdot {\hat b}  
\label{eq:CQM}
\mbox{.}
\end{equation} 
In the case of the predetermined values,
the average of the product of the two spin components 
$\sigma^{(1)}_{\hat{a}}\sigma^{(2)}_{\hat{b}}$  
is obtained by taking an average over $\lda$:
\beq
P(\hat{a},\hat{b}) = \int d \lda \rho(\lda) A(\lda,\hat{a}) 
B(\lda,\hat{b})
\mbox{,}
\label{eq:CHV}                        
\eeq
where  $\rho (\lambda)$ is the probability 
distribution over $\lda$.
$\rho (\lambda)$ is normalized by: 
\beq
\int d \lda \mbox{  } \rho(\lda) =1   
\label{eq:Norm}
\mbox{.}
\eeq
We will now examine the question of whether the correlations function
given by \ref{eq:CHV} is
compatible with the quantum mechanical prediction \ref{eq:CQM} 
for this function.  

Crucial to the EPR analysis is the fact that there 
is a perfect correlation between the results of the
measurement of any component of particle $1$'s spin in a given direction  
 with the measurement of the same component of particle $2$'s spin, such 
that the results are of opposite sign.
 To account for this, the correlation function must give 
\beq
P(\hat{a},\hat{a}) = -1 \mbox{ }\forall \hat{a}
\mbox{.}
\eeq
It is easy to see that the quantum correlation function satisfies this 
condition. If the prediction derivable using the predetermined values 
is to reflect this, we must have 
\beq
A(\lda,\hat{a}) = -B(\lda,\hat{a}) \mbox{   } \forall \hat{a},\lda
\mbox{.}
\label{eq:PC}
\eeq

At this point, we have enough information to derive the conclusion
of the theorem. Using \ref{eq:CHV} together with \ref{eq:PC} and the fact 
that $[A(\lda,\hat{a})]^{2} = 1$, we write 
\begin{eqnarray}
P(\hat{a},\hat{b}) - P(\hat{a},\hat{c}) & = & 
- \int d \lda \rho(\lda)[A(\lda,\hat{a})A(\lda,\hat{b})
-A(\lda,\hat{a})A(\lda,\hat{c})] \\ 
 & = & - \int d \lda \rho(\lda)A(\lda,\hat{a})A(\lda,\hat{b})
[1-A(\lda,\hat{b})A(\lda,\hat{c})]                     
\nonumber
\end{eqnarray}
Using $A,B = \pm 1$, we have that
\beq
|P(\hat{a},\hat{b})-P(\hat{a},\hat{c})|
\leq 
\int d \lda \rho(\lda) [1-A(\lda,\hat{b})A(\lda,\hat{c})] 
\mbox{;}
\eeq
then using the normalization \ref{eq:Norm}, and \ref{eq:PC} we 
have
\beq 
|P(\hat{a},\hat{b})-P(\hat{a},\hat{c})| \leq 
1 + P(\hat{b},\hat{c})                           
\label{eq:Bell}
\mbox{,}
\eeq
and this relation, which is commonly referred to as ``Bell's 
inequality'', is the theorem's conclusion. 

Thus, the general framework of Bell's theorem is as follows. 
The definite values of the various components of the two particles' spins are 
represented by the mathematical functions $A(\lda,\hat{a})$,  and
$B(\lda,\hat{b})$. 
The condition
\beq
A(\lda,\hat{a}) = -B(\lda,\hat{a}) \mbox{   } \forall \hat{a},\lda
\eeq 
(equation \ref{eq:PC}), placed 
on the functions $A(\lda,\hat{a})$, 
$B(\lda,\hat{b})$ ensures the agreement of these functions with
the perfect correlations. Bell's theorem tells us that
from these conditions it follows that the theoretical prediction 
for the correlation function, $P(\hat{a},\hat{b})$, must satisfy
the Bell inequality, \ref{eq:Bell}. 

Based on the fact that the Bell inequality is not satisfied by the
quantum mechanical correlation function \ref{eq:CQM} (as we shall see 
below), some authors\footnote{This
is the conclusion reached by the following authors: 
\cite{Bethe},\cite[p. 172]{Gell-Mann}, \cite[Wigner p. 291]{Big Red}} 
have concluded that Bell's theorem proves the impossibility of
hidden variables. As we shall see, however, this conclusion does not follow.

\section{The EPR paradox, Bell's theorem, and nonlocality 
\label{eprbellnon}} 
Recall our discussion of the EPR paradox. We found 
that for a system described by the spin singlet
state, each component of the spin of one
particle is perfectly correlated with the same component of the spin
of the other. Such perfect correlations seem to give the appearance 
of nonlocality, 
since the measurement of one spin seems capable of immediately influencing 
the result of a measurement of its distant partner. The conclusion of 
nonlocality can be avoided only if all components of the
spins of each particle possess definite values. This much is developed
from the EPR analysis. According to Bell's theorem, these definite values
lead to the prediction that the
correlation function $P$ must satisfy the inequality \ref{eq:Bell}.
Since Bell's theorem assumes nothing beyond what can be concluded from the
spin singlet EPR paradox, one can deduce from the {\em conjunction}
of EPR with Bell that {\em any local theoretical description} which 
accounts for the perfect correlations of the spin singlet state 
leads to a correlation function satisfying Bell's inequality. 

Consider now the quantum mechanical prediction for the correlation 
function \ref{eq:CQM}.
If we examine this function, we find that it does {\em not} in 
general satisfy Bell's inequality. Suppose, for example, that we have defined 
some angular orientation such that $\hat{a},\hat{b},\hat{c}$ all lie in the 
$x,y$ plane (so that $\theta=90^{\circ}$), with
$\hat{a}$ along $\phi=60^{\circ}$, $\hat{b}$ along
 $\phi=0^{\circ}$, and $\hat{c}$ along
$\phi=120^{\circ}$. With this, we have
 $P_{QM}(\hat{a},\hat{b})=\frac{1}{2}$, $P_{QM}(\hat{a},\hat{c})=
\frac{1}{2}$ and $P_{QM}(\hat{b},\hat{c})=
-\frac{1}{2}$, so that $|P_{QM}(\hat{a},\hat{b})-P_{QM}(\hat{a},\hat{c})| 
= 1$
 and $1+P_{QM}(\hat{b},\hat{c}) = \frac{1}{2}$, which is in violation of 
 \ref{eq:Bell}.
From what we have found above, the disagreement of the quantum
mechanical prediction for this correlation function with Bell's 
inequality implies that {\em quantum mechanics
must disagree with any local theoretical description.} In the words of
Bell: \cite{EPW} ``It is known that with Bohm's example of EPR correlations, 
involving particles with spin, there is an irreducible nonlocality.'' 
 
Concerning the claim that Bell's theorem is an `impossibility proof',
the falsity of this conclusion is already indicated by the success  
of Bohmian mechanics (section \ref{Bohm}). On the other hand, Bell's theorem
shows that from the existence of definite values for the spins 
follows a conclusion that is in conflict with the quantum mechanical 
predictions. This prompts the question of just what feature Bohmian
mechanics possesses which allows it to succeed where the quite general 
looking formulation of hidden variables analyzed by Bell does not.
If we examine the functions $A(\lda,\hat{a})$ and $B(\lda,\hat{b})$
analyzed by Bell, we see that they do not possess the feature 
we have just seen 
is present in the quantum theory itself: nonlocality. This 
follows since the value of $A$ does not depend on the setting $\hat{b}$ 
of the apparatus measuring particle $2$, nor does $B$ depend on the 
setting $\hat{a}$ of
the apparatus measuring particle $1$. Thus, Bell analyzes a {\em local} 
theory of hidden 
variables\footnote{This feature is not ``accidental'': the hidden variables
analyzed by Bell are precisely those given by the spin singlet EPR 
analysis---which itself is based on the 
locality assumption. Moreover, Bell did not regard his analysis as a
hidden variables impossibility proof, since he was aware that: 
\cite{Bell's theorem} ``\ldots a hidden variable interpretation 
(Bohmian mechanics) has been explicitly constructed''. Instead, Bell
viewed his theorem as a proof that the ``grossly nonlocal structure''
inherent in Bohmian mechanics ``is characteristic \ldots of any such 
(hidden variables) theory which reproduces exactly the quantum 
mechanical predictions''.}.
Bohmian mechanics, on the other hand, is nonlocal, and it is precisely this
that allows it to ``escape'' disproof by Bell's theorem. 
Hence, Bell's theorem does not constitute a disproof of hidden
variables {\em in general}, but only of {\em local} hidden
variables.

\setcounter{chapter}{3}
\chapter{\Erwins paradox and nonlocality}
\section{\Erwins paradox}
In addition to his seminal role in the development of quantum theory 
itself, Erwin \Erwin made important contributions \cite{Present Sit, 
Camb1, Camb2} to its interpretation, both in his development of the
paradox of ``\Erwins cat'' and in his generalization of the 
Einstein--Podolsky--Rosen paradox. In this latter analysis, \Erwin 
demonstrated that the state considered by EPR leads to a much more 
general result than these authors had concluded. According to
the incompleteness argument given by Schr{\"o}dinger, for any
system described by a certain class of quantum state\footnote{A 
`maximally entangled state'.}, {\em all} observables 
of both particles must be elements of reality.
As was the case with EPR, \Erwins argument was based on the
existence of perfect correlations exhibited by the state.
In this chapter, we 
show that \Erwins result can be developed in a simpler way which allows
one to determine which observables are perfectly correlated with one 
another using the form of the maximally entangled state in question. 
Moreover, we derive a wide variety of new quantum nonlocality proofs based on
\Erwins generalization of EPR in conjunction with
 the theorems discussed in chapter 2, 
Gleason's theorem, Kochen and Specker's theorem, and Mermin's 
theorem. We show that {\em any} such ``spectral-incompatibility'' theorem
(see section \ref{Spec}) when taken together with the \Erwin paradox 
provides a proof of quantum nonlocality. Since the 
spectral-incompatibility theorems involve the quantum predictions for
individual measurements, the 
nonlocality proofs to which they lead are of
a {\em deterministic}, rather than statistical, character. A further
noteworthy feature of this type of quantum nonlocality 
is that its experimental confirmation need involve only the verification 
of the perfect correlations, with no further observations required.
Before addressing these matters, we discuss the original form of the
EPR paradox.

\subsection{The Einstein--Podolsky--Rosen quantum state \label{EPR orig}}
The Bohm version of the EPR paradox presented in chapter $3 $
was addressed to the spin components of two particles represented 
by the spin singlet state. While this argument was concerned with perfect 
correlations in the spin components, the original form of the EPR paradox
involved perfect correlations of a slightly different
form for the positions and momenta. For a system described by the original 
EPR state, we find that 
measurements of the positions $x_{1}$ and $x_{2}$ of the two particles 
give equal results\footnote{EPR discuss a form in which the difference between 
the positions is equal to a {\em constant} they call $d$. The
case we consider differs from this only in that the points of origin 
from which the positions of the two particles are measured are 
different, so that the quantity $x_{2}-x_{1} +d$ for EPR becomes 
$x_{2}-x_{1}$ for our case.} 
\footnote{It is clear that the Einstein--Podolsky--Rosen quantum state 
\ref{eq:EPR state} is not normalizable. Nevertheless, as we saw in the example
of the spin singlet state, it is possible to carry
out a similar argument for states which {\em can} be normalized. The
`maximally entangled states', which are the subject of the \Erwin
paradox, contain among them a large class of normalizable states,
as we shall see.}, while 
measurements of the momenta give results which sum to zero. 

To develop the perfect correlations in position and momentum, we first
recall how the spin singlet state leads to the spin correlations.       
The spin singlet state takes the form \ref{eq:thetass}:
\beq
\psi_{ss} = |\!\upa \theta,\phi \rangle  \,
\otimes \, |\!\dwna \theta,\phi \rangle
  \,\, -\,\, |\!\dwna \theta, \phi \rangle  
\, \otimes \, |\!\upa \theta, \phi \rangle
\label{eq:thetasss}
\mbox{,}
\eeq
where we have suppressed the normalization for simplicity.
We have here a sum of 
two terms, each being a product of an eigenfunction of 
$\sigma^{(1)}_{\theta,\phi}$
with an eigenfunction of $\sigma^{(2)}_{\theta,\phi}$ such that the
corresponding eigenvalues sum to zero. In the first term of
\ref{eq:thetasss} for example, the factors are eigenvectors corresponding
to $\sigma^{(1)}_{\theta,\phi}=\frac{1}{2}$ and $\sigma^{(2)}_{\theta,\phi}
=-\frac{1}{2}$. 
With this, it is clear that we will have perfect correlations between
the results of measurement of $\sigma^{(1)}_{\theta,\phi}$ and 
$\sigma^{(2)}_{\theta,\phi}$, i.e. measurements of these observables will
give results which sum to zero. By analogy with this result,
the perfect correlations in position and momentum emphasized by Einstein, 
Podolsky, and Rosen will follow 
if the wave function assumes the forms:
\begin{eqnarray}
\psi & = & \int^{\infty}_{-\infty} dp \mbox{ }| \phi_{-p}\ra 
\otimes |\phi_{p}\ra 
\label{eq:xpcorr}
\\
\psi & = & \int^{\infty}_{-\infty} dx \mbox{ } 
|\varphi_{x}\ra \otimes |\varphi_{x}\ra
\nonumber
\mbox{.}
\end{eqnarray}
Here $|\phi_{p}\ra$ is the eigenvector of momentum
operator corresponding to a momentum
of $p$.  $|\varphi_{x}\ra$ is the
eigenvector of position. We now show that 
the the wave function given by the first equation in \ref{eq:xpcorr}:
\beq
\psi_{EPR}=\frac{1}{2\pi\hbar} \int^{\infty}_{-\infty} dp 
\mbox{ } |\phi_{-p}\ra \otimes |\phi_{p}\ra
\label{eq:EPR state}
\eeq   
 assumes also the form of the second equation in \ref{eq:xpcorr}. 

 To 
see this 
we expand the first factor in the summand of
$\psi_{EPR}$ in terms of $|\varphi_{x}\ra$:
\beq
\psi_{EPR}=\int^{\infty}_{-\infty} dp 
\left( 
\int^{\infty}_{-\infty} dx 
|\varphi_{x} \ra \la \varphi_{x} | \phi_{-p} \ra  
\right) 
\otimes |\phi_{p}\ra
\mbox{.}
\eeq
Using $\la \varphi_{x}  | \phi_{-p} \ra = \la \phi_{p}  | \varphi_{x} \ra$,
this becomes
\beq
\psi_{EPR}=\int^{\infty}_{-\infty} dx | \varphi_{x} \ra \otimes 
\int^{\infty}_{-\infty} dp |\phi_{p} \ra \la \phi_{p} | \varphi_{x} \ra
\mbox{.}
\eeq
Since $\int^{\infty}_{-\infty} dp |\phi_{p} \ra \la \phi_{p} |$
is a unit operator, it follows that
\beq
\psi_{EPR} = \int^{\infty}_{-\infty} dx  | \varphi_{x} \ra \otimes | \varphi_{x} \ra 
\eeq 
This is the result we had set out to obtain.

\subsection{Schr{\"o}dinger's generalization \label{parad}}
\subsubsection{Maximal perfect correlations \label{Perf}}
Schr{\"o}dinger's work essentially revealed the full potential of the
quantum state which Einstein, Podolsky, and Rosen had considered.
 In his analysis, \Erwin demonstrated that 
the perfect correlations the EPR state exhibits are not limited to those
in the positions and momenta. For two particles described by the EPR state,
{\em every} observable of each particle will exhibit perfect correlations 
with an observable of the other. What we present here and in subsequent 
sections is a simpler way to develop such perfect correlations than that
given by \Erwin\footnote{See \cite{Camb1}}
This result may be 
seen as follows. The EPR state \ref{eq:EPR state} is rewritten 
as\footnote{The reader may object that for this state, the two
particles lie `on top of one another', i.e., the probability
that $x_{1} \neq x_{2}$ is identically $0$. However, in the
following sections, we show that the same conclusions drawn for
the EPR state also follow for a more general class of `maximally
entangled states', of which many do {\em not} restrict the
positions in just such a way.}
\beq
\la x_{1},x_{2}| \psi_{EPR} \ra 
= \int^{\infty}_{-\infty} dp_{1}dp_{2}
\la x_{1},x_{2}|p_{1},p_{2} \ra \la p_{1},p_{2}|\psi_{EPR}\ra
=\delta(x_{2}-x_{1})
\label{eq:delta}
\mbox{.}
\eeq
We may then use 
a relationship known as the {\em completeness} relationship. According to 
the completeness relation, if $\{\phi_{n}(x)\}$ is any basis for the 
Hilbert space $L_{2}$, then we have 
$\sum^{\infty}_{n=1} \phi^{*}_{n}(x_{1}) \phi_{n}(x_{2}) = 
\delta(x_{2}-x_{1})$. The EPR state may be rewritten using this 
relation as:
\beq
\psi_{EPR}(x_{1},x_{2})=\sum^{\infty}_{n=1} \phi^{*}_{n}(x_{1}) \phi_{n}(x_{2})
\label{eq:SCHR}
\mbox{,}
\eeq
where $\{\phi_{n}(x)\}$ is an {\em arbitrary} basis of $L_{2}$. 
Since the form \ref{eq:SCHR} resembles the spin 
singlet form \ref{eq:thetass}, one might anticipate that it will lead to the 
existence of perfect correlations between the observables of particles 
$1$ and $2$. As we shall see, this is true not only for quantum 
systems described by \ref{eq:SCHR}, but also for a more general class 
of states as well. 

Let us consider a Hermitian operator $A$ (assumed to possess a 
discrete spectrum) on the Hilbert space $L_{2}$. At this point,
we postpone defining an observable of the EPR system in terms of $A$---
we simply regard $A$ as an abstract operator. Suppose 
that $A$ can be written as 
\beq
A  =  \sum^{\infty}_{n=1}\mu_{n}\left|\phi_{n}\right>\left<\phi_{n}\right| 
\label{eq:eprpc} 
\mbox{,}
\eeq
where $\left|\phi_{n}\right>\left<\phi_{n}\right|$ 
is the one-dimensional projection
operator associated with the vector $|\phi_{n}\ra$. The eigenvectors 
and eigenvalues of $A$ are respectively the sets $\{|\phi_{n}\ra\}$, and 
$\{\mu_{n}\}$. Suppose that another Hermitian operator, called $\tilde{A}$, 
is defined by the relationship
\beq
\tilde{A}_{x,x^{'}} = A^{*}_{x,x^{'}}
\label{eq:EPRstar} 
\mbox{,}
\eeq
where $\tilde{A}_{x,x^{'}}$, and $A_{x,x^{'}}$ are respectively the
matrix elements of $\tilde{A}$ and $A$ in the position basis
$\{\varphi_{x}\}$, and the superscript `$*$' denotes complex conjugation.
Note that for any given $A$, the `complex conjugate' operator 
$\tilde{A}$ defined by \ref{eq:EPRstar} is {\em unique}. 
Writing out $A_{x,x^{'}}$ gives
\begin{eqnarray}
A_{x,x^{'}} & = &  \la\varphi_{x}|
\left( \sum^{\infty}_{n=1} \mu_{n}\left|\phi_{n}\ra\la\phi_{n}\right|
\right) | \varphi_{x^{'}}\ra \\
&  =  &  \sum^{\infty}_{n=1}\mu_{n} 
\la\varphi_{x}|\phi_{n}\ra
\la\phi_{n}|\varphi_{x^{'}}\ra
\nonumber
\mbox{.}
\end{eqnarray}
Using \ref{eq:EPRstar}, we find  
\begin{eqnarray}
\tilde{A}_{x,x^{'}}  & =  & \sum^{\infty}_{n=1}\mu_{n} 
\la\varphi_{x}|\phi_{n}\ra^{*}
\la\phi_{n}|\varphi_{x^{'}}\ra^{*} \label{eq:positel} \\
 & = & \sum^{\infty}_{n=1}\mu_{n} 
\la\varphi_{x}|\phi^{*}_{n}\ra
\la\phi^{*}_{n}|\varphi_{x^{'}}\ra
 \nonumber
\mbox{,}
\end{eqnarray}
where $|\phi^{*}_{n}\ra$ is just the Hilbert space vector 
corresponding\footnote{More formally, we develop the correspondence of
Hilbert space vectors $|\phi\ra$ to functions $\phi(x)$ by expanding
$|\psi\ra$ in terms of the position eigenvectors $|\varphi_{x}\ra$:
\beq
|\phi\ra = \int^{\infty}_{-\infty}dx|\varphi_{x}\ra \la\varphi_{x}|\phi\ra
\mbox{.}
\eeq
The function $\phi(x)$ can then be identified with the inner 
product $\la\varphi_{x}|\phi\ra$.
Then the Hilbert space vector corresponding to $\phi^{*}(x)$ is just
\begin{eqnarray}
|\phi^{*}\ra & = & 
\int^{\infty}_{-\infty}dx|\varphi_{x}\ra \la\varphi_{x}|\phi\ra^{*} \\
& = & \int^{\infty}_{-\infty}dx|\varphi_{x}\ra \la\phi|\varphi_{x}\ra
\nonumber
\mbox{.}
\end{eqnarray} } 
to the function $\phi^{*}_{n}(x)$. From the second equation in 
\ref{eq:positel}, it follows that $\tilde{A}$ takes the form
\beq
\tilde{A}=\sum^{\infty}_{n=1} \mu_{n} |\phi^{*}_{n}\ra
\la\phi^{*}_{n}|
\mbox{.}
\eeq
The eigenvectors and eigenvalues of $\tilde{A}$ are respectively the sets 
$\{|\phi^{*}_{n}\ra\}$, and $\{\mu_{n}\}$.

We now consider the observables ${\bf 1}\otimes A$
and $\tilde{A} \otimes {\bf 1}$ on the Hilbert space of the EPR 
state, $L_{2}\otimes L_{2}$, where ${\bf 1}$ is the identity operator
on $L_{2}$. Put less formally, we consider $A$ to be an observable of
particle $2$, and $\tilde{A}$ as an observable of particle $1$, just as
$\sigma^{(1)}_{\hat{a}}$ represented `the spin component of 
particle 1' in our above discussion of the spin singlet state.
Examining \ref{eq:SCHR}, we see that each term
is a product of an eigenvector of $A$
with an eigenvector of $\tilde{A}$ such that the eigenvalues are equal. 
For these observables, we have that $ \psi_{EPR} \mbox{ {\em is an eigenstate 
of} } A-\tilde{A}$ {\em of eigenvalue zero}, i.e., 
$(A-\tilde{A})\psi_{EPR}=0$.
Thus, $\psi_{EPR}$ exhibits {\em perfect correlation} between $A$ and 
$\tilde{A}$ in that the measurements of $A$ and $\tilde{A}$ 
give results that are equal. 

Recall that the $L_{2}$ basis $\{\phi_{n}(x)\}$ appearing in
\ref{eq:SCHR} is
an arbitrary $L_{2}$ basis. With this, and the fact that
the eigenvalues $\mu_{n}$ chosen for $A$ are arbitrary, it follows
that the operator $A$ can represent {\em any}
observable of particle $2$. We can thus conclude that
for any observable $A$ of
particle $2$, there exists a unique observable $\tilde{A}$ 
(defined by \ref{eq:EPRstar}) of particle $1$
which exhibits perfect correlations with the former. 

We may interchange the roles of particles $1$ and $2$ in the above 
argument, if we note that the EPR state is {\em symmetric} in
$x_{1}$ and $x_{2}$. This latter is proved as follows. Since the Dirac 
delta function is an even function, we have
\beq
\delta(x_{2} -x_{1})
=\delta(x_{1} -x_{2})
\label{eq:SCHRsymm}
\mbox{.}
\eeq
Thus the EPR state assumes the form
\beq
\psi_{EPR} 
=\delta(x_{2} -x_{1})
=\delta(x_{1} -x_{2})
=\sum^{\infty} _{n=1} \phi_{n} (x_{1}) \phi^{*}_{n} (x_{2})
\mbox{,}
\eeq
where $\{\phi_{n}(x)\}$ is an arbitrary basis of $L_{2}$, and the second
equality is the completeness relation. This establishes the desired symmetry.

Suppose now that we consider the observables $A \otimes {\bf 1}$ and ${\bf 1} 
\otimes \tilde{A}$, i.e., we consider $A$ as an observable of 
particle $1$, and $\tilde{A}$ as an observable of particle $2$. Then
from the symmetry of the EPR 
state, it follows that we can reverse the roles of the particles in 
the above discussion to show that for any observable $A$ of particle $1$,
there exists a unique observable $\tilde{A}$ of particle $2$ defined by
\ref{eq:EPRstar}, which exhibits perfect correlations with the former.

The perfect correlations in position and momentum originally noted by
Einstein, Podolsky and Rosen are a special case of the perfectly 
correlated observables $A$ and $\tilde{A}$ given here. EPR found that 
the measurement of the positions $x_{1} $ and $x_{2} $ of the two particles 
must give results that are {\em equal}. The momenta $p_{1} $ and 
$p_{2} $ showed slightly different perfect correlations in which their
measurements give results which sum to zero. To assess 
whether the EPR perfect correlations in position and momentum are 
consistent with the scheme given above, we derive using \ref{eq:EPRstar} the 
forms of $\tilde{x}$ and $\tilde{p}$. 
Since the position observable $x$ is diagonal in the position
eigenvectors and has real matrix elements, \ref{eq:EPRstar} implies 
that $\tilde{x}=x$. As for the momentum, this operator is
represented in the position basis by the differential operator
$-i\hbar\frac{d}{dx}$. Using \ref{eq:EPRstar} we see that
$\tilde{p}$ is just equal to the {\em negative} of $p$ itself.
Hence the perfect correlations developed in the present section, 
according to which $A$ and $\tilde{A}$ are equal, imply that 
measurements of the
positions of the two particles must give equal results, whereas 
measurements of the
momenta $p_{1} $ and $p_{2} $ must give results which sum to zero. 
This is just what EPR had found. 

\subsubsection{Incompleteness argument \label{inc}}
The existence of such ubiquitous perfect correlations allows one to
develop an argument demonstrating that all observables of 
particle $1$ and all observables of particle $2$ are ``elements of reality'', 
i.e., they possess definite values. This argument is similar to the 
incompleteness argument given above for the spin singlet EPR case. We consider 
the possibility of separate measurements of observables $\tilde{A}$, $A$
being performed respectively on particles $1$ and $2$ of
the EPR state. As in the spin singlet EPR analysis, we assume
locality, so that the properties of each particle must be regarded as
independent of those of its spatially separated partner. 
Suppose we
perform an experimental procedure ${\cal E}(\tilde{A})$ which
constitutes a measurement of the observable $\tilde{A}$ of particle $1$. 
Since $\tilde{A}$ is perfectly correlated with $A$, the
result we find allows us to predict with certainty the result of any
subsequent experiment ${\cal E}(A)$ measuring $A$ of
particle $2$. For example, if ${\cal E}(\tilde{A})$ gives the result 
$\tilde{A}=\mu_{a}$ then
we can predict with certainty that ${\cal E}(A)$ will give
$A=\mu_{a}$. Since we have assumed {\em locality}, 
$A$ must be an element of reality, i.e., there exists
some definite value $E(A)$.
Furthermore, since the {\em same} number is
predicted no matter what experimental procedure 
${\cal E}(A)$ is used in $A$'s measurement, $E(A)$ cannot 
depend on the choice of procedure, i.e., $E(A)$ must be
non-contextual. 

We saw in the discussion above that for {\em any} observable $A$ of
particle $2$, there exists an observable $\tilde{A}$ of particle $1$
with which the former shows perfect correlation. Hence, the above 
argument may be applied to any observable of particle $2$,
and we can therefore conclude that {\em all} observables of particle $2$ are 
elements of reality, and that the definite value of each is 
non-contextual. Of course, if we consider the set of all observables
of particle $2$, there are some pairs among the set which are
non-commuting. 
Since, from the quantum formalism's description of state, one can 
deduce, at most, one observable of such a pair, we may conclude that this 
description of state is incomplete.
Thus, any theoretical
structure which hopes to capture these definite values must contain 
a state description which extends that of quantum mechanics.

As we noted above, it is possible to interchange the roles of 
the two particles when deriving the perfect correlations. Thus,
to every observable $A$ of particle $1$, there is a unique observable
$\tilde{A}$ of particle $2$ with which the former is perfectly 
correlated. This being the case, one can construct a similar 
incompleteness argument to show that all observables of particle $1$
possess non-contextual definite values.

Finally, note that if we consider the possibility of an
experiment measuring a {\em commuting 
set} of observables of either particle, the quantum theory predicts 
that the result is one of the joint-eigenvalues of that set. For 
agreement with this prediction, the values  $E(O)$ assigned to the 
observables of this particle must map each commuting set to a 
joint-eigenvalue. 

The \Erwin paradox is seen to have the following structure. The analysis 
begins with the observation that the EPR state \ref{eq:SCHR} exhibits 
maximal perfect
correlations such that for any observable $A$ of either particle,
there is a unique observable $\tilde{A}$ of the other with which $A$
is perfectly correlated.
If we assume locality, then these perfect correlations imply
the existence of a non-contextual value map on all observables of both 
particles. These value maps must assign to every commuting set a
joint-eigenvalue. 
 
The key to the above incompleteness argument is that one
can, by measurement of an observable of particle
$2$, predict with certainty the result of any measurement of a 
particular observable of particle $1$. 
In his presentation of the EPR incompleteness argument (which concerns 
position and momentum) \Erwin uses a colorful 
analogy to make the situation clear. He imagines particles $1$ and $2$
as being a student and his instructor. The measurement of particle $2$ 
corresponds to the instructor consulting a textbook to check the answer to an 
examination question, and the measurement of particle $1$ to the 
response the student gives to this question. Since he always gives 
the correct answer, i.e., the same as that in the instructor's textbook, 
the student must have known the answer beforehand. 
\Erwin presents the situation as follows: \cite{Present Sit} 
(emphasis by original author)

\begin{quotation}
Let us focus attention on the system labeled with small letters
$p,q$ and call it for brevity the ``small'' system. Then things stand as 
follows. I can direct {\em one} of two questions to the small system,
either that about $q$ or that about $p$. Before doing so I can, if I
choose, procure the answer to {\em one} of these questions by a measurement
on the fully separated other system (which we shall regard as auxiliary
apparatus), or I may take care of this afterwards. My small system, like
a schoolboy under examination, {\em cannot possibly know} whether I have
done this or for which questions, or whether or for which I intend to do
it later. From arbitrarily many pretrials I know that the pupil will
correctly answer the first question I put to him. From that it follows
that in every case, he {\em knows} the answer to {\em both} questions.
\ldots No school principal would judge otherwise \ldots He would not come
to think that his, the teacher's, consulting a textbook first suggests
to the pupil the correct answer, or even, in the cases when the 
teacher chooses to consult it only after ensuing answers from the
pupil, that the pupil's answer has changed the text of the notebook
in the pupil's favor.
\end{quotation}  

\subsubsection{Generalized form of the EPR state \label{csection}}
Consideration of the structure of the 
EPR state \ref{eq:SCHR} suggests the possibility that
a more general class of states might also exhibit maximal
perfect correlations. Before developing this, we first note that 
the formal way to write \ref{eq:SCHR} is such that each term consists of a
tensor product of a vector of particle $1$'s Hilbert space
with a vector of particle $2$'s Hilbert space:
\beq
\psi_{EPR} =\sum^{\infty} _{n=1} |\phi_{n}^{*} \ra  \otimes 
|\phi_{n} \ra
\mbox{.}
\eeq
We shall often use this convenient notation in expressing the
maximally entangled states.
The operation of complex conjugation in the first factor of
the tensor product may be regarded as a special
case of a class of operators known as ``anti-unitary
involutions''---operators
which we shall denote by $C$, to suggest complex conjugation. We
will find that any state of the form
\beq
\psi_{C} =\sum^{\infty} _{n=1} C|\phi_{n} \ra \otimes |\phi_{n} \ra
\mbox{,}
\label{eq:SCHRC}
\eeq
will exhibit ubiquitous perfect
correlations leading to the conclusion of definite values on all
observables of both subsystems.

An {\em anti-unitary involution} operation represents the 
generalization of complex conjugation from scalars
to vectors. The term
``involution'' refers to any operator $C$ whose square is equal to the 
identity operator, i.e. $C^{2}={\bf 1}$. 
Anti-unitarity\footnote{This term may appear confusing for the following
reason. It does not not refer to an operator which
is ``not unitary'', but instead the prefix ``anti'' refers to 
anti-linearity.} 
entails two conditions, the first of which is anti-linearity:
\beq
C(c_{1}|\psi_{1}\ra+c_{2}|\psi_{2}\ra+\ldots)
=c^{*}_{1}C|\psi_{1}\ra+c^{*}_{2}C|\psi_{2}\ra+\ldots
\mbox{,}
\eeq
where $\{c_{i}\}$ are constants and $\{|\psi_{i}\ra\}$ are vectors. The
second condition is the anti-linear counterpart of unitarity:
\beq
\la C\psi| C\phi\ra
=\la\psi|\phi\ra^{*} 
=\la\phi|\psi\ra
\mbox{ } \forall \psi,\phi
\mbox{,}
\label{eq:anti-lin unit}
\eeq
which tells us that under the 
operation of $C$, inner products are replaced by their complex conjugates. 
Note that this property is sufficient to guarantee that if the set
$\{|\phi_{n} \ra\}$ is a basis, then the vectors 
$\{C|\phi_{n} \ra\}$ in \ref{eq:SCHRC} must also form a basis.
For each anti-unitary involution $C$, there is a special Hilbert 
space basis whose elements are invariant under $C$. The operation
of $C$ on any given vector $|\psi\ra$ can be easily obtained by expanding
the vector in terms of this basis. If $\{\varphi_{n} \}$ is this special
basis then we have
\beq
C|\psi\ra=C\sum^{\infty} _{i=1} |\varphi_{i} \ra\la \varphi_{i} |\psi\ra
=\sum^{\infty} _{i=1} |\varphi_{i} \ra\la \varphi_{i} |\psi\ra^{*} 
\label{eq:var}
\mbox{.}
\eeq
When one is analyzing any given state of the form \ref{eq:SCHRC},
it is convenient to express the state and observables using this
special basis. The EPR state is a special case of the state 
\ref{eq:SCHRC} in which the anti-unitary involution $C$ is such 
that the position basis $\{|\varphi_{x} \ra\}$ plays this
role.   

The state \ref{eq:SCHRC} shows an invariance similar to that we
developed for the EPR state: the basis $\{|\phi_{n} \ra\}$ in terms of
which the state is expressed, is {\em arbitrary}.
We now develop this result. Note that the expression \ref{eq:SCHRC} 
takes the form
\beq
\psi_{C} = \sum^{\infty}_{n=1} C \left( \sum^{\infty}_{i=1} |\chi_{i} \ra 
\la \chi_{i}|\phi_{n}\ra \right) 
\otimes \left( \sum^{\infty}_{j=1} |\chi_{j} \ra 
\la \chi_{j}|\phi_{n}\ra \right)
\mbox{,}
\eeq
if we expand the vectors in terms of an arbitrary basis $\{\chi_{i}\}$.
Applying the $C$ operation in the first factor and rearranging the 
expression, we obtain
\begin{eqnarray}
\psi_{C}  & = &  \sum^{\infty}_{n=1} \left( \sum^{\infty}_{i=1} 
C |\chi_{i} \ra 
\la \phi_{n} | \chi_{i} \ra  \right)
\otimes \left( \sum^{\infty}_{j=1} |\chi_{j} \ra 
\la \chi_{j}|\phi_{n}\ra \right)  \label{eq:SCHRexp} \\
 & = &  \sum^{\infty}_{i=1} \sum^{\infty}_{j=1} 
\sum^{\infty}_{n=1}  \la \chi_{j}|\phi_{n}\ra \la \phi_{n} | \chi_{i} \ra 
   C |\chi_{i} \ra \otimes |\chi_{j} \ra \nonumber \\
 & = & \sum^{\infty}_{i=1} \sum^{\infty}_{j=1} 
  \la \chi_{j}|\left(\sum^{\infty}_{n=1} |\phi_{n}\ra \la \phi_{n}| \right)
  | \chi_{i} \ra 
   C |\chi_{i} \ra \otimes |\chi_{j} \ra
\nonumber
\mbox{,}
\end{eqnarray}
where the first equality follows from the anti-linearity of $C$.
Since the expression $\sum^{\infty}_{n=1} |\phi_{n}\ra \la \phi_{n}|$ is the
identity operator, the orthonormality of the set $\{\chi_{i}\}$ implies 
that
\beq
\psi_{C}  =  \sum^{\infty}_{i=1} \sum^{\infty}_{j=1} \delta_{ij}
C |\chi_{i} \ra \otimes |\chi_{j} \ra 
= \sum^{\infty}_{i=1} C |\chi_{i} \ra \otimes |\chi_{i} \ra
\label{eq:SCHdelta}
\mbox{.}
\eeq
Thus, the form of the state \ref{eq:SCHRC} is invariant under 
any change of basis, and we have the desired result.

Note then the role played by the properties of anti-linear
unitarity and anti-linearity: from the former followed the result
that the the vectors $\{C|\phi_{n}\ra\}$ form a basis if the
vectors $\{|\phi_{n}\ra\}$ do so, and from the latter followed the
invariance just shown.   

Consider a Hermitian operator $A$ on $L_{2}$  
which can be written as
\beq
A  =  \sum^{\infty}_{n=1}\mu_{n}\left|\phi_{n}\right>\left<\phi_{n}\right| 
\label{eq:cpc} 
\mbox{.}
\eeq
Note that $A$'s eigenvectors and eigenvalues are given by the sets
 $\{|\phi_{n}\ra\}$ and $\{\mu_{n}\}$, respectively. 
If we define the observable $\tilde{A}$ by the relationship  
\beq
\tilde{A}=CAC^{-1}
\label{eq:cacistar}
\mbox{,}
\eeq 
then $\tilde{A}$'s eigenvalues are the same as $A$'s, i.e., 
$\{\mu_{n}\}$, and its eigenvectors are given by 
$\{C|\phi_{n}\ra\}$. To see this, note that
\beq
CAC^{-1}C|\phi_{n}\ra = C\mu_{n}|\phi_{n}\ra=\mu_{n}C|\phi_{n}\ra
\mbox{,}
\eeq
where the first equality follows from $CC^{-1}={\bf 1}$ and 
$A|\phi_{n}\ra=\mu_{n}\phi_{n}$. Since $C^{2}={\bf 1}$, it follows 
that $C=C^{-1}$, and we may rewrite \ref{eq:cacistar} 
as\footnote{Note that to evaluate $\tilde{A}$, one can express 
it using its matrix elements with respect to 
the invariant basis $\{\varphi_{n}\}$ 
of $C$. If we evaluate the matrix element $CAC_{ij}$, we find
\beq
\la\varphi_{i}|CAC\varphi_{j}\ra 
= \la \varphi_{i}|CA\varphi_{j}\ra 
= \la\varphi_{i}|A\varphi_{j}\ra^{*}
\mbox{,}
\label{eq:calc}
\eeq
where the second equality follows from \ref{eq:var}.
From \ref{eq:calc} follows the relationship 
\beq
\tilde{A}_{ij}=A^{*}_{ij}
\label{eq:vstar}
\mbox{,}
\eeq
which is a convenient form one may use to evaluate $\tilde{A}$, as we will 
see in section \ref{spinmax}. Note that \ref{eq:vstar} reduces to 
\ref{eq:EPRstar} when the invariant basis $\{\varphi_{n}\}$ of $C$ is the
position basis $|\varphi_{x}\ra$.}:
\beq
\tilde{A}=CAC
\label{eq:cacstar}
\mbox{.}
\eeq
Note that for any given $A$, the observable $\tilde{A}$ defined by 
\ref{eq:cacstar} is unique.

If we identify $A$ as an observable of subsystem $2$, and $\tilde{A}$
as an observable of subsystem $1$, then examination of the state 
\ref{eq:SCHRC} shows that these exhibit {\em perfect correlations} 
such that their measurements always yield results that are equal. From
the 
invariance of the state, we can then conclude that for {\em any} observable 
$A$ of 
subsystem $2$, there is a unique observable $\tilde{A}$ of subsystem
$1$ defined 
by \ref{eq:cacstar} which exhibits perfect correlations with $A$. 
Since, as can easily be proved\footnote{The proof of this result follows 
similar lines as the invariance proof given above. One
expands the vectors of the state \ref{eq:SCHRC} in terms of the 
invariant basis $\{\varphi_{n}\}$. Doing so leads to an expression 
similar to the first equality in \ref{eq:SCHdelta}, with  
$\varphi_{i}$ and $\varphi_{j}$ replacing $\chi_{i}$ and $\chi_{j}$,
respectively. The delta function is then replaced using the 
relationship 
\beq
\delta_{ij}=\la\varphi_{i}|
\left(\sum^{\infty}_{n=1}|\phi_{n}\ra\la\phi_{n}|\right)|\varphi_{j}\ra
\mbox{,}
\eeq 
and one can then easily develop \ref{eq:SCHRCP}}, 
the state \ref{eq:SCHRC} assumes the form
\beq
\psi_{C} = \sum^{\infty} _{n=1} |\phi_{n} \ra \otimes C |\phi_{n} \ra
\label{eq:SCHRCP}
\mbox{,}
\eeq  
one may develop the same results with the roles of the subsystems 
reversed, i.e., for any observable $A$ of subsystem $1$, there exists 
a unique observable $\tilde{A}$ of subsystem $2$ which exhibits 
perfect correlations with $A$.
 
An incompleteness argument similar to that given above can be
given, and one can show the existence of 
a value map $E(O)$ on all observables of both subsystems.

\subsubsection{The general form of a maximally entangled state}
States exhibiting ubiquitous perfect correlations are not limited to those 
of the form \ref{eq:SCHRC}. If we examine any composite system  
whose subsystems are of the same 
dimensionality\footnote{The derivations of
this section can be carried 
out for an entangled system whose subsystems are either finite or 
infinite dimensional. Infinite sums may be substituted for the finite 
sums written here to develop the same 
results for the infinite dimensional case.} 
and which is represented by a 
state\footnote{This is equivalent to the form
\beq
\psi_{ME}=\sum^{N}_{n=1}c_{n}|\psi_{n}\ra \otimes |\phi_{n}\ra
\label{eq:MEC}
\mbox{,}
\eeq
where $\{|\psi_{n}\ra\}$ is any basis of subsystem $1$ and 
$\{|\phi_{n}\ra\}$ is any basis of subsystem $2$ and 
the $|c_{n}|^{2}=1 \mbox{ } \forall n$.}
\beq
\psi_{ME}=\sum^{N}_{n=1} |\psi_{n}\ra \otimes |\phi_{n}\ra
\label{eq:ME}
\mbox{,}
\eeq
where $\{|\psi_{n}\ra\}$ is any basis of subsystem $1$ and 
$\{|\phi_{n}\ra\}$ is any basis of subsystem $2$,
we find that each observable of either subsystem 
exhibits perfect correlations with some observable of the other
subsystem. To examine the properties of the state 
\ref{eq:ME}, we note that it may be 
rewritten\footnote{At this point, one may address the objection
that the EPR state \ref{eq:SCHR} makes the positions of the two
particles coincide. For example, one can consider an anti-unitary
operator $U_{d}$ defined by 
\beq
U_{d} | \psi \ra = \int^{\infty}_{\infty} dx |\varphi_{x} \ra \la \psi |
\varphi_{x+d} \ra
\mbox{,}
\eeq
where $d$ is an arbitrary constant. Then, in the state \ref{eq:ME2},
 the two particles are separated by a distance $d$.} 
as
\beq
\psi_{ME}=\psi=\sum^{N}_{n=1} U |\phi_{n}\ra \otimes |\phi_{n}\ra
\label{eq:ME2}
\mbox{,}
\eeq 
where $U$ is the anti-unitary operator defined by $U|\phi_{n}\ra 
= |\psi_{n}\ra$. Recall from the above discussion that (anti-linear)
unitarity and anti-linearity are sufficient to guarantee both that 
$\{U|\phi_n\ra\}$ is a basis if $\{|\phi_{n}\ra\}$ is, and that
the state shows the invariance we require.
We conclude our presentation of
the \Erwin paradox by discussing this general form.

We consider a Hermitian operator $A$ on ${\cal H}_{N}$ which can be 
written as
\beq
A  =  \sum^{N}_{n=1}\mu_{n}\left|\phi_{n})\right>\left<\phi_{n}\right| 
\label{eq:upc} 
\mbox{.}
\eeq
We then define the observable $\tilde{A}$ 
by\footnote{For comparison with the form \ref{eq:EPRstar}, and 
\ref{eq:vstar}, we note that if we express the operator $U$ as 
$U=C\bar{U}$ with $C$ an anti-unitary involution, and $\bar{U}$ a 
unitary matrix, then \ref{eq:uaustar} leads to 
\beq
\tilde{A}_{ij}=(\bar{U}A\bar{U}^{-1})^{*}_{ij}
\mbox{,}
\eeq 
where the $ij$ subscript indicates the $ij$th matrix element in terms 
of $\{\varphi_{n}\}$, the basis that is invariant under $C$.} 
\beq
\tilde{A}=UAU^{-1}
\label{eq:uaustar}
\mbox{.}
\eeq
For any given $A$, this defines a unique operator $\tilde{A}$.
Using $A|\phi_{n}\ra=\mu_{n}|\phi_{n}\ra$ and $UU^{-1}={\bf 1}$, we 
have
\beq
UAU^{-1}U|\phi_{n}\ra=\mu_{n}U\phi_{n}
\mbox{,}
\label{eq:Cosmos}
\eeq
so that the eigenvectors and eigenvalues of $\tilde{A}$ are given by
$\{U\phi_{n}\}$ and $\{\mu_{n}\}$.

Identifying $A$ as an observable of subsystem $2$ and $\tilde{A}$ as an
observable of subsystem $1$, we can see from the form of \ref{eq:ME2}
that these exhibit perfect correlations for such a state. 
As in the EPR case and its generalization, the basis-invariance and symmetry 
of the state imply that for {\em any} observable $A$ of either 
subsystem, there is a unique observable $\tilde{A}$ of the other with 
which $A$ is perfectly correlated.
An argument similar to that given for the EPR state leads to
 the existence of a non-contextual value map $E(O)$
on all observables of subsystem $1$ and all observables of subsystem $2$.

\subsection{The spin singlet state as a maximally entangled state 
\label{spinmax}}
It is instructive to make an explicit 
comparison of the properties given above for a general maximally entangled 
state with those of the spin singlet state.
The spin singlet state takes the form 
\beq
\psi_{ss}=|\!\upa \theta,\phi \rangle \, \otimes \,
|\!\dwna \theta,\phi \rangle
  \,\, -\,\, |\!\dwna \theta, \phi \rangle \, \otimes 
\, |\!\upa \theta, \phi \rangle
\mbox{,}
\label{eq:MEss2}
\eeq
when expressed in terms of the eigenvectors of 
$\sigma_{\theta,\phi}$. We have suppressed the normalization
for simplicity. Inspection of this form leads one to 
conclude that the state must exhibit correlations 
such that the results of measurement of $\sigma^{(1)}_{\theta,\phi}$ and 
$\sigma^{(2)}_{\theta,\phi}$ will give results which sum to zero. These
might be better named ``perfect anti-correlations'', since the 
measurement results are the negatives of one another.
On examining the form given in \ref{eq:MEC},
one can see immediately that the spin singlet state \ref{eq:MEss2} is
a maximally entangled state. 

This being the case, it follows that the 
spin singlet state must assume the form \ref{eq:ME2}. To see this,
we write the anti-unitary operator $U=C\bar{U}$, where $\bar{U}$
is a unitary operator given by 
$\bar{U} = \left( 
\begin{array}{rr} 
0 & 1 \\
-1 & 0 
 \end{array}  \right)$ 
and $C$ is an anti-unitary involution 
under which the $\sigma_{z}$ eigenvectors are 
invariant, i.e.,
$C|\uupa\ra=|\uupa\ra$ and $C|\ddwna\ra=|\ddwna\ra$.
Using $|\uupa\ra,|\ddwna\ra$ as the basis in the maximally entangled 
state expression \ref{eq:ME2}, we obtain
\beq
\psi=U|\uupa\ra\otimes|\uupa\ra+U|\ddwna\ra\otimes|\ddwna\ra
\mbox{.}
\eeq
This 
reduces\footnote{To within an overall minus sign, which can be ignored.} 
to the familiar spin singlet form
\beq
\psi = |\!\upa \rangle \, \otimes \,
|\!\dwna  \rangle
  \,\, -\,\, |\!\dwna  \rangle \, \otimes 
\, |\!\upa \rangle
\mbox{,}
\eeq
when we write out the operation of $\bar{U}$ as a matrix multiplication.
 
 The perfect correlations between spin components
of the two particles can be regarded as a special case of the maximally
entangled state perfect correlations, which hold between any observables 
$A$ of one 
subsystem and $\tilde{A}$ of the other, where $\tilde{A}=UAU^{-1}$.
To develop this, we recall that the observable $\tilde{A}$ exhibits
perfect correlations with the observable $A$ such that 
 measurements of $A$ and $\tilde{A}$ give results that are equal. In the
case of the spin singlet state, we have what one might call ``perfect 
anti-correlations'', i.e., measurements of
$\sigma^{(1)}_{\theta,\phi}$ and $\sigma^{(2)}_{\theta,\phi}$ give
results which sum to zero. Thus, for the case of the spin singlet state, we 
expect that $\widetilde{\sigma_{\theta,\phi}}=-\sigma_{\theta,\phi}$, or
\beq
U\sigma_{\theta,\phi}U^{-1}=-\sigma_{\theta,\phi}
\mbox{,}
\label{eq:Sbar}
\eeq
where $U$ is the operator described above. 
Using $C^{-1}=C$, one can see that the left hand side of 
\ref{eq:Sbar} reduces
to $C\bar{U}\sigma_{\theta,\phi}\bar{U}^{-1}C$.
In deriving \ref{eq:Sbar}, we begin by evaluating the expression 
$\bar{U}\sigma_{\theta,\phi}\bar{U}^{-1}$.

Recall that the form of the observable $\sigma_{\theta,\phi}$ is given by
\beq
\sigma_{\theta,\phi}=
\left(
\begin{array}{rr}
\cos(\theta) & e^{-i\phi}\sin(\theta) \\
e^{i\phi}\sin(\theta) & -\cos(\theta)
\end{array}
\right)
\mbox{.}
\eeq
Note that one can write this as $\sigma_{\theta,\phi}=
\left(
\begin{array}{rr}
a & b \\
b^{*} & -a
\end{array}
\right)
\mbox{,}$
where $a=\cos(\theta)$ and $b=e^{-i\phi}\sin(\theta)$.
A simple calculation involving matrix multiplication shows that 
\beq
\bar{U}\left(
\begin{array}{rr}
a & b \\
b^{*} & -a
\end{array}
\right)\bar{U}^{-1}
= \left(
\begin{array}{rr}
-a & -b^{*} \\
-b & a
\end{array}
\right)
\mbox{.}
\eeq  

To complete the calculation, we must evaluate the expression 
$CAC$, where $A$ is given by the matrix 
$\left(
\begin{array}{rr}
-a & -b^{*} \\
-b & a
\end{array}
\right)$.
We saw in section \ref{csection} an expression of the form $CAC$ 
can be evaluated in terms of
its matrix elements in $\{\varphi_{n}\}$, the basis that is invariant under 
$C$. To do so, we use the relation \ref{eq:calc}:
\beq
(CAC)_{ij}=A^{*}_{ij}
\mbox{.}
\label{eq:ssstar}
\eeq
Using \ref{eq:ssstar}, it follows that $C\left(
\begin{array}{rr}
-a & -b^{*} \\
-b & a
\end{array}
\right)C=
\left(
\begin{array}{rr}
-a & -b \\
-b^{*} & a
\end{array}
\right)
\mbox{.}$ 
In comparing this relation to the form of $\sigma_{\theta,\phi}$, one can 
see that 
$C\bar{U}\sigma_{\theta,\phi}\bar{U}^{-1}C=-\sigma_{\theta,\phi}$, and
we have arrived at \ref{eq:Sbar}.

\subsection{Bell's theorem and the maximally entangled 
states \label{Bellext}}
 We have seen, in chapter 3, how quantum nonlocality may
be proved for one particular maximally entangled state, which is the 
spin singlet state. Here, we show how Bell's theorem may be applied in 
proofs of the nonlocality of a larger class of maximally entangled 
states. A more general form of the proof we give here is
given by Popescu and Rohrlich \cite{Popescu}, who demonstrate that the
nonlocality of {\em any} maximally entangled state follows from Bell's 
theorem.

Consider a general maximally entangled state given by
\beq
\psi_{ME}=\sum^{N}_{n=1} |\phi_{n}\ra \otimes |\psi_{n}\ra
=|\phi_{1}\ra \otimes |\psi_{1}\ra
+ |\phi_{2}\ra \otimes |\psi_{2}\ra + 
\sum^{N}_{n=3}  |\phi_{n}\ra \otimes |\psi_{n}\ra
\label{eq:MEP}
\mbox{.}
\eeq 
Let us define ${\cal H}_{1}$  and ${\cal H}_{2}$ as the subspaces
of particle $1$ and $2$'s Hilbert spaces spanned respectively by 
$\phi_{1}, \phi_{2}$ and $\psi_{1},\psi_{2}$. We define the sets of
observables $\{\xi^{(1)}_{\theta,\phi}\}$ and 
$\{\xi^{(2)}_{\theta,\phi}\}$ as follows. The set 
$\{\xi^{(1)}_{\theta,\phi}\}$ is formally 
identical\footnote{Since any two Hilbert spaces of the same dimension 
are isomorphic, it is possible to define such formally identical 
observables.} 
to the set $\{\sigma_{\theta,\phi}\}$ on ${\cal H}_{1}$, and 
all its members give zero when operating on any vector in the orthogonal 
complement of ${\cal H}_{1}$. 
Similarly, the set $\{\xi^{(2)}_{\theta,\phi}\}$ is formally identical to 
$\{\sigma_{\theta,\phi}\}$ on ${\cal H}_{2}$, and its members give zero when 
operating on
any vector in its orthogonal complement.
We now select a class of states smaller than
that given by \ref{eq:MEP} by making the following
substitutions: the vectors $|\uupa\ra$ and $|\ddwna\ra$ replace 
$\phi_{1}$ and $\phi_{2}$ and the vectors $|\ddwna\ra$ and $-|\uupa\ra$
replace $\psi_{1}$ and $\psi_{2}$. Here, $|\uupa\ra$ and $|\ddwna\ra$
are the eigenvectors of $\sigma_{z}$.
The state \ref{eq:MEP} then becomes
\beq
\psi
= |\uupa\ra \otimes |\ddwna \ra
- |\ddwna\ra \otimes |\uupa \ra
+ \sum^{N}_{n=3}  \phi_{n} \otimes \psi_{n}
\label{eq:Bellext}
\mbox{.}
\eeq 
The sets of observables $\{\xi^{(1)}_{\theta,\phi}\}$ and
$\{\xi^{(2)}_{\theta,\phi}\}$ are zero on every term of
\ref{eq:Bellext} with the exception of the first two. Since the
first two terms are identical to the spin singlet state, it follows 
that $\psi$ is an eigenstate of $\xi^{(1)}_{\theta,\phi}+
\xi^{(2)}_{\theta,\phi}$ of eigenvalue zero, $\forall \theta,\phi$, and 
therefore perfect correlations between all such observables. We may exploit 
this situation to derive a nonlocality proof with some further effort,
as we now show.

Let $P^{(1)}$ and $P^{(2)}$ be the projections operators which project 
respectively onto ${\cal H}_{1}$ and ${\cal H}_{2}$. 
Since the state \ref{eq:Bellext} exhibits maximal perfect correlations,
one can give an incompleteness argument to show that
$P^{(1)}, P^{(2)}$, and the sets $\{\xi^{(1)}_{\hat{a}}\}$ and
$\{\xi^{(2)}_{\hat{b}}\}$ must all possess definite values.
However, the existence of such values  
cannot agree with the statistical predictions of quantum 
mechanics, as one may see. We now examine what 
might be called the ``conditional correlation function'', which we 
define as follows. As was done in the presentation of Bell's theorem,
we represent the predetermined values of the observables 
$\{\xi^{(1)}_{\theta,\phi}\}$ and $\{\xi^{(2)}_{\theta,\phi}\}$
using the mathematical functions
$A(\lda ,\hat{a})$ and $B(\lda ,\hat{b})$. Suppose that we consider 
measurements
of the set $\{\xi^{(1)}_{\hat{a}},P^{(1)}\}$ of subsystem $1$ and
$\{\xi^{(2)}_{\hat{b}},P^{(2)}\}$ of subsystem $2$.
In every such case,
we discard those results for which either $P^{(1)}$ or $P^{(2)}$ (or both)
equals zero. We consider the correlation in measurements of 
$\xi^{(1)}_{\hat{a}}$ and $\xi^{(2)}_{\hat{b}}$ in
those cases for which $P^{(1)}=P^{(2)}=1$. Under these conditions, the
quantum mechanical prediction for the correlation function 
$P_{QM}(\hat{a},\hat{b})$ is given by 
\beq
P_{QM}(\sigma^{(1)}_{\hat{a}}\sigma^{(2)}_{\hat{b}})=\la \psi_{ss}
|\xi^{(1)}_{\hat{a}}\xi^{(2)}_{\hat{b}} | \psi_{ss} \ra = 
-\hat{a} \cdot \hat{b}
\mbox{.}
\eeq
On the other hand, the prediction derived from the predetermined values
$A(\lda ,\hat{a})$ and $B(\lda ,\hat{b})$ 
can be shown to satisfy Bell's inequality \ref{eq:Bell}. Therefore, the
statistics of these values---which themselves follow from the 
assumption of locality---are in conflict with the predictions of 
quantum mechanics and so we must conclude quantum nonlocality. 

\section{\Erwin nonlocality, and a discussion of
experimental verification \label{Meat}} 
\subsection{\Erwin nonlocality \label{Meatus}}
We have seen that the \Erwin paradox gives a more general result
than follows from either version of the EPR paradox. Because of 
this greater generality, it is
possible to construct a nonlocality proof using \Erwins paradox
in conjunction with any one of a wide variety of 
theorems besides that of Bell. This is due mainly to the
fact that the \Erwin paradox predicts
definite values for such a large class of observables that the theorems 
required need not address more than one particle or subsystem---even 
value map impossibility proofs which are concerned with {\em single 
particle} systems may be sufficient. Consider the case of Kochen and Specker's 
theorem. Suppose that some system is described by
a maximally entangled state whose subsystems are of dimensionality three:
\beq
\sum^{3}_{n=1}\phi_{n}\otimes\psi_{n}
\label{eq:schks}
\mbox{.}
\eeq
Among the set of all observables on both ${\cal H}_{1}$ and ${\cal H}_{2}$
are the squares of the 
various spin components\footnote{Of course, such a system need not consist of 
two 
spin $1$ particles. If it does not, then the same conclusion as is
given here holds for those observables which are {\em formally 
equivalent} to the sets $\{s^{2}_{\theta,\phi}\}$ and 
$\{\widetilde{s^{2}_{\theta,\phi}}\}$. Since any two Hilbert spaces
of the same dimension are isomorphic, it follows that such formally
equivalent observables must exist.}
of a spin $1$ 
particle, $\{s^{2}_{\theta,\phi}\}$. The state \ref{eq:schks} exhibits perfect 
correlations between all the observables 
$\{s^{2}_{\theta,\phi}\}$ of one subsystem,
and their counterparts $\{\widetilde{s^{2}_{\theta,\phi}}\}$ of the
other. As we saw in the incompleteness argument 
given in section \ref{inc}, if we assume locality, then the perfect 
correlations imply the existence of a non-contextual 
value map $E(O)$ on the sets $\{s^{2}_{\theta,\phi}\}$
and $\{\widetilde{s^{2}_{\theta,\phi}}\}$. However, we know from the
Kochen and Specker theorem that there exists no value map 
$\{E(s^{2}_{\theta,\phi})\}$ such that a joint-eigenvalue is assigned to 
every commuting set. This conclusion {\em contradicts} the
quantum mechanical prediction that the measurements
of any commuting set always give one of its joint-eigenvalues. 
Thus, if we demand that the perfect correlations of the state 
\ref{eq:schks} are to be explained through a {\em local} theory, we are
led to a conclusion that is in conflict with quantum mechanics. 
Therefore the quantum description of any system described by a state
such as \ref{eq:schks} must entail nonlocality. Note that the development of 
the contradiction between the quantum predictions and those of
the definite values is expressed using {\em the observables of 
one subsystem}, namely the set $\{s^{2}_{\theta,\phi}\}$.

Moreover, if we consider any maximally entangled state of dimensionality $N$ 
of at least three:
\beq
\psi=\sum^{N}_{n=1} \phi_{n} \otimes \psi_{n}
\mbox{,}
\label{eq:shelly}
\eeq
then we can evidently develop a nonlocality proof for this 
state by using the Kochen and Specker
theorem. To see this, we re-write \ref{eq:shelly}
as 
\beq
\psi=\phi_{1} \otimes \psi_{1}
+ \phi_{2} \otimes \psi_{2} + \phi_{3} \otimes \psi_{3} +
\sum^{N}_{n=4}  \psi_{n} \otimes \psi_{n}
\label{eq:MEKSgen}
\mbox{,}
\eeq  
and we define ${\cal H}_{2}$ to be the subspace
of particle $2$'s Hilbert space spanned by the vectors
$\psi_{1},\psi_{2},\psi_{3}$, and the operator $P$ as the 
projection operator of ${\cal H}_{2}$. Let us define the set of observables 
$\{\zeta_{\theta,\phi}\}$ such
that they are formally identical to the squares of the spin components 
$\{\zeta_{\theta,\phi}\}$ on ${\cal H}_{2}$ and give zero
when operating on any vector in its orthogonal complement. Then the 
Kochen and Specker
theorem tells us that the existence of definite values for the set
$\{P,\{\zeta^{2}_{\theta,\phi}\}\}$
must conflict with quantum mechanics. To see this, consider
those joint-eigenvalues of $\{P,\{\zeta_{\theta,\phi}\}\}$
for which $P=1$. The corresponding values of the set 
$\{\zeta_{\theta,\phi}\}$ in this case are equal to
joint-eigenvalues on the spin observables themselves. Since
the \ks theorem implies the impossibility of an assignment of values
to the spin components, then the same follows for the observables
$\{\zeta_{\theta,\phi}\}$ given that $P=1$.

If we consider the conjunction of \Erwins paradox
with either Gleason's theorem or Mermin's
theorem, we again find that quantum nonlocality 
follows. From Gleason's theorem we have 
that a value map on the set of all projections $\{P\}$ must 
conflict with quantum mechanics.
One can show that for any maximally entangled state 
\beq
\sum^{N}_{n=1}\phi^{(1)}_{n}\otimes\psi^{(2)}_{n}
\mbox{,}
\label{eq:thcase}
\eeq
where $N$ is at least three, we have quantum nonlocality.
This result also holds true for for maximally entangled states of 
infinite-dimensionality, in which case the sum in \ref{eq:thcase} is
replaced by an infinite sum. In the case of Mermin's theorem, we can 
show that for any maximally entangled state whose
subsystems are at least four dimensional, the quantum predictions
must entail nonlocality. 

Thus, the nonlocality of the maximally entangled 
states may be proved not only by the analyses involving Bell's theorem 
(see section \ref{Bellext}),  but also by consideration of any of the 
theorems of Gleason, Kochen and Specker, or Mermin, taken together 
with the \Erwin paradox. We have seen that the latter type of argument,
which we have called ``\Erwin nonlocality'', differs from that 
involving Bell's theorem in several ways.
First, the \Erwin nonlocality it is of a deterministic, rather than statistical
character. This is due to the fact that the conflict of the \Erwin 
incompleteness with quantum mechanics is in 
terms of the quantum prediction for individual measurements, i.e., 
the prediction that measurements of a commuting set always give one of 
that set's joint-eigenvalues. Second, such quantum nonlocality proofs can be 
developed from these theorem's implications regarding the observables 
of just {\em one} subsystem. Finally, the
\Erwin nonlocality can be proved for a larger class of observables 
than can the EPR/Bell nonlocality. 
Note that beyond the arguments presented here, there remains the possibility 
of further instances of 
\Erwin nonlocality, since any `spectral incompatibility theorem' (see section 
\ref{Spec})  leads to such a demonstration. If the theorem in question 
concerns a class of observables on an $N$-dimensional Hilbert space, 
then the proof can be applied to any maximally 
entangled states whose subsystem's are at least $N$-dimensional.

\subsection{Discussion of the experimental tests of EPR/Bell
and \Erwin nonlocality} 
In chapter three, we reviewed the demonstration that the spin singlet 
version of the EPR paradox in conjunction with Bell's theorem 
leads to the conclusion of quantum nonlocality. That quantum theory entails 
such
an unusual feature as nonlocality invited physicists to perform laboratory 
experiments to test whether the quantum predictions for the relevant
phenomena are actually borne out. The 
experimental tests\footnote{See for example, Freedman and 
Clauser \cite{They is us}, Fry and Thompson \cite{Fry}, and Aspect, 
et. al. \cite{Aspect 1, Aspect 2, Aspect 3}. For the two-photon 
systems studied by these authors, an analysis first 
given by Clauser, Horne, Holt and Shimony \cite{CHHS} plays the role of Bell's 
theorem: the assumption of locality together with the existence of the
perfect correlations in the photon polarizations leads to the CHHS 
inequality, which relationship is not generally satisfied by
the quantum mechanical predictions.} 
inspired by the discovery of quantum nonlocality have 
focused mainly\footnote{The experiment of Lamehi-Rachti, and 
Mittig \cite{singlet experiment} involves a pair of protons
described by the spin singlet state. These results are in support of 
the quantum predictions.} 
on a maximally entangled system of two {\em photons} rather 
than two spin $\frac{1}{2}$ particles. The results of these 
experiments are in {\em agreement} with the quantum mechanical 
predictions\footnote{For a discussion of the
experimental results, see following: \cite{Bell experiments, Bell cascade
photons, Baby Snakes}.}. Since we have described the spin singlet version of 
the EPR/Bell nonlocality in some detail (chapter 3), we address this case
rather than the two-photon maximally entangled state. 

Let us briefly recall how the EPR argument and Bell's theorem
give rise to the conclusion of quantum nonlocality.
According to the EPR argument, the perfect correlations 
exhibited by the spin singlet state can be
explained under locality {\em only} if there exist definite values
for all components of the spins of both particles. Bell's theorem
shows that any such definite values lead to the prediction that the
`correlation function' $P(\hat{a},\hat{b})$ must satisfy Bell's
inequality. Thus, if we combine the EPR 
paradox with Bell's theorem we obtain the argument that any local theoretical 
explanation of the perfect correlations must give a prediction for
$P(\hat{a},\hat{b})$ that satisfies the Bell inequality. On the 
other hand, the quantum mechanical predictions for $P(\hat{a},\hat{b})$ 
{\em violate} this inequality. Thus, any local theoretical
description of the spin singlet state 
must conflict with the quantum mechanical one. 

We now consider what such a laboratory test of quantum nonlocality would 
entail. The EPR/Bell analysis depends on the correctness of the
quantum mechanical predictions of perfect correlations.  
If these quantum predictions are {\em not} confirmed, i.e., 
the experimental test does not reveal perfect 
correlations, one could not interpret the experiment in terms of
EPR/Bell. It is only with the observation of the perfect correlations that
further tests can be performed to make an experimental judgment between
quantum mechanics and the family of local theories.
Such a judgment can be made based on the appropriate
measurements of the correlation function $P$, and the comparison of
the results 
with the Bell inequality. One must measure the
correlation function for just those angles 
$\hat{a},\hat{b},\hat{c}$ for which the quantum mechanical prediction 
violates the Bell inequality (see section \ref{eprbellnon}). 
If the results of these tests {\em agree} with the Bell inequality,  
then the quantum predictions are refuted, and 
the experiment may be interpreted in terms of a theory 
consistent with the concept of locality.
If the results disagree with the Bell inequality, this 
would confirm the quantum mechanical predictions and would 
support the notion of nonlocality. 

\subsubsection{Test of \Erwin nonlocality by perfect correlations only}
If one examines the various proofs of quantum nonlocality, one finds
that all have a somewhat similar structure. The first part of such a proof 
consists of an 
EPR-like incompleteness argument, according to which the perfect correlations 
exhibited by some system together with the assumption of locality are 
shown to imply the existence definite values for certain observables. The 
second part of a nonlocality proof is a demonstration
that such definite values must conflict with the 
predictions of quantum mechanics. The two parts can be 
combined to produce a proof of quantum nonlocality. When we come to 
consider the 
laboratory confirmation of such quantum nonlocality, it is natural to 
expect that the two part structure of its proof will
be reflected in the various stages of the experiment itself. As we have 
seen above, this is indeed true for the test of the EPR/Bell nonlocality:
one must observe the perfect correlations, {\em and} the violation of 
the Bell inequality to verify that the spin singlet state reflects 
quantum nonlocality. When we examine the experimental test of what we 
have called `Schr{\"o}dinger nonlocality', however, we find that it is 
possible to perform the experiment in such a way that the perfect 
correlations {\em by themselves} are sufficient to verify that the 
system being studied shows nonlocal effects of this sort.
We now turn to the question of such an experimental test. 

For the sake of definiteness, let us consider a particular example of 
Schr{\"o}dinger nonlocality, namely that which arises from the
conjunction of Schr{\"o}dinger's paradox with Mermin's theorem
(see section \ref{Mermin}). This proof is constructed as follows. 
 The
Schr{\"o}dinger paradox incompletness argument implies that the 
perfect correlations exhibited by the maximally entangled state
lead to the existence of a non-contextual value map
on all observables of both subsystems. If the subsystems of the state in 
question are associated with Hilbert spaces that are at least $4$ 
dimensional (see section \ref{Meatus}), then there must exist 
such a map on the Mermin observables (or a set of formally equivalent 
observables). Mermin's theorem, however, contradicts the possibility of 
such a map. If we {\em combine} this theorem with
the Schr{\"o}dinger incompleteness argument the result is a quantum
nonlocality proof.

For the reader's convenience, we recall Mermin's theorem here. The observables 
addressed by this theorem are the $x$ and
$y$ components of two spin $\frac{1}{2}$ particles, 
and the observables $A,B,C,X,Y,Z$ which are defined in terms of these 
through the relations
\begin{eqnarray}
A & = & \sigma^{(\alpha )}_{x} \sigma^{(\beta )}_{y} 
\label{eq:Pducto} \\
B & = & \sigma^{(\alpha )}_{y} \sigma^{(\beta )}_{x} \nonumber \\ 
X & = & \sigma^{(\alpha )}_{x} \sigma^{(\beta )}_{x} \nonumber \\
Y & = & \sigma^{(\alpha )}_{y} \sigma^{(\beta )}_{y} \nonumber 
\mbox{,}
\end{eqnarray}
and 
\begin{eqnarray}
C & = & AB \label{eq:SPducto}  \\
Z & = & XY \nonumber
\mbox{.}
\end{eqnarray}
Note that the symbols $\sigma^{(\alpha )}$ and $\sigma^{(\beta )}$ are used 
here\footnote{In the present section, we consider the 
entire set
of Mermin observables as being associated with {\em one} subsystem of
a maximally entangled state. Use of the notation $\sigma^{(1)}$,
$\sigma^{(2)}$ suggests observables belonging to separate subsystems,
and might have led to confusion.},
 rather than
$\sigma^{(1)}$ and $\sigma^{(2)}$, which we employed in the 
presentation of Mermin's theorem given in section \ref{Mermin}.
We will refer to these observables as the `Mermin observables', and
the notation $M_{i}\mbox{ }i=1 \ldots 10$ shall refer to 
an arbitrary member of the set.
In the discussion of Mermin's theorem given in section
\ref{Mermin}, we showed that the commutation relationships among the
spin components led to the relation
\beq
\sigma^{(\alpha )}_{x}\sigma^{(\beta )}_{y}\sigma^{(\alpha )}_{y}
\sigma^{(\beta )}_{x}
\sigma^{(\alpha )}_{x}\sigma^{(\beta )}_{x}\sigma^{(\alpha )}_{y}
\sigma^{(\beta )}_{y}=-1
\label{eq:PMo}
\mbox{.}
\eeq
From this and the definitions \ref{eq:Pducto} and \ref{eq:SPducto}, it is 
easy to see $C$ and $Z$ satisfy
\beq
CZ=-1
\mbox{.}
\label{eq:CZducto}
\eeq
Mermin's theorem implies that there exists no function
$E(O)$ on the observables $M_{i}$ that satisfies the all
relationships constraining the commuting observables. 

On inspection of the first equation in \ref{eq:Pducto}, we see that
the three observables
involved are a commuting set: 
$[\sigma^{(\alpha )}_{x},\sigma^{(\beta )}_{y}]=0$, $[\sigma^{(\alpha )}_{x},
A]=[\sigma^{(\alpha )}_{x},\sigma^{(\alpha )}_{x}\sigma^{(\beta )}_{y}]=0$ and 
$[\sigma^{(\beta )}_{y},A]=[\sigma^{(\beta )}_{y},\sigma^{(\alpha
)}_{x}\sigma^{(\beta
)}_{y}]=0$. 
Examination of
the other equations in \ref{eq:Pducto} shows that 
the same holds true for 
these, i.e. the observables in each form a commuting set. Repeated
application of the commutation rules may be used to show that the sets 
$\{C,A,B\}$, $\{Z,X,Y\}$, and $\{C,Z\}$ are also commuting sets. 
For convenience, we list these sets here:
\begin{eqnarray}
\{A, \sigma^{(\alpha )}_{x}, \sigma^{(\beta )}_{y}\} 
& \{B, \sigma^{(\alpha )}_{y}, \sigma^{(\beta )}_{x}\} 
& \{X, \sigma^{(\alpha )}_{x}, \sigma^{(\beta )}_{x}\} 
\label{eq:Johnlennon} \\
\{Y, \sigma^{(\alpha )}_{y}, \sigma^{(\beta )}_{y}\} & \{C,A,B\} 
& \{Z,X,Y\} \nonumber \\
\{C,Z\} &  & \nonumber
\end{eqnarray}
The relationships 
which Mermin's theorem requires the function $E(O)$ to satisfy are
the defining equations in \ref{eq:Pducto}, \ref{eq:SPducto} and
\ref{eq:CZducto}.

We now recall the Schr{\"o}dinger incompleteness 
argument in detail. In this argument, one considers the possibility
of separate experimental procedures being performed
to measure observables of the two subsystems of any given
maximally entangled state. Suppose that one measures
$A$ of subsystem $2$ and $\tilde{A}$ of subsystem $1$, through 
experimental procedures ${\cal E}(A)$, and ${\cal E}(\tilde{A})$.
The perfect correlation
between these observables implies that from the value of $A$ found in
any procedure ${\cal E}(A)$, we can predict with certainty
the value of $\tilde{A}$ found in any procedure ${\cal E}(\tilde{A})$. 
With this and the assumption of locality, 
we must conclude that $\tilde{A}$ possesses a definite value 
$V(\tilde{A})$, which cannot depend on its measurement context. 
 The 
symmetry of the system allows one to argue in a similar fashion to 
show that $A$ also possesses a non-contextual value $V(A)$. The
invariance and symmetry
of the maximally entangled state then imply that such an argument can 
be performed to show that {\em all} observables of subsystems $1$ 
and $2$ must possess definite values.

To confirm the `Schr{\"o}dinger-Mermin' nonlocality, one must perform
perfect correlation tests in such a way that---in light of 
the above argument---it follows that all of the Mermin observables 
possess non-contextual values. Let us examine the 
following series of experiments. For each of the Mermin observables, 
perform the following series of experiments. First, we simultaneously
peform ${\cal E}(M_{i})$ on subsystem $2$ and ${\cal
E}(\widetilde{M_{i}})$ on subsystem $1$ (The precise experiments
required for each $M_{i}$ are to be defined below.). 
Second, we simultaneously perform ${\cal E}^{'}(M_{i})$
on subsystem $2$ and ${\cal E}(\widetilde{M_{i}})$. If the
perfect correlations are verified in both cases and for
all Mermin observables then it follows that all of the
Mermin observables of subsystem $2$ must have definite
values, and that these must be noncontextual. The latter follows
since the perfect correlations are observed under conditions
where the context of $M_{i}$ is varied (i.e., from ${\cal E}(M_{i})$
to ${\cal E}^{'}(M_{i})$) while that of $\widetilde{M_{i}}$
is not.

In this way, one performs that part of the
experiment which corresponds to the incompleteness argument of the nonlocality
proof. One might suppose that what is required next is the performance 
of a separate series of tests to judge between the existence of the
definite values implied by the perfect correlations, and the quantum 
mechanical predictions. As we saw above, Bell's inequality provides
for the empirical difference between these two approaches in the case
of EPR/Bell nonlocality. In the present case, we have that Mermin's theorem 
provides such a difference: the noncontextual values must {\em fail} to
satisfy the relationships \ref{eq:Pducto}, \ref{eq:SPducto}, and 
\ref{eq:CZducto}, while quantum mechanics predicts that such 
relationships {\em are} satisfied. To confirm the quantum mechanical 
predictions and the presence of nonlocality in the maximally 
entangled state in question, it would appear that one must check
whether or not these relationships are satisfied. However, as we 
shall see, if the measurements involved in the perfect correlations 
test are done in a particular way, then {\em the existence of
perfect correlations is sufficient in itself} to prove such a 
conclusion. We now demonstrate this.

Consider the measurement of the commuting set 
$\{A,\sigma^{(\alpha )}_{x},\sigma^{(\beta )}_{y}\}$.
These observables are related by the first equation in 
\ref{eq:Pducto}, which we repeat here for convenience:
\beq
A=\sigma^{(\alpha )}_{x}\sigma^{(\beta )}_{y}
\label{eq:PductoFir}
\mbox{.}
\eeq
Suppose that to measure the set in question, we {\em first}
measure the set $\{\sigma^{(\alpha )}_{x},\sigma^{(\beta )}_{y}\}$. 
Then the values for $\sigma^{(\alpha )}_{x}$ and $\sigma^{(\beta )}_{y}$ 
so obtained are simply {\em multiplied together}
to determine the value of $A$. One can, of course measure any 
commuting set that obeys a constraining relationship in this way, 
i.e., any commuting set $\{O,O_{1},O_{2},\ldots\}$ where
\beq
O=f(O_{1},O_{2},\ldots)
\mbox{,}
\eeq
can be measured by first performing an experiment to measure 
$\{O_{1},O_{2},\ldots\}$, and then evaluating $f$ of the resulting
values obtained, to obtain the value of $O$. If the set 
$\{A,\sigma^{(\alpha )}_{x},\sigma^{(\beta )}_{y}\}$ is measured in such
a way then it is {\em a~priori} that the relationship \ref{eq:PductoFir} 
will be satisfied by the measurement results, since this relationship is 
``built into'' the very 
procedure itself, i.e., in such a procedure, we {\em use} \ref{eq:PductoFir} 
in determining these results. 

Now let us suppose that in all of procedures ${\cal E}(M_{i})$
involved in the perfect correlation tests 
discussed above, the above described method is followed. That it is 
possible to use this method in each case is clear from the fact that every 
commuting set
in \ref{eq:Johnlennon} is constrained by one of the relationships
of \ref{eq:Pducto}, \ref{eq:SPducto}, and \ref{eq:CZducto}. If we follow
such a method to measure the perfect correlations, then it
is {\em a~priori} that the measurement results will obey the 
commuting relationships \ref{eq:Pducto}, \ref{eq:SPducto}, and 
\ref{eq:CZducto}. Since, as we have mentioned, it is the judgment of whether
or not these relationships are obeyed which is required to
complete our laboratory test of nonlocality, it follows that the 
perfect correlation tests by themselves are sufficient for this test.

If, in fact, the perfect correlation test described above gives a
`positive' result, then the conclusion of quantum nonlocality 
necessarily follows. The only {\em local} theoretical interpretation of 
such a result---the existence of definite values---is immediately 
ruled out since (through Mermin's theorem) it follows that such values
cannot satisfy the commuting relationships, which are {\em a~priori}
obeyed when one performs the commuting set measurements in the fashion
just discussed.

Moreover, the \Erwin nonlocality that follows from either of the
other theorems studied in chapter 2 (Gleason's, and Kochen and Specker's) 
can also be tested through the confirmation of perfect correlations. This 
follows since every observable among the set addressed by each 
theorem is a 
member of two `incompatible commuting sets', where each set 
obeys a relationship of the form $O=f(O_{1},O_{2},\ldots)$.
Thus, each observable can be 
measured by two distinct procedures ${\cal E}(O_{i})$ and 
${\cal E}^{'}(O_{i})$ for
which the commuting relationships are {\em a~priori}. 

\section{\Erwins paradox and von Neumann's no hidden variables argument}
Lastly, we would like to make a few observations which are of  
historical interest. We now focus on the line of thought \Erwin followed 
subsequent to his 
generalization of the Einstein--Podolsky--Rosen paradox. In addition, 
we consider the question of the possible consequences had he 
repeated the mistake made by a well-known contemporary, or 
had he anticipated any of several mathematical theorems that were developed 
somewhat later.

Having concluded the existence of definite 
values on the observables, \Erwin considered the question of the type 
of relationships which might govern these values.
As we discussed in section \ref{Schrvon}, he was able to show that no 
such values can obey the same relationships that constrain the 
observables themselves. In particular, he observed that the relationship
\beq H=p^{2}+a^{2}q^{2} 
\mbox{,}
\label{eq:HrO2} 
\eeq
is not generally obeyed by the eigenvalues of the observables 
$H,p^{2},q^{2}$, so that no value map $V(O)$ can satisfy this 
equation. From this it 
follows\footnote{To show this, one need only note that the value of any 
observable
$f(O)$ will be $f$ of the value of $O$, where $f$ is any mathematical 
function (see section \ref{Schrvon}).} 
immediately that there exists 
no value map which is {\em linear} on the 
observables. 
Thus we see that \Erwins argument essentially leads to 
the same conclusion regarding hidden variables as von Neumann's 
theorem. \Erwin did not, however, consider this as proof of the impossibility 
of hidden variables.

Instead, the fact that the definite values of 
his EPR generalization must fail to obey such 
relationships prompted \Erwin to consider the possibility that  
{\em no} relationship whatsoever serves to constrain them: 
\cite{Present Sit} (emphasis due to original author) 
``Should one now think that because we are so 
ignorant about the relations among the variable-values held ready in
{\em one} system, that none exists, that far-ranging arbitrary 
combination can occur?'' Note, however, that if the `variable-values' obey 
no constraining relationship, each must be an independent 
parameter of the system. Thus, \Erwin continues with the statement: ``That 
would mean that a system of `{\em one} degree of freedom' would need 
not merely {\em two} numbers for adequately describing it, as in 
classical mechanics, but rather many more, perhaps infinitely many.''

Recall the discussion of the hidden variables theory known as Bohmian 
mechanics presented in section 
\ref{UncleAlb}. There we saw that the state description of a system is 
given in this theory by the wave function $\psi$ and the system 
configuration ${\bf q}$. The mathematical form of $\psi$
is the same as in the quantum formalism, i.e., $\psi$ is a 
vector in the Hilbert space associated with the system. Consider a spinless 
particle constrained to move in $1$ dimension. The
Bohmian mechanics state description would consist of the  
wave function $\psi(x) \in L_{2}$ and position $x \in \R$. Since 
any $L_{2}$ 
function $\psi(x)$ is infinite dimensional, i.e. it is an 
assignment of numbers to the points $x \in (-\infty ,\infty )$, the 
Bohmian mechanics state description is infinite-dimensional.
Thus, a theory of hidden variables such as Bohmian mechanics provides just 
the type of description that \Erwins 
speculations had led him to conclude.

Let us now suppose that rather than reasoning as he did, \Erwin 
instead committed the same error as von Neumann. In other words, we 
consider the possible consequences had \Erwin regarded the failure of
the definite values to satisfy the same relations as the observables 
as proof that no such values can possibly agree with quantum mechanics. 
This false result would appear to refute the conclusion of \Erwins 
generalization of EPR; it would seem to imply that the \Erwin paradox 
leads to a conflict with the quantum theory. The \Erwin paradox, like the EPR 
paradox, assumes only that the perfect correlations can 
be explained in terms of a local theoretical model. Hence, if \Erwin 
had concluded with von Neumann that hidden variables must conflict 
with quantum mechanics, then he would have been led to deduce quantum 
nonlocality. 

As we saw above, such a conclusion actually follows from 
the combination of \Erwins paradox with any of the spectral 
incompatibility theorems, for example, those of Gleason, Kochen and
Specker, and 
Mermin. Since Gleason's theorem involves the proof of the trace 
relation \ref{eq:Gltracedude}, it seems reasonable to regard it 
as being the most similar of these theorems to that of von Neumann, which 
features a derivation of this same formula\footnote{Von Neumann's 
derivation is based on quite different assumptions, as we saw.}.
Note that to whatever degree one might regard Gleason's theorem as being 
similar to von 
Neumann's, one must regard \Erwin as having come to within that same 
degree of a proof of quantum nonlocality. 
  
\section{Summary and conclusions}
Our investigation of the hidden variables issue
was motivated by several mathematical results that have been interpreted
as proofs either of the incompatibility of hidden variables with
the quantum theory, or of the existence of serious limitations
on such theories. We have reviewed the arguments first presented
by J.S. Bell, according to which not only do these theorems fail to demonstrate
the impossibility of hidden variables, but the restrictions they place on such 
theories---contextuality and nonlocality---are quite similar to particular 
features of the quantum theory itself. When considering the theorems of \gl 
and of Kochen and Specker, we 
found that they imply essentially that within any hidden variables 
theory, the way in which values are assigned each observable $O$ 
must allow for contextuality---an attribute reflecting
the quantum formalism's rules of measurement.
Nonlocality is certainly surprising and unexpected, yet it has been 
proved as a feature intrinsic to quantum mechanics from the combination 
of Bell's
theorem with the spin singlet version of EPR. 
We found also that from Erwin \Erwins extensive generalization of
the EPR paradox, it follows that every spectral-incompatibility 
theorem (such as the theorems of Gleason, Kochen and Specker, and 
Mermin) will give 
a new proof of ``nonlocality without inequalities.''

We began our exposition by discussing the earliest work on hidden variables,  
John von Neumann's. Von Neumann considered the possibility of a theory whose 
state description would supplement that of the quantum formalism with a
parameter we called $\lda$. He regarded this scheme as a way to introduce 
determinism into the quantum phenomena. 
Mathematically, such determinism is represented by requiring that for each 
$\psi$ and $\lda$, i.e., for 
each state, there exists a function assigning to each observable
its value. Von Neumann showed that no function $E(O)$ on the observables 
satisfying his assumptions can be such a
value map. From this, he concluded that empirical agreement between any
hidden variables theory and the quantum theory is impossible. 
That this conclusion is unjustified
follows since one of von Neumann's assumptions---that
$E(O)$ be {\em linear} on the observables---is quite unreasonable.
There is no basis for the demand that $E(O)$ obey $O=X+Y$ where 
$[X,Y] \neq 0$, since each of these observables is measured by a 
{\em distinct} procedure.

The theorems of \gl, \ks, and Mermin seem at first to 
succeed where von Neumann failed, i.e., to prove
the impossibility of hidden variables. Nevertheless, as Bell has shown 
\cite{Bell Eclipse, Bell Imposs Pilot}, these theorems also fail as 
arguments against hidden variables, since they do not account for 
contextuality. This concept is easily illustrated by 
examination of the quantum formalism's rules of measurement. We find
that the `measurement of an observable $O$' can be performed using
distinct experimental procedures ${\cal E}(O)$ and ${\cal E}^{'}(O)$.
That ${\cal E}(O)$ and ${\cal E}^{'}$ are distinct is especially obvious if 
these measure the commuting sets ${\cal C}$ and ${\cal C}^{'}$ 
where ${\cal C}$ and ${\cal C}^{'}$ both contain $O$, but the members of 
${\cal C}$ fail to commute with those of ${\cal C}^{'}$. 
It is therefore quite reasonable to expect that a hidden variables 
theory should allow for the possibility that different procedures for 
some observable's measurement might yield different results for an 
individual system. That 
it is necessary to account for the detailed experimental arrangement 
recalls the views of Niels Bohr, who warns us
of \cite[page 210]{Einstein impeachment} ``the 
impossibility of any sharp separation between the behavior of atomic objects 
and the interaction with the measuring instruments which serve to define the 
conditions under which the phenomena appear.''

Examples of just such 
incompatible commuting sets are found among the observables in
each of the theorems we addressed in chapter 2: Gleason's, Kochen and 
Specker's, and Mermin's. For example, in the theorem of Kochen
and Specker, the commuting sets are simply of the form 
$\{s^{2}_{x},s^{2}_{y},s^{2}_{z}\}$, i.e., the squares of the 
spin components of a spin $1$ particle taken with respect to
some Cartesian axis system $x,y,z$. Here one can see that a given
observable $s^{2}_{x}$ belongs to both $\{s^{2}_{x},s^{2}_{y},
s^{2}_{z}\}$, and $\{s^{2}_{x},s^{2}_{y^{'}},s^{2}_{z^{'}}\}$, where
the $y^{'}$ and $z^{`}$ axes are oblique relative to the $y$, and $z$ axes.
Since the theorems of Gleason, Kochen and Specker, and Mermin consider
a function $E(O)$ which {\em assigns a single value to each 
observable}, they cannot account for the possibility of 
incompatible measurement procedures which the 
quantum formalism's rules of measurement allows us. Clearly, the 
approach taken by these theorems falls far short of addressing the 
hidden variables issue properly. Thus, we 
come to concur with J.S. Bell's assessment that \cite{Bell 
Imposs Pilot} ``What is proved by impossibility proofs 
\ldots  \ldots is lack of imagination.'' 

To address the general question of hidden variables, one must 
allow for this important feature of contextuality. The theorems of 
Gleason, Kochen and Specker, and Mermin may be
seen as explicit proofs that this simple and natural feature is
necessary in any hidden variables theory. In particular, any attempt to 
construct a value map that neglects this feature will fail in that
it cannot satisfy the requirement that the relationships  
constraining the commuting sets must be obeyed. We have seen that
there is a simpler way to express this result: there exists no value map 
on the observables mapping every commuting set to one of its joint-eigenvalues.

We were able to gain some measure of additional insight into
contextuality in examining Albert's example.
As one can see from the experiments considered by Albert, the Hermitian 
operators cannot
be considered as representing properties intrinsic to the quantum system
itself. Instead, the results of the ``measurement of a quantum observable''
must be considered as the joint-product of system {\em and} measuring
apparatus. 

When we come to consider Bell's theorem, we must do so in the context
of the spin singlet version of the Einstein--Podolsky--Rosen paradox.
This argument essentially shows that locality necessarily leads to
the existnce of definite values for all components of the spin of both
particles of the spin singlet state. 
Since the quantum mechanical description of the
state does not account for such values, EPR
conclude that this description is incomplete.
Bell's theorem essentially continues where the spin singlet EPR analysis 
concluded. The fixed values for the spin components of the two particles are 
represented by
functions $A(\lda,\hat{a})$, $B(\lda,\hat{b})$ 
where $\hat{a}$, $\hat{b}$ are unit vectors in the directions of the
axis of the spin component for particle $1$ and $2$ respectively.
In his analysis, Bell considers the statistical correlation between spin
component measuring experiments carried out on the two particles. 
According to Bell's theorem, the theoretical prediction
for this correlation derivable from the variables $A(\lda,\hat{a})$
and $B(\lda,\hat{b})$ must satisfy `Bell's inequality'. The prediction
given by quantum mechanics does {\em not} generally agree with this
inequality. Bell's theorem in itself provides a proof
that local hidden variables must conflict with quantum mechanics. 

It is important to note that what might {\em appear} to be Bell's 
assumption---the existence of definite values for all spin components---is
identical to the conclusion of the spin singlet version of the 
Einstein--Podolsky--Rosen argument. In fact, Bell assumes nothing beyond
what follows from EPR.  Therefore the proper way to assess the implications
of these arguments is to combine them into
a single analysis that will begin with the assumptions of the
spin singlet EPR paradox, and end with the conclusion of Bell's 
theorem; i.e., the assumption of locality leads to the conclusion of Bell's 
inequality.
Since this inequality, as we saw, disagrees with
the quantum theory, we finally have an argument that {\em locality
implies disagreement with the predictions of quantum mechanics}. 
This agrees with Bell's own assertion in the matter: \cite{Cosmologists} 
``It now seems that the non-locality is deeply rooted in
quantum mechanics itself and will persist in any completion.''   

We have now come to understand the issues of contextuality and nonlocality 
as features of a hidden variables interpretation of quantum theory.
Contextuality in hidden variables is a 
natural feature to expect since it reflects the possibility
of distinct experimental procedures for measurement of a single 
observable. As J.S. Bell 
expresses it: \cite{Bell Eclipse} ``The result of an observation may
reasonably 
depend not only on the state of the system (including hidden 
variables) but also on the complete disposition of the apparatus.'' 
According to the above results, the fact that nonlocality is required of 
hidden variables does not in any way diminish the prospect of these types 
of theories.  As we have 
mentioned, there has existed a successful theory of hidden variables since 
1952, and there is no reason not to consider this theory (Bohmian mechanics) 
as a serious interpretation of quantum
mechanics. We noted in section \ref{Bohm} that Bohmian mechanics possesses the 
advantages of objectivity and determinism.

In the fourth and final chapter, we addressed Erwin \Erwins 
generalization of the EPR paradox. Besides greatly extending
the incompleteness argument of EPR, \Erwins analysis provides for
a new set of ``nonlocality without inequalities'' proofs, which 
have several important features. The \Erwin paradox
concerns the perfect correlations exhibited not just by a 
single quantum state, but for a general class of states  
called the maximally entangled states. A maximally entangled state is
any state of the form 
\beq
\sum^{N}_{n=1}|\phi_{n}\ra\otimes|\psi_{n}\ra
\mbox{,}
\eeq
 where
$\{|\phi_{n}\ra\}$ and $\{|\psi_{n}\ra\}$ are bases of the 
($N$-dimensional) Hilbert
spaces of subsystems $1$ and $2$, respectively. As in the
EPR analysis (both the spin singlet and the original version) \Erwin
gives an incompleteness argument, according to which there must
exist precise values for all observables of both subsystems. 
It may be shown using Gleason's theorem, Kochen and Specker's theorem, 
Mermin's theorem, or any other `spectral incompatibility' proof, that 
the definite values concluded in the \Erwins paradox
must {\em conflict} with the empirical predictions of quantum 
mechanics. The implication of such a disagreement is quantum 
nonlocality. This conflict in empirical predictions differs from that 
developed within Bell's theorem, in that it involves predictions 
for individual measurements, rather than the statistics of a 
series of measurements. Thus, combining \Erwins paradox with any 
spectral incompatibility theorem provides a `nonlocality without 
inequalities' proof. Moreover, as we 
observed in section \ref{Meatus}, the conflict between the \Erwin 
incompleteness and such a theorem exists even 
when one considers only the observables of one of the two subsystems. 
We have 
referred to this type of proof by the name ``\Erwin nonlocality.''

When we consider the experimental verification of \Erwin nonlocality, 
we find a curious result. The measurement of a commuting 
set satisfying an equation of constraint may be carried out
in such a way that the set of values obtained will 
satisfy this constraint {\em a~priori}. Since the definite values 
concluded in 
the \Erwin paradox cannot satisfy these constraining relationships, performing
an experimental test using such measurement procedures provides that the 
perfect correlations {\em themselves} are sufficient to imply nonlocality.   

All this leads us to inquire how \Erwin himself regarded his results, and
what further conclusions he drew within his remarkable paper. Clearly, 
it would have been possible for him to argue for quantum nonlocality
had he anticipated the results of any of a wide variety of theorems including 
at least Bell's (see section \ref{Bellext}), Gleason's, Kochen and Specker's, 
Mermin's, or any other spectral incompatibility theorem. 
Instead, as we saw in chapter 1, \Erwin essentially reproduced the
von Neumann argument against hidden variables, in his observation of
a set of observables that obey a linear relationship not satisfied by the 
set's eigenvalues. What may be seen from von Neumann's result is just what
\Erwin noted---relationships constraining the observables do 
not necessarily constrain their values. \Erwin continued this line of 
thought by speculating on the case for which {\em no} relation 
whatsoever constrained the values of the various observables. In light of 
his generalization 
of the EPR paradox, this line of thought led \Erwin to the idea that the
quantum system in question might possess an infinite number of degrees of 
freedom, which concept is actually quite similar to that of
Bohmian mechanics. 

If, on the other hand, \Erwin {\em had} made von 
Neumann's error, i.e., had concluded the impossibility of a map from 
observables to values, this mistaken line of reasoning would have 
permitted him to arrive at the concept of quantum nonlocality. Thus,
insofar as one might regard the von Neumann proof as ``almost'' 
leading to the type of conclusion that 
follows from Gleason's theorem, one must consider \Erwin as having 
come precisely that close to a proof of quantum nonlocality.

It is quite interesting to see that many of the issues related to 
quantum mechanical incompleteness and hidden variables were addressed 
in the 1935 work \cite{Present Sit, Camb1, Camb2} of Erwin
\Erwin. \Erwins work seems to be the most far 
reaching of the early analyses addressed to the subject.
Not only did he see deeper into the problem than 
did von Neumann, but \Erwin developed results beyond those
of the Einstein, Podolsky, Rosen paper---an extension of their 
incompleteness argument, and an analysis of incompleteness in terms of the 
implication of von Neumann's theorem.
It seems clear that the field of foundations of 
quantum mechanics 
might have been greatly advanced had these features
of \Erwins paper 
been more widely appreciated at the time it was first published.

 \end{document}